\documentclass[twocolumn,superscriptaddress,showpacs]{revtex4}
 \usepackage{graphicx}

\usepackage{bm,color}

\newcommand {\be}{\begin{equation}}
\newcommand {\ee}{\end{equation}}

\begin{document}

\title{Phase transitions in self-gravitating
systems and  bacterial populations \\ with a screened attractive potential}
%\title{Chemotaxis and self-gravitating systems}
% First author block:
\author{P.H. Chavanis and L. Delfini}
%\email{chavanis@irsamc.ups-tlse.fr}
% \homepage{An author's web page; optional}
\affiliation{Laboratoire de Physique Th\'eorique (IRSAMC), CNRS and UPS, Universit\'e de Toulouse, F-31062 Toulouse, France}
% You may list several affiliation, using separate commands for each:
%\affiliation{The third affiliation is shared by both co-authors}

% For other authors please repeat the author block as needed
%\author{Second Author}
% Note how REVTeX 4 deals with identical affiliations
%\affiliation{The third affiliation is shared by both co-authors}

\begin{abstract}

We consider a system of particles interacting via a screened Newtonian potential and study phase transitions between homogeneous and inhomogeneous states in the microcanonical and canonical ensembles.  Like for
other systems with long-range interactions, we obtain a great
diversity of microcanonical and canonical phase transitions depending on the dimension of space and on the importance of the screening length.  We also consider a system of particles in Newtonian interaction in the presence of a ``neutralizing background''. By a proper interpretation of the parameters, our study describes (i) self-gravitating systems in a cosmological setting, and (ii) chemotaxis of bacterial populations in the original Keller-Segel model.
\end{abstract}

% Insert suggested PACS numbers (up to 4) in braces.
% The PACS (Physics and Astronomy Classification Scheme)
% can be accessed on the web at http://www.aip.org/pacs/
\pacs{47.10.A-,47.15.ki}

% Insert keywords (up to about 4) in braces; optional.
% \keywords{Up to four keywords}

\maketitle

% Here the text of your article begins

\section{Introduction}

Many biological species like bacteria, amoebae, endothelial cells, or
even ants interact through the phenomenon of chemotaxis
\cite{murray}. These organisms secrete a chemical substance (like a
pheromone) that has an attractive (or sometimes repulsive) action on
the organisms themselves. This phenomenon is responsible for the
self-organization and morphogenesis of many biological species. It has
also been proposed as a leading mechanism for the formation of blood
vessels during embriogenesis \cite{gamba}. On a theoretical point of
view, chemotaxis can be described by the Keller-Segel \cite{ks} model
or its generalizations \cite{nfp}. The Keller-Segel model consists in
a drift-diffusion equation for the evolution of the density of
bacteria $\rho({\bf r},t)$ coupled to a reaction-diffusion equation
for the evolution of the secreted chemical $c({\bf r},t)$. In certain
approximations, the reaction-diffusion equation is replaced by a
Poisson equation. In that case, the Keller-Segel (KS) \cite{ks} model
becomes isomorphic to the Smoluchowski-Poisson (SP) system
\cite{critique} describing self-gravitating Brownian particles (see,
e.g., \cite{crrs} for a description of this analogy). The KS model
and SP system have been studied thoroughly in applied mathematics (see
Refs.  in \cite{perthame}) and in theoretical physics (see Refs. in
\cite{critique}).

However, the original KS model \cite{ks} also allows for the
possibility that the chemical suffers a degradation process which has
the effect of reducing the range of the interaction.  In that case,
the Poisson equation is replaced by a screened Poisson equation
\cite{nagai}.
In the gravitational analogy, this amounts to replacing the
gravitational potential by a screened gravitational potential, i.e. an
attractive Yukawa potential.  In that case, there exists interesting
phase transitions between spatially homogeneous and spatially
inhomogeneous equilibrium distributions. This is a physical motivation
to consider the thermodynamics of N-body systems interacting via an
attractive Yukawa potential \cite{hb1}. This will be called the {\it
screened Newtonian model}.  We shall also consider a related model
where the interaction is not screened but the Poisson equation is
modified so as to allow for the existence of spatially homogeneous
distributions at equilibrium. This will be called the {\it modified
Newtonian model}. In that model, the source of the potential is the
deviation between the actual density $\rho({\bf r},t)$ and the average
density $\overline{\rho}$.  This is similar to the effect of a
``neutralizing background'' in plasma physics \cite{nicholson}. This
model can be derived from the Keller-Segel model in the limit of
vanishing degradation of the chemical \cite{jl}. It also appears in
cosmology, due to the expansion of the universe, when we work in the
comoving frame \cite{peebles}.  It is therefore interesting to
consider this form of interaction at a general level and study the
corresponding phase transitions. We shall also compare them with the
ones obtained within the ordinary Newtonian model
\cite{antonov,lbw,ah,hk1,hk2,katz,klb,kl,kiessling,paddy,stahl,ap,vg1,vg2,aaiso,prefermi,sc,found} 
(see a review in \cite{ijmpb}).

The paper is organized as follows.  In Sec. \ref{sec_ke}, we discuss
several kinetic models taken from astrophysics, plasma physics and
biology for which our study applies. We consider either isolated
systems described by the microcanonical ensemble (fixed energy $E$) or
dissipative systems described by the canonical ensemble (fixed
temperature $T$). We characterize their equilibrium states in the mean
field approximation: in the microcanonical ensemble (MCE), they
maximize the entropy at fixed mass and energy and in the canonical
ensemble (CE) they minimize the free energy at fixed mass. In
Sec. \ref{sec_modif}, we specifically consider the case of a Newtonian
interaction with a neutralizing background. We study phase transitions
between homogeneous and inhomogeneous states depending on the
dimension of space. In $d=1$, the system presents canonical and
microcanonical second order phase transitions. In $d=2$, the system
presents an isothermal collapse in CE (zeroth order phase transition)
and a first order phase transition in MCE. In $d=3$, the system
presents an isothermal collapse in CE and a gravothermal catastrophe
in MCE (zeroth order phase transitions).  In Sec. \ref{sec_yuk}, we
perform a similar study for the attractive Yukawa potential with
screening length $k_0^{-1}$. In $d=1$, there exists a canonical
tricritical point $(k_{0})_c R=\sqrt{2}\pi\simeq 4.44$ and a
microcanonical tricritical point $(k_{0})_m R\simeq 11.8$, where $R$
is the system size. If $k_0<(k_{0})_c$, the system presents canonical
and microcanonical second order phase transitions. In that case, the
ensembles are equivalent. If $(k_{0})_c<k_0<(k_{0})_m$, the system
presents a canonical first order phase transition and a microcanonical
second order phase transition. In that case, there exists a region of
negative specific heats in MCE and the ensembles are inequivalent. If
$k_0>(k_{0})_m$, the system presents canonical and microcanonical
first order phase transitions. In $d=2$ and $d=3$, the phase
transitions are similar to those reported for the modified Newtonian
model.  In Sec. \ref{sec_stab}, we study the dynamical stability of
the homogeneous phase and analytically determine the critical point
$(E_c^*,T_c^*)$ that marks the onset of instability of the homogeneous
branch and the starting point of the bifurcated inhomogeneous
branch. Direct numerical simulations associated with these phase
transitions will be reported in a forthcoming paper.

Finally, it may be noted that the phase transitions reported in this
paper share analogies (but also differences) with phase transitions
observed in the Hamiltonian mean field (HMF) model \cite{inagaki,ar,cvb,tri,reentrant}, the spherical mass shell (SMS) model \cite{miller},
 the Blume-Emery-Griffiths (BEG) model \cite{bmr}, the infinite-range attactive interaction (IRAI) model \cite{art}, the self-gravitating
Fermi gas (SGF) model \cite{prefermi}, the self-gravitating ring (SGR)
model \cite{tbdr} and the one-dimensional static cosmology (OSC) model
\cite{valageas}.

\section{Kinetic models and statistical equilibrium states}
\label{sec_ke}

\subsubsection{Isolated systems}
\label{sec_h}

We consider an isolated system of $N$ particles in interaction
described by the Hamiltonian equations
\begin{eqnarray}
\label{h0}
m\frac{d{\bf r}_i}{dt}=\frac{\partial H}{\partial {\bf v}_i}, \quad m\frac{d{\bf v}_i}{dt}=-\frac{\partial H}{\partial {\bf r}_i},
\end{eqnarray}
where
\begin{eqnarray}
\label{h0b}
H=\sum_{i}\frac{1}{2}mv_i^2+m^2\sum_{i<j}u({\bf r}_i,{\bf r}_j)+m\sum_i V({\bf r}_i).
\end{eqnarray}
We assume that the particles interact through a binary potential
$u({\bf r},{\bf r}')$ that is symmetric with respect to the
interchange of ${\bf r}$ and ${\bf r}'$, and that they also evolve in
a fixed external potential $V({\bf r})$. Since the system is isolated,
with strict conservation of energy and mass, the proper statistical
ensemble is the microcanonical ensemble \cite{hb1}. In this paper, we shall use a
mean field approach \footnote{It is known that this approximation
becomes exact for systems with long-range interactions in a proper
thermodynamic limit $N\rightarrow +\infty$ \cite{ms}. For systems with
short-range interactions (e.g. a screened Newtonian potential), we
shall still use a mean field approximation although it may not be
exact. One motivation of our approach is that the Keller-Segel model
in biology is formulated in the mean field approximation even if the
degradation of the chemical is large. The incorrectness of the mean
field approximation as the interaction becomes short-range is
interesting but will not be considered in this paper.}.  In the
microcanonical ensemble, the statistical equilibrium state is obtained
by maximizing the Boltzmann entropy at fixed mass and energy. We thus
have to solve the maximization problem
\begin{eqnarray}
\label{mce1}
\max_f\left\lbrace S\lbrack f\rbrack\, |\, E\lbrack f\rbrack=E, \, M\lbrack f\rbrack=M\right\rbrace,
\end{eqnarray}
with
\begin{eqnarray}
\label{mce2}
S=-k_B\int \frac{f}{m}\ln\frac{f}{m}\, d{\bf r}d{\bf v},
\end{eqnarray}
\begin{eqnarray}
\label{mce4}
M=\int \rho \, d{\bf r},
\end{eqnarray}
\begin{eqnarray}
\label{mce3}
E=\int f \frac{v^2}{2} \, d{\bf r}d{\bf v}+\frac{1}{2}\int \rho({\bf r},t) u({\bf r},{\bf r}')\rho({\bf r}',t)\, d{\bf r}d{\bf r}'\nonumber\\
+\int \rho V\, d{\bf r},\qquad\qquad
\end{eqnarray}
where $\rho({\bf r},t)=\int f({\bf r},{\bf v},t)\, d{\bf v}$ is the spatial density.
Introducing the mean field potential
\begin{eqnarray}
\label{mce7}
\Phi({\bf r})=\int u({\bf r},{\bf r}')\rho({\bf r}')\, d{\bf r}'+V({\bf r}),
\end{eqnarray}
the energy  can also be written
\begin{eqnarray}
\label{h5}
E=\frac{1}{2}\int f v^2\, d{\bf r}d{\bf v}+\frac{1}{2}\int \rho (\Phi+V)\, d{\bf r}.
\end{eqnarray}

We shall be interested in global and local entropy maxima.  Let us first determine the  critical points of entropy at fixed mass and energy which cancel the first order variations. Introducing Lagrange multipliers, they satisfy
\begin{eqnarray}
\label{mce5}
\delta S-\frac{1}{T}\delta E-\alpha\delta M=0.
\end{eqnarray}
The variations are straightforward to evaluate and we obtain the mean field Maxwell-Boltzmann distribution
\begin{eqnarray}
\label{mce6}
f=Ae ^{-\beta m\left (\frac{v^2}{2}+\Phi\right )},
\end{eqnarray}
where $\beta=1/k_B T$ and  $\Phi({\bf r})$ is given by Eq. (\ref{mce7}). Integrating over the velocity, the find that the density is given by the mean field Boltzmann distribution
\begin{eqnarray}
\label{mce18}
\rho=A'e^{-\frac{m\Phi}{k_B T}}.
\end{eqnarray}
This critical point is a (local) entropy maximum at fixed mass and energy iff
\begin{eqnarray}
\label{mce8}
\delta^2 J=-\int \frac{(\delta f)^2}{2m f}\, d{\bf r}d{\bf v}-\frac{1}{2}\beta\int\delta\rho\delta\Phi\, d{\bf r}\le 0,
\end{eqnarray}
for all perturbations $\delta f$ that conserve mass and energy at first order. In Appendix \ref{sec_simpler}, we provide an equivalent but simpler condition of stability in the microcanonical ensemble [see inequality (\ref{mce18b})].

The time evolution of the distribution function $f({\bf r},{\bf v},t)$ is governed by a kinetic equation of the form
\begin{eqnarray}
\label{h1}
\frac{\partial f}{\partial t}+{\bf v}\cdot \frac{\partial f}{\partial {\bf v}}-\nabla\Phi\cdot \frac{\partial f}{\partial {\bf r}}=\left (\frac{\partial f}{\partial t}\right )_{coll},
\end{eqnarray}
where
\begin{eqnarray}
\label{h2}
\Phi({\bf r},t)=\int u({\bf r},{\bf r}')\rho({\bf r}',t)\, d{\bf r}'+V({\bf r}),
\end{eqnarray}
is the time-dependent mean field potential. The l.h.s. is an advective operator (Vlasov) in phase space. The r.h.s. is a ``collision'' operator like the Boltzmann operator in the kinetic theory of gases or like the Landau (or Lenard-Balescu) operator in plasma physics or stellar dynamics.
The ``collision'' operator in Eq. (\ref{h1}) takes into account the development of correlations between particles. It can have a more or less complicated form but it satisfies general properties associated with the first and second principles of thermodynamics: (i) it  conserves mass and energy; (ii) it satisfies an $H$-theorem for the Boltzmann entropy (\ref{mce2}), i.e. $\dot S\ge 0$ with an equality iff $f$ is the Maxwell-Boltzmann distribution (\ref{mce6}). Furthermore, the Maxwell-Boltzmann distribution is dynamically stable iff it is a (local) entropy maximum at fixed mass and energy. These general properties can be checked directly for the Boltzmann equation, for the Landau equation, for the Lenard-Balescu equation and for the BGK operator.  Therefore, the kinetic equation (\ref{h1})  is consistent with the maximization problem (\ref{mce1}) describing the statistical equilibrium state of the system in  MCE. If we neglect the collisions for sufficiently short times, Eq. (\ref{h1}) reduces to the Vlasov equation which can  experience a complicated process of collisionless violent relaxation towards a quasi stationary state (QSS) \cite{lb}.

\subsubsection{Dissipative systems in phase space}
\label{sec_dps}

We consider a dissipative system of $N$ Brownian particles in interaction described by the Langevin equations
\begin{eqnarray}
\label{b1a}
m\frac{d{\bf r}_i}{dt}=\frac{\partial H}{\partial {\bf v}_i},
\end{eqnarray}
\begin{eqnarray}
\label{b1b}
\frac{d{\bf v}_i}{dt}=-\frac{1}{m}\frac{\partial H}{\partial {\bf r}_i}-\xi {\bf v}_i+\sqrt{2D} {\bf R}_i(t),
\end{eqnarray}
where $H$ is the Hamiltonian defined by Eq. (\ref{h0b}), $-\xi{\bf v}_i$ is a friction force and ${\bf R}_i(t)$ is a white noise satisfying $\langle {\bf R}_i(t)\rangle=0$ and $\langle {R}_i^{\mu}(t){R}_j^{\nu}(t)\rangle=\delta_{ij}\delta_{\mu\nu}\delta(t-t')$. The diffusion coefficient $D$ and the friction coefficient $\xi$ are related to each other according to the Einstein relation $\xi=D\beta m$ where $\beta=1/(k_B T)$ is the inverse temperature. Since this system is dissipative, the proper statistical ensemble is the  canonical ensemble \cite{hb1}. In the canonical ensemble, the statistical equilibrium state is obtained by minimizing the Boltzmann free energy $F[f]=E[f]-TS[f]$ at fixed mass. We thus have to solve the minimization problem
\begin{eqnarray}
\label{ce1}
\min_f\left\lbrace F\lbrack f\rbrack\, |\, M\lbrack f\rbrack=M\right\rbrace
\end{eqnarray}
with
\begin{eqnarray}
\label{ce2}
F=\int f \frac{v^2}{2} \, d{\bf r}d{\bf v}+\frac{1}{2}\int \rho({\bf r},t) u({\bf r},{\bf r}')\rho({\bf r}',t)\, d{\bf r}d{\bf r}'\nonumber\\
+\int \rho V\, d{\bf r}+ k_B T\int \frac{f}{m}\ln\frac{f}{m}\, d{\bf r}d{\bf v}.\qquad
\end{eqnarray}
We shall be interested by global and local minima of free energy.  Let us first determine the  critical points of free energy at fixed mass which cancel the first order variations. Introducing a Lagrange multiplier, they satisfy
\begin{eqnarray}
\label{ce3}
\delta F+\alpha T\delta M=0.
\end{eqnarray}
The variations are straightforward to evaluate and we obtain the mean field Maxwell-Boltzmann distribution (\ref{mce6}) and  the mean field Boltzmann distribution (\ref{mce18}) as in the microcanonical ensemble. This critical point is a (local) minimum of free energy iff
\begin{eqnarray}
\label{ce3bq}
\delta^2 F=\frac{1}{2}\int\delta\rho\delta\Phi\, d{\bf r}+\frac{k_B T}{m}\int \frac{(\delta f)^2}{2 f}\, d{\bf r}d{\bf v}\ge 0,
\end{eqnarray}
for all perturbations $\delta f$ that conserve mass.

In the mean field approximation, the evolution of the distribution function $f({\bf r},{\bf v},t)$ is governed by a kinetic equation of the form
\begin{eqnarray}
\label{b2}
\frac{\partial f}{\partial t}+{\bf v}\cdot \frac{\partial f}{\partial {\bf v}}-\nabla\Phi\cdot \frac{\partial f}{\partial {\bf r}}=\frac{\partial}{\partial {\bf v}}\left (D\frac{\partial f}{\partial {\bf v}}+\xi f {\bf v}\right ),
\end{eqnarray}
coupled to the mean field potential (\ref{h2}). This is called the mean field Kramers equation.
The mean field Kramers equation conserves mass and satisfies an $H$-theorem for the Boltzmann free energy (\ref{ce2}), i.e. $\dot F\le 0$ with an equality iff $f$ is the Maxwell-Boltzmann distribution (\ref{mce6}).  Furthermore, the Maxwell-Boltzmann distribution is dynamically stable iff it is a (local) minimum of free energy  at fixed mass.  Therefore, the kinetic equation (\ref{b2})  is consistent with the minimization problem (\ref{ce1}) describing the statistical equilibrium state of the system in CE.

{\it Remark:} the critical points in MCE and CE are the same because the variational problems (\ref{mce1}) and (\ref{ce1}) are equivalent at the level of the first order variations (\ref{mce5}) and (\ref{ce3}). However, they are not equivalent at the level of the second order variations (\ref{mce8}) and (\ref{ce3bq}) because of the different class of perturbations to consider. Therefore, we can have {\it ensembles inequivalence} \cite{paddy,ellis,bb,ijmpb}. In fact, the condition of canonical stability  (\ref{ce1}) provides a {\it sufficient} condition of  microcanonical stability (\ref{mce1}). Indeed, if inequality (\ref{ce3bq}) is satisfied for all perturbations that conserve mass, then it is a fortiori satisfied for perturbations that conserve mass {\it and} energy, so that inequality (\ref{mce8}) is satisfied.  Therefore, canonical stability implies microcanonical stability:
\begin{eqnarray}
\label{ineq}
(\ref{ce1}) \Rightarrow (\ref{mce1}).
\end{eqnarray}
However, the converse is wrong in case of ensembles inequivalence.

\subsubsection{Dissipative systems in physical space}
\label{sec_dp}

In the strong friction limit $\xi\rightarrow +\infty$, we can formally neglect the inertial term $d{\bf v}_i/dt$ in Eq. (\ref{b1b}) and  we obtain the overdamped  Langevin equations
\begin{eqnarray}
\label{b7}
\xi\frac{d{\bf r}_i}{dt}=-\frac{1}{m}\frac{\partial H}{\partial {\bf r}_i}+\sqrt{2D}{\bf R}_i(t).
\end{eqnarray}
The statistical equilibrium state of this system (described by the canonical ensemble \cite{hb1}) is obtained by solving the minimization problem
\begin{eqnarray}
\label{ce6}
\min_\rho\left\lbrace F\lbrack \rho\rbrack\, |\,  M\lbrack \rho\rbrack=M\right\rbrace,
\end{eqnarray}
with
\begin{eqnarray}
\label{b8}
F=\frac{1}{2}\int \rho({\bf r},t) u({\bf r},{\bf r}')\rho({\bf r}',t)\, d{\bf r}d{\bf r}'\nonumber\\
+\int \rho V\, d{\bf r}+k_B T\int \frac{\rho}{m}\ln \frac{\rho}{m}\, d{\bf r}.
\end{eqnarray}
Writing the variational principle as
\begin{eqnarray}
\label{ce7}
\delta F+\alpha T\delta M=0,
\end{eqnarray}
we obtain the mean field Boltzmann distribution (\ref{mce18}).  This critical point is a (local) minimum of free energy at fixed mass iff
\begin{eqnarray}
\delta^2F=\frac{1}{2}\int\delta\rho\delta\Phi\, d{\bf r}+\frac{k_B T}{m}\int \frac{(\delta\rho)^2}{2\rho}\, d{\bf r}\ge 0,
\label{ce3b}
\end{eqnarray}
for all perturbations $\delta\rho$ that conserves mass.

In the mean field approximation, the evolution of the  density profile $\rho({\bf r},t)$ is governed by a kinetic equation of the form
\begin{eqnarray}
\label{b6}
\frac{\partial \rho}{\partial t}=\nabla \cdot \left\lbrack\frac{1}{\xi} \left (\frac{k_B T}{m}\nabla \rho + \rho \nabla \Phi
\right )\right\rbrack,
\end{eqnarray}
coupled to the mean field equation (\ref{h2}). This is called the mean field Smoluchowski equation.
The mean field Smoluchowski equation (\ref{b6}) conserves mass and satisfies an $H$-theorem for the Boltzmann free energy (\ref{b8}), i.e. $\dot F\le 0$ with an equality iff $\rho$ is the Boltzmann distribution (\ref{mce18}).  Furthermore, the Boltzmann distribution is dynamically stable iff it is a (local) minimum of free energy  at fixed mass.  Therefore, the kinetic equation (\ref{b6})  is consistent with the minimization problem (\ref{ce6}) describing the statistical equilibrium state of the system in CE.

{\it Remark 1}: the Smoluchowski equation (\ref{b6}) can also be deduced from the Kramers equation (\ref{b2}) in the strong friction limit \cite{risken}. For $\xi,D\rightarrow +\infty$ and $\beta=\xi/Dm$ finite, the time-dependent distribution function $f({\bf r},{\bf v},t)$ is  Maxwellian
\begin{eqnarray}
\label{b5}
f({\bf r},{\bf v},t)=\left (\frac{\beta m}{2\pi}\right )^{d/2}\rho({\bf r},t) e^{-\beta m\frac{v^2}{2}}+O(1/\xi),
\end{eqnarray}
and the time-dependent density  $\rho({\bf r},t)$ is solution of the Smoluchowski equation (\ref{b6}). Using Eq. (\ref{b5}), we can express the free energy (\ref{ce2}) as a functional of the density and we obtain the free energy (\ref{b8}) up to some unimportant constants.

{\it Remark 2:} it is shown in Appendix \ref{sec_simpler} that the maximization problems (\ref{ce1}) and (\ref{ce6})  are equivalent in the sense that $f({\bf r},{\bf v})$ is solution of (\ref{ce1}) iff $\rho({\bf r})$ is solution of  (\ref{ce6}). Thus, we have
\begin{eqnarray}
\label{imp}
(\ref{ce1})\Leftrightarrow (\ref{ce6}).
\end{eqnarray}
As a consequence, the Maxwell-Boltzmann distribution $f({\bf r},{\bf v})$ is dynamically stable with respect to the mean field  Kramers equation (\ref{b2}) iff the corresponding Boltzmann distribution  $\rho({\bf r})$ is dynamically stable with respect to the mean field  Smoluchowski equation (\ref{b6}). On the other hand, according to the implication (\ref{ineq}),  the Maxwell-Boltzmann distribution $f({\bf r},{\bf v})$ is dynamically stable with respect to the kinetic  equation (\ref{h1}) if it is stable with respect to the mean field Kramers  equation (\ref{b2}), but the reciprocal is wrong in case of ensembles inequivalence.

\subsubsection{The Keller-Segel model of chemotaxis}
\label{sec_ksc}

The Keller-Segel model \cite{ks} describing the chemotaxis of biological populations can be written as
\begin{eqnarray}
\frac{\partial \rho}{\partial t}=\nabla \cdot \left (D\nabla \rho -\chi \rho \nabla c
\right ),
\label{ks1}
\end{eqnarray}
\begin{eqnarray}
\frac{1}{D'}\frac{\partial c}{\partial t}=\Delta c-k^2 c+\lambda\rho,
\label{ks2}
\end{eqnarray}
where $\rho$ is the concentration of the biological species (e.g. bacteria) and $c$ is the concentration of the secreted chemical. The bacteria diffuse with a diffusion coefficient $D$ and undergo a chemotactic drift with strength $\chi$ along the gradient of chemical. The chemical is produced by the bacteria at a rate $D'\lambda$, is degraded at a rate $D' k^2$ and diffuses with a diffusion coefficient $D'$. We adopt Neumann boundary conditions \cite{ks}:
\begin{eqnarray}
\nabla c\cdot {\bf n}=0, \qquad  \nabla \rho\cdot {\bf n}=0,
\label{ksbc}
\end{eqnarray}
where ${\bf n}$ is a unit vector normal to the boundary of the domain. The drift-diffusion equation (\ref{ks1}) is similar to the mean field  Smoluchowski equation (\ref{b6}) where the concentration of chemical $-c({\bf r},t)$ plays the role of the potential $\Phi({\bf r},t)$. Therefore, there exists  many analogies between chemotaxis and  Brownian particles in interaction \cite{crrs}. In particular, the effective statistical ensemble associated with the Keller-Segel model is the canonical ensemble. The steady states of the Keller-Segel model are of the form
\begin{eqnarray}
\rho=Ae^{\frac{\chi}{D}c},
\label{kseq}
\end{eqnarray}
which is similar to the Boltzmann distribution (\ref{mce18}) with an effective temperature $T_{eff}=D/\chi$. The Lyapunov functional associated with the KS model is \cite{nfp}:
\begin{eqnarray}
\label{ks3}
F=\frac{1}{2\lambda}\int \left \lbrack (\nabla c)^2+k^2 c^2\right\rbrack\, d{\bf r}-\int \rho c\, d{\bf r}\nonumber\\
+T_{eff}\int \rho\ln\rho\, d{\bf r}.
\end{eqnarray}
It is similar to a free energy $F=E-T_{eff}S$ in thermodynamics, where $E$ is the energy and $S$ is the Boltzmann entropy. The KS model conserves mass and satisfies an $H$-theorem for the free energy (\ref{ks3}), i.e. $\dot F\le 0$ with an equality iff $\rho$ is the Boltzmann distribution (\ref{kseq}). Furthermore, the Boltzmann distribution is dynamically stable iff it is a (local) minimum of free energy  at fixed mass.
In that context, the minimization problem
\begin{eqnarray}
\label{ks4}
\min_{\rho,c}\left\lbrace F\lbrack \rho,c\rbrack\, |\, M\lbrack \rho\rbrack=M\right\rbrace,
\end{eqnarray}
determines a steady state of the KS model that is dynamically stable. This is similar to a condition of thermodynamical stability in the canonical ensemble.

Let us consider some simplified forms of the Keller-Segel model that have been introduced in the literature:

(i) In the limit of large diffusivity of the chemical $D'\rightarrow +\infty$ at fixed $k^2$ and $\lambda$, the reaction-diffusion equation (\ref{ks2}) takes the form of a screened Poisson equation \cite{nagai}:
\begin{eqnarray}
\Delta c-k^2 c=-\lambda\rho,
\label{ks5}
\end{eqnarray}
and the free energy becomes
\begin{eqnarray}
\label{ks5b}
F=-\frac{1}{2}\int \rho c\, d{\bf r}+T_{eff}\int \rho\ln\rho\, d{\bf r}.
\end{eqnarray}
In that case, the KS model is isomorphic to the Smoluchowski equation (\ref{b6}) with an attractive Yukawa potential (\ref{ay1}).

(ii) In the limit of large diffusivity of the chemical $D'\rightarrow +\infty$ and a vanishing  degradation rate $k^2=0$, the reaction-diffusion equation (\ref{ks2}) takes the form of a modified Poisson equation \cite{jl}:
\begin{eqnarray}
\Delta c=-\lambda(\rho-\overline{\rho}),
\label{ks6}
\end{eqnarray}
where $\overline{\rho}=M/V$ is the average density, and the free energy becomes
\begin{eqnarray}
\label{ks7}
F=-\frac{1}{2}\int (\rho-\overline{\rho}) c\, d{\bf r}+T_{eff}\int \rho\ln\rho\, d{\bf r}.
\end{eqnarray}
In that case, the KS model is isomorphic to the Smoluchowski equation (\ref{b6}) with a modified Poisson equation (\ref{mn1}).

(iii) Some authors have also considered a simple model of chemotaxis where the reaction-diffusion equation (\ref{ks2}) is replaced by the Poisson equation \cite{herrero}:
\begin{eqnarray}
\Delta c=-\lambda\rho.
\label{ks8}
\end{eqnarray}
This is valid in the absence of degradation of the chemical and for sufficiently large densities $\rho\gg\overline{\rho}$. This model can be used in particular to study chemotactic collapse. The corresponding  free energy is
\begin{eqnarray}
\label{ks9}
F=-\frac{1}{2}\int \rho c\, d{\bf r}+T_{eff}\int \rho\ln\rho\, d{\bf r}.
\end{eqnarray}
In that model, the boundary conditions (\ref{ksbc}) must be modified \footnote{We cannot impose the boundary conditions (\ref{ksbc}) for the Poisson equation (\ref{ks8}) since the integration of this equation $\oint\nabla c\cdot d{\bf S}=-\lambda M\neq 0$ implies  that $\nabla c\cdot {\bf n}\neq 0$ on the boundary of the domain.} and we must impose that $c\rightarrow 0$ at infinity like for the gravitational potential in astrophysics. Furthermore, we must impose that the normal component of the current vanishes on the boundary: $(D\nabla \rho -\chi \rho \nabla c)\cdot {\bf n}=0$ so as to conserve mass. In that case, the KS model is isomorphic to the Smoluchowski-Poisson (SP) system describing self-gravitating Brownian particles in the overdamped limit \cite{sc}.

\subsubsection{Physical justification of the canonical ensemble for systems with long-range interactions}
\label{sec_phys}

In statistical mechanics, the canonical distribution  is usually derived by considering a subpart of a large system and assuming that the rest of the system plays the role of a thermostat \cite{huang}. However, this justification implicitly assumes that energy is additive. Since energy is {\it non-additive} for systems with long-range interactions,  it is sometimes concluded that the canonical ensemble has no foundation  to describe
systems with long-range interactions \cite{gross}. In fact, this  is not quite true \cite{hb1}. We can give two justifications of the canonical ensemble for systems with long-range interactions:

(i) The canonical ensemble is relevant to describe a system of particles in contact with a thermal bath of a {\it different} nature \cite{hb1}. This is the case  if we consider a system of Brownian particles in interaction described by the stochastic equations (\ref{b1a})-(\ref{b1b}). The particles interact through a potential $u({\bf r},{\bf r}')$ that can be long-range, but they also undergo a friction force and a stochastic force that are due to other types of interaction (they model in general short-range interactions). As we have seen, this system is described by the  canonical ensemble. It does not correspond to a subsystem of a larger system, but simply to a system as a whole with long-range and short-range interactions \footnote{This interpretation also holds for the chemotactic problem. We have seen that the (mean field) Keller-Segel model has an effective thermodynamical structure associated with the canonical ensemble. Furthermore, we can derive kinetic models of chemotaxis in which the evolution of the cells (or bacteria) is described in  terms of coupled stochastic equations \cite{grima,cskin}. In that sense, the cells behave as Brownian particles in interaction as in  Secs. \ref{sec_dps} and \ref{sec_dp}, and the canonical structure of this model is clear.}.

(ii) Since canonical stability implies microcanonical stability
\cite{ellis}, the condition of canonical stability provides a {\it
sufficient} condition of microcanonical stability. In this sense, the
canonical stability criterion (see Secs. \ref{sec_dps} and
\ref{sec_dp}) can be useful even for an isolated Hamiltonian system
(see Sec. \ref{sec_h}) because if we can prove that this system is
canonically stable, then it is granted to be microcanonically
stable. This remark also applies to other ensembles (grand canonical,
grand microcanonical,...).

\section{The modified Newtonian model}
\label{sec_modif}

In this section, we discuss phase transitions that appear in the modified Newtonian model.

\subsection{Physical motivation of the model}

We consider a system of particles interacting via a mean field potential $\Phi({\bf r},t)$ that is  solution of the modified Poisson equation
\begin{eqnarray}
\label{mn1}
\Delta\Phi=S_d G(\rho-\overline{\rho}),
\end{eqnarray}
where $\overline{\rho}=M/V$ is the average density (conserved quantity). At statistical equilibrium, the density is given by the Boltzmann distribution
\begin{eqnarray}
\label{mn2}
\rho=A e^{-\beta m\Phi}.
\end{eqnarray}
We have used the notations of astrophysics (where $G$ is the constant of gravity and $S_d$ the surface of a unit sphere in $d$ dimensions) in order to make the connection with ordinary self-gravitating systems where Eq. (\ref{mn1}) is replaced by the Poisson equation $\Delta\Phi=S_d G\rho$. However, this model can have application in other contexts as explained below. We assume that the system is confined in a finite domain (box) and we impose the Neumann boundary conditions
\begin{eqnarray}
\nabla \Phi\cdot {\bf n}=0, \qquad  \nabla \rho\cdot {\bf n}=0,
\label{ksbcb}
\end{eqnarray}
where ${\bf n}$ is a unit vector normal to the boundary of the box (the explicit expression of the potential in $d=1$ is given in Appendix \ref{sec_exp}). This model admits spatially homogeneous solutions ($\rho=\overline{\rho}$ and $\Phi=0$) at any temperature. It also admits spatially inhomogeneous solutions at sufficiently low temperatures. We shall study this model in arbitrary dimensions of space $d$ with explicit computations for $d=1,2,3$.  This model has different physical applications:

(i) It describes self-gravitating systems in a cosmological setting \cite{peebles}. Due to the expansion of the universe, when we work in the comoving frame, the Poisson equation takes the form of Eq. (\ref{mn1}) where the potential is produced by the deviation between the actual density $\rho({\bf r},t)$ and  the mean density $\overline{\rho}$. In cosmology, we must also account for the scale factor $a(t)$ but if we consider timescales that are short with respect to the Hubble time $H^{-1}=a/\dot a$, we can ignore this time dependence.  This model has been studied by Valageas \cite{valageas} in $d=1$ with periodic boundary conditions. In that context, the relevant ensemble is the MCE since the system is isolated.

(ii) By  a proper reinterpretation of the parameters, the field equation (\ref{mn1}) describes the relation between the concentration of the chemical and the density of bacteria in the Keller-Segel model (\ref{ks6}). In that case, the most physical dimension is $d=2$ and the boundary conditions are of the form (\ref{ksbcb}). Furthermore, the relevant ensemble is the CE since the KS model has a canonical structure.
This model has been studied by applied mathematicians, starting with J\"ager \& Luckhaus \cite{jl}, but they have not performed the type of study that we are developing in this paper.

In view of these different applications, we shall study this model in the microcanonical and canonical ensembles in any dimension of space.

\subsection{The modified Emden equation}
\label{sec_modifemden}

\begin{figure}
\begin{center}
\includegraphics[clip,scale=0.3]{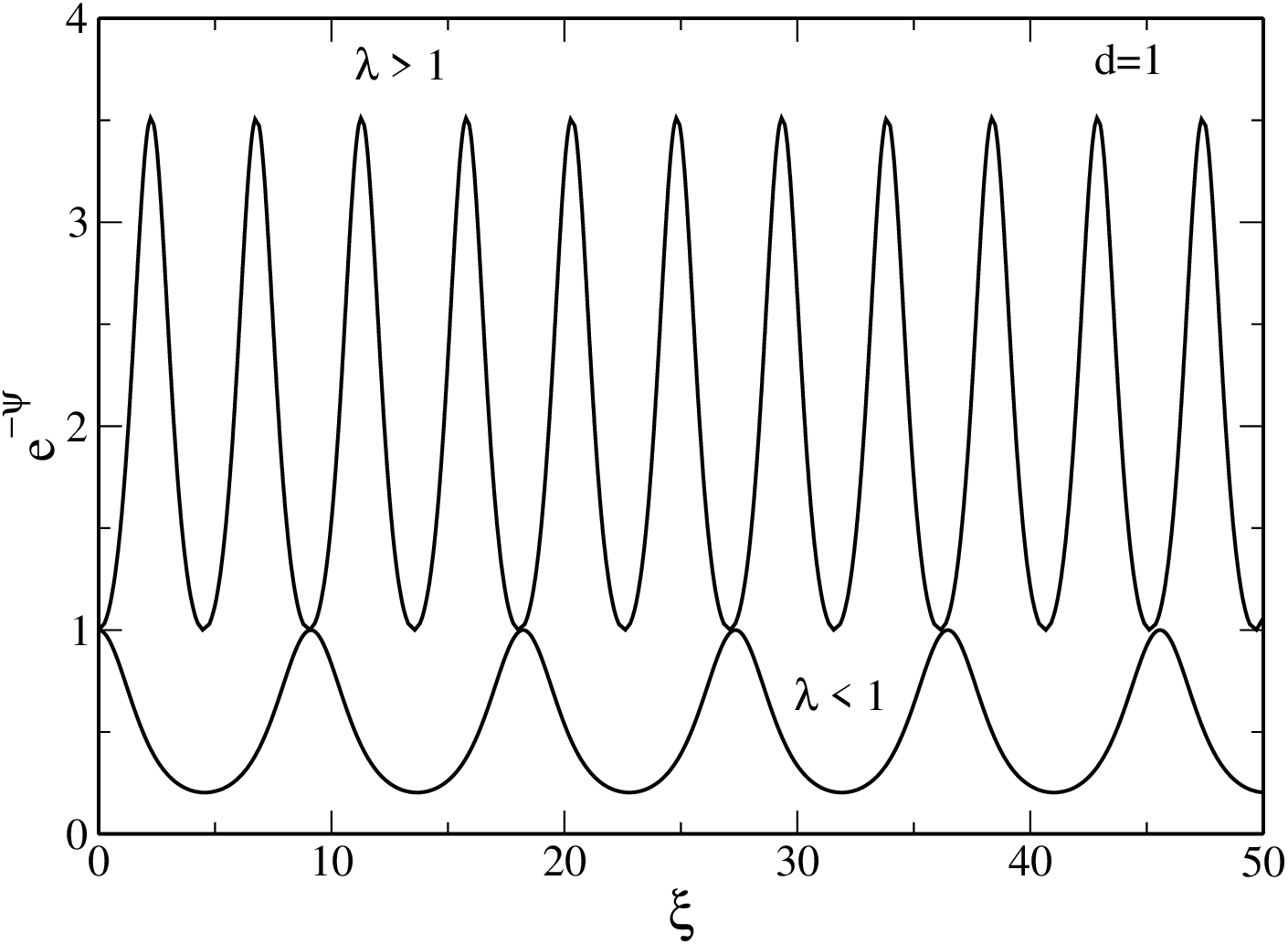}
\caption{The function $e^{-\psi}$  for $d=1$ and $\lambda=0.5<1$ (bottom) or $\lambda=2>1$ (top). In $d=1$, the oscillations are undamped.}
\label{s1-d1}
\end{center}
\end{figure}

\begin{figure}
\begin{center}
\includegraphics[clip,scale=0.3]{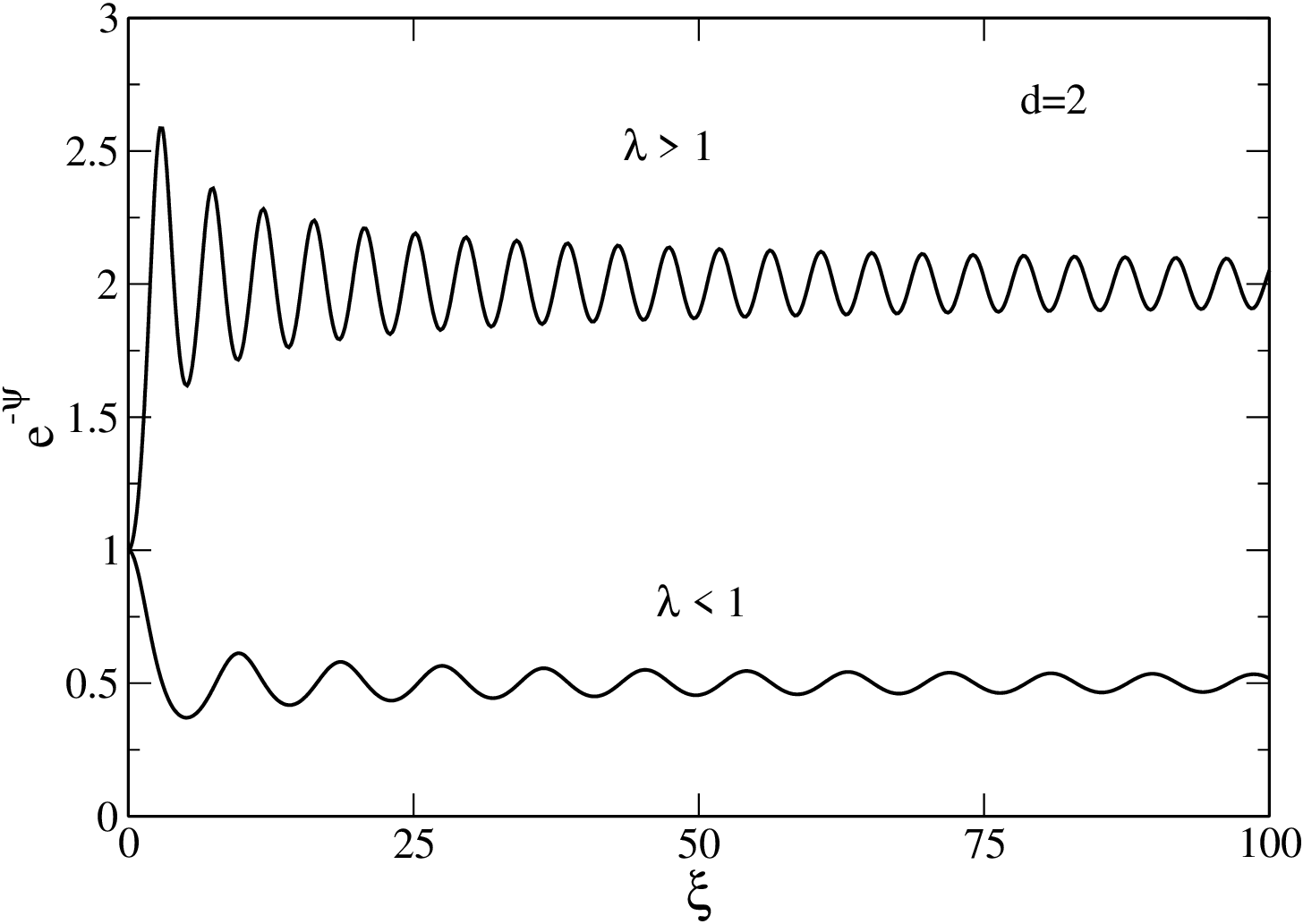}
\caption{The function $e^{-\psi}$ for $d=2$ and $\lambda=0.5<1$ (bottom) or $\lambda=2>1$ (top). In $d\ge 2$, the oscillations are damped. The case $d=3$ (not represented) is similar.}
\label{s2-d2}
\end{center}
\end{figure}

In the modified Newtonian model, the statistical equilibrium state is given by the Boltzmann distribution
(\ref{mn2}) coupled to the modified Poisson equation (\ref{mn1}). We look for spherically symmetric solutions because, for non rotating systems, entropy maxima (or minima of free energy) are spherically symmetric. Introducing the central density $\rho_0=\rho(0)$, the central potential $\Phi_0=\Phi(0)$, the new field $\psi = \beta m (\Phi - \Phi_0)$ and the scaled distance $\xi = (S_dG\beta m \rho_0)^{1/2} r$, the Boltzmann distribution (\ref{mn2}) can be rewritten
\begin{eqnarray}
\label{mn3}
\rho=\rho_0 e^{-\psi(\xi)}.
\end{eqnarray}
Substituting this relation in the modified Poisson equation (\ref{mn1}), we obtain the modified Emden equation
\begin{eqnarray}
\label{mn4}
\frac{1}{\xi^{d-1}}\frac{d}{d\xi}\left (\xi^{d-1}\frac{d\psi}{d\xi}\right )=e^{-\psi} - \lambda,
\end{eqnarray}
where $\lambda=\overline{\rho}/\rho_0$ plays the role of the inverse central density. Since $\Phi'(0)=0$ for a spherically symmetric system, the boundary conditions at the origin are
\begin{eqnarray}
\label{mn5}
\psi (0)=\psi^{\prime}(0)=0.
\end{eqnarray}
The ordinary Emden equation \cite{chandra} is recovered for $\lambda=0$, i.e. for very large central densities with respect to the average density. The function  $e^{-\psi(\xi)}$ is plotted in Figs. \ref{s1-d1} and \ref{s2-d2} for different values of $\lambda$ and different dimensions of space $d$. It presents an infinity of oscillations.  For  $d=1$, the oscillations are undamped and their period is given by Eq. (\ref{app11}).  For $d\ge 2$, the oscillations are damped and the function $\psi(\xi)$ tends to the asymptotic value $-\ln\lambda$ for $\xi\rightarrow +\infty$.

We assume that the system is enclosed in a spherical box of radius $R$. The normalized box radius $\alpha= (S_dG\beta m \rho_0)^{1/2} R$ is determined by the boundary condition $\Phi^{\prime} (R)=0$ that becomes
\begin{eqnarray}
\label{mn7}
\psi^{\prime}(\alpha)=0.
\end{eqnarray}
For a given value of $\lambda$, we need to integrate the modified Emden equation (\ref{mn4})-(\ref{mn5}) until the point $\xi=\alpha$ such that $\psi^{\prime}(\alpha)=0$.  Since the function $\psi(\xi)$ presents an infinite number of oscillations, this determines an infinity of solutions $\alpha_1(\lambda)$, $\alpha_2(\lambda)$,... that will correspond to different branches in the following diagrams.
Once $\alpha_n(\lambda)$ is determined, the density profile is given by Eq. (\ref{mn3}). The density profile is extremum at the center and at the boundary. On the $n$-th branch, the density profile shows $n$ ``clusters'' corresponding to the oscillations of $e^{-\psi(\xi)}$.   Close to the origin, the density increases for $\lambda>1$ while it decreases for $\lambda<1$. The homogeneous state $\psi=0$ corresponds to $\lambda = 1$. This solution is degenerate because the boundary condition (\ref{mn7}) is satisfied for any $\alpha$.

{\it Remark:} When $\lambda\rightarrow 0$, corresponding to large values of the central density, we expect to obtain results similar to those obtained for the usual Newtonian model since the differential equation (\ref{mn4}) reduces to the ordinary Emden equation. However, the results are different because the boundary conditions are not the same. In the Newtonian model, the force at the boundary is non zero (for a spherically symmetric system, according to the Gauss theorem, we have $\Phi'(R)=GM/R^{d-1}$) while in the modified Newtonian model the force at the boundary is zero ($\Phi'(R)=0$).  Therefore, strictly speaking, the Newtonian and the modified Newtonian models behave differently even when $\rho_0\rightarrow +\infty$.
Nevertheless, for large central concentrations, the Newtonian solution provides a good approximation of the modified Newtonian solution in the core (see Appendix \ref{sec_app}).

\subsection{The temperature}
\label{sec_modifT}

\begin{figure}
\begin{center}
\includegraphics[clip,scale=0.3]{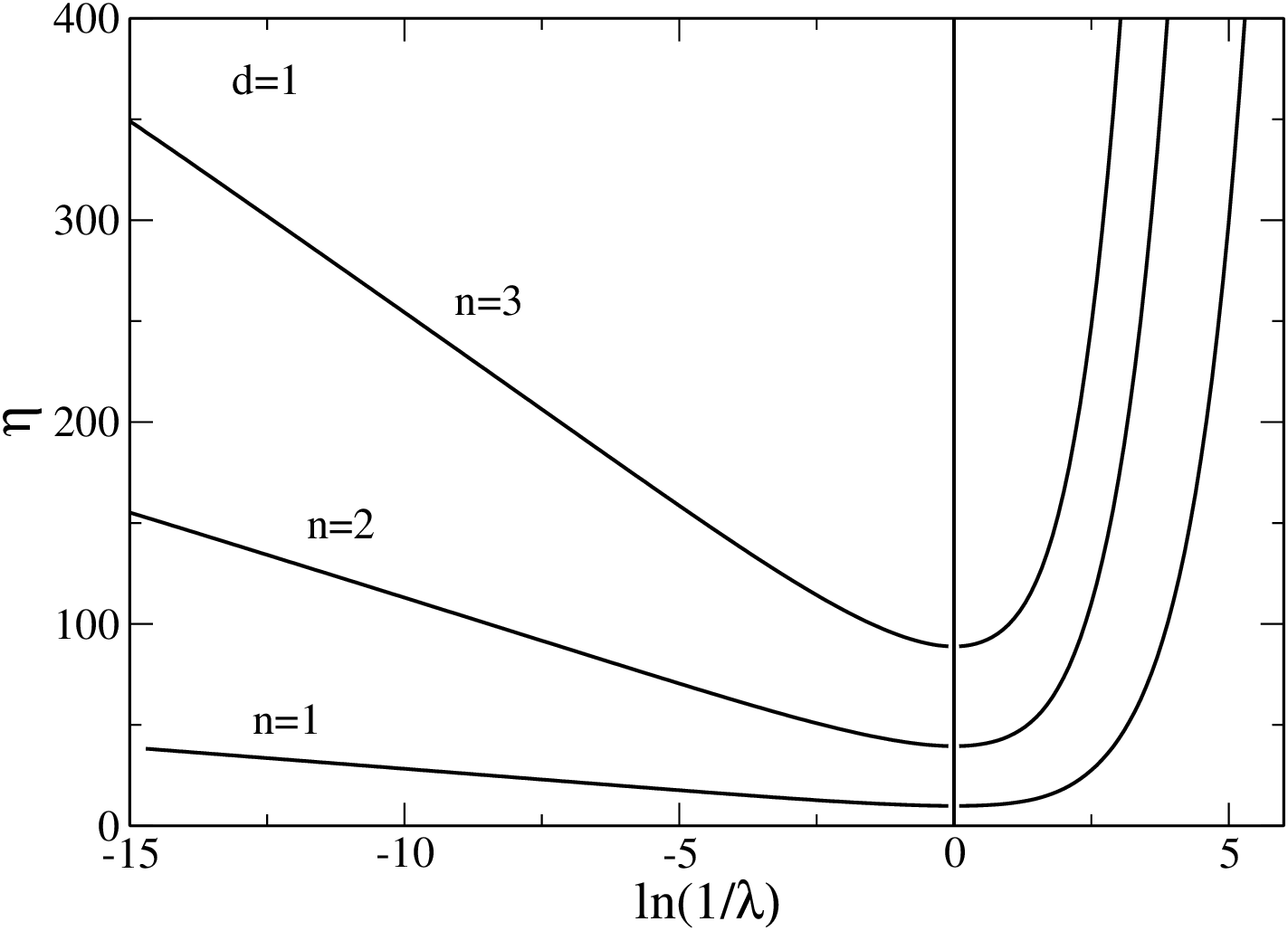}
\caption{Inverse temperature $\eta$ as a function of the central density  $1/\lambda$ for the first three branches in  $d=1$.}
\label{eta1}
\end{center}
\end{figure}

\begin{figure}
\begin{center}
\includegraphics[clip,scale=0.3]{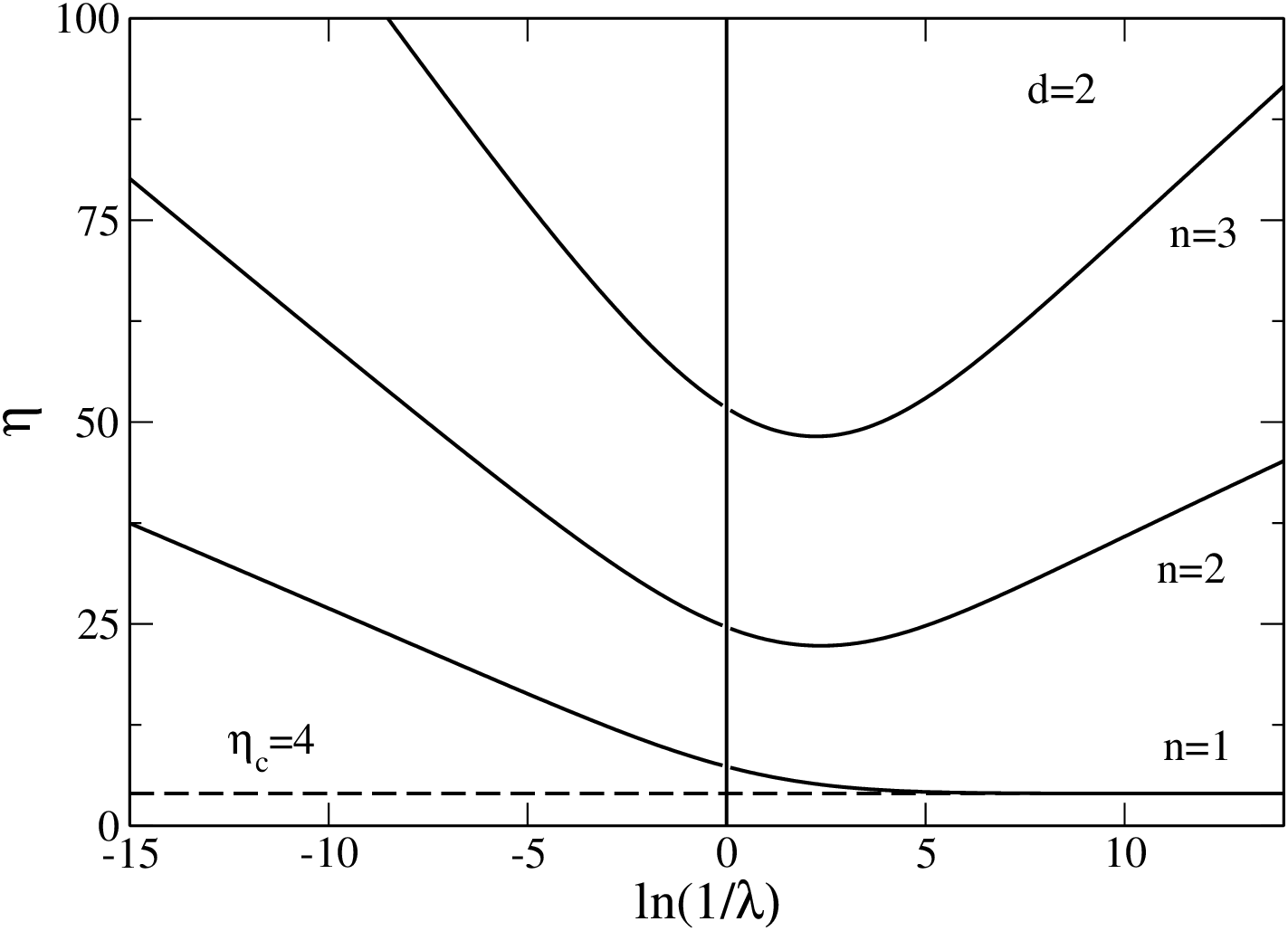}
\caption{Inverse temperature $\eta$ as a function of the central density  $1/\lambda$ for the first three branches in  $d=2$. }
\label{eta2}
\end{center}
\end{figure}

\begin{figure}
\begin{center}
\includegraphics[clip,scale=0.3]{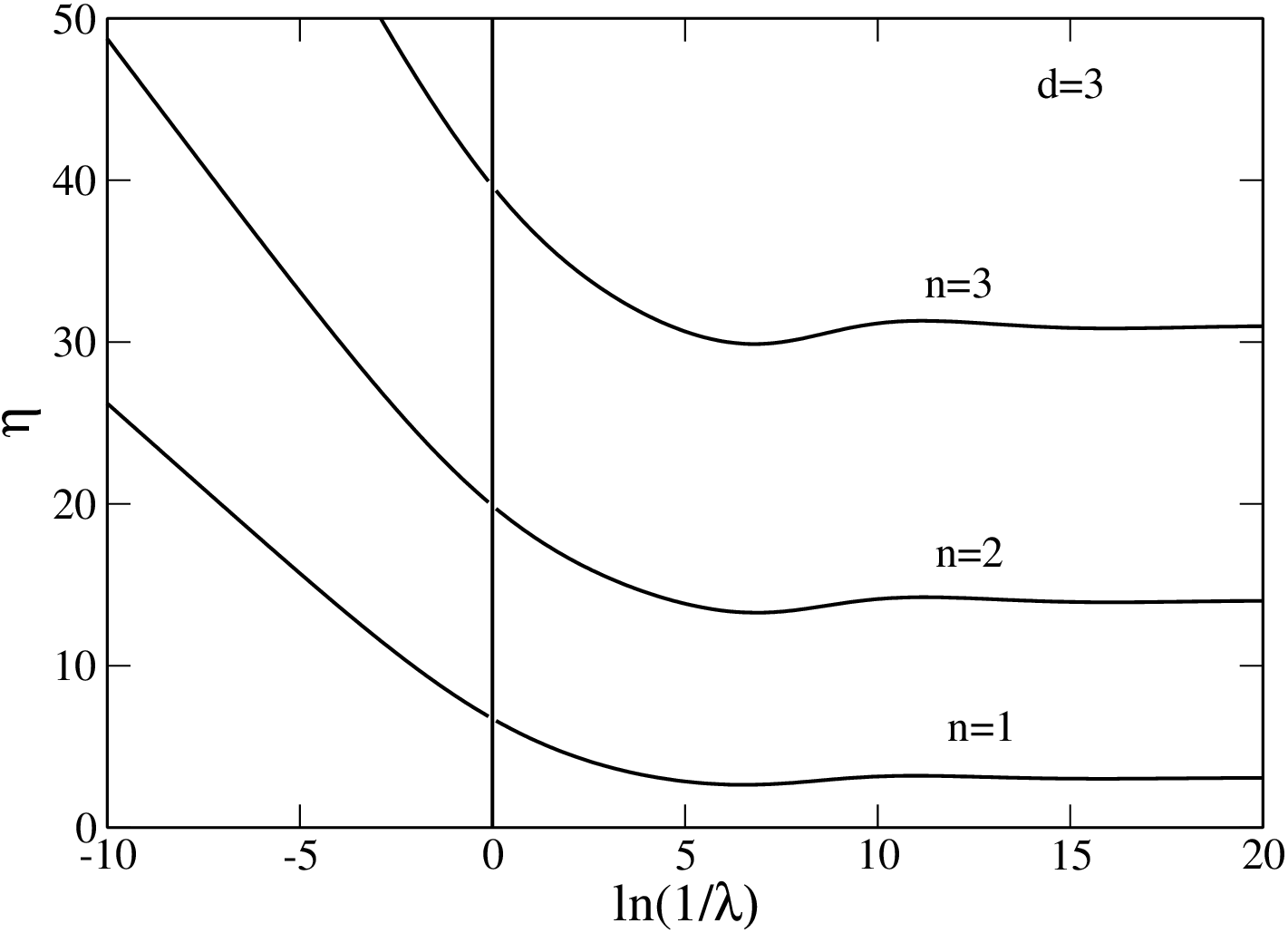}
\caption{Inverse temperature $\eta$ as a function of the central density  $1/\lambda$ for the first three branches in  $d=3$.}
\label{eta3}
\end{center}
\end{figure}

We must now relate the normalized central density $1/\lambda$ to the temperature $T$. Recalling that $\overline{\rho}=M/V$ with $V=\frac{1}{d}S_d R^d$, we obtain
\begin{eqnarray}
\label{mn8}
\lambda=\frac{\overline{\rho}}{\rho_0}=\frac{dM}{S_dR^d}\frac{1}{\rho_0}
=d\frac{GMm\beta}{R^{d-2}}\frac{1}{\alpha^2}.
\end{eqnarray}
Introducing  the normalized temperature
\begin{eqnarray}
\label{mn9}
\eta\equiv \frac{\beta GMm}{R^{d-2}},
\end{eqnarray}
we find the relation
\begin{eqnarray}
\label{mn10}
\eta=\frac{1}{d}\lambda \alpha^2.
\end{eqnarray}
Recalling that $\alpha=\alpha_n(\lambda)$ for the $n$-th branch, this equation gives the relation between the inverse temperature $\eta$ and the central density  $1/\lambda$ for the $n$-th
branch.  In Figs. \ref{eta1}, \ref{eta2} and \ref{eta3}, we plot the inverse temperature  $\eta$ as a function of the central density $1/\lambda$ for the first three branches $n=1,2,3$ in different dimensions of space $d=1,2,3$.

Let us discuss the asymptotic behaviors of the temperature (we only describe the first branch $n=1$) and compare with the Newtonian model (see, e.g., \cite{sc}):

$\bullet$ In $d=1$: for the ordinary Newtonian model, the series of equilibria is parameterized by $\alpha$, which is a measure of the central density. When $\alpha\rightarrow +\infty$, the distribution tends to a Dirac peak $\rho=M\delta(x)$ and the inverse temperature $\eta\rightarrow +\infty$. When $\alpha\rightarrow 0$, the distribution is homogeneous and the inverse temperature $\eta\rightarrow 0$.  For the modified Newtonian model, the series of equilibria is parameterized by the central density $1/\lambda$. When $1/\lambda\rightarrow +\infty$, the distribution tends to a Dirac peak $\rho=M\delta(x)$ and $\eta\rightarrow +\infty$ with the same asymptotic behavior as in the Newtonian model (see Appendix \ref{sec_app}).  When $\lambda=1$, the distribution is homogeneous and $\eta=\eta_c^*=\pi^2\simeq 9.8696044$  (see Appendix \ref{sec_bif}). When $1/\lambda\rightarrow 0$, the distribution tends to a Dirac peak $\rho=\frac{M}{2}(\delta(x-R)+\delta(x+R))$ concentrated at the box and $\eta\rightarrow +\infty$.

$\bullet$ In $d=2$: for the ordinary Newtonian model, the
series of equilibria is parameterized by $\alpha$.  When $\alpha\rightarrow +\infty$,
the distribution tends to a Dirac peak $\rho=M\delta({\bf r})$ and the
inverse temperature tends to $\eta_c=4$.  When
$\alpha\rightarrow 0$, the distribution is homogeneous and the inverse
temperature $\eta\rightarrow 0$. For the modified Newtonian
model, the series of equilibria is parameterized by $1/\lambda$. When
$1/\lambda\rightarrow +\infty$, the distribution tends to a Dirac peak
$\rho=M\delta({\bf r})$ and $\eta\rightarrow \eta_c=4$ (since the density
is very much concentrated, the boundary conditions do not matter and
we recover the same results as in the Newtonian case). When $\lambda=1$, the distribution is
homogeneous and $\eta=\eta_c^*=\frac{1}{2} j_{11}^2\simeq 7.3410008$
(see Appendix \ref{sec_bif}). When $1/\lambda\rightarrow 0$, the
distribution is concentrated at the boundary and $\eta\rightarrow
+\infty$.

$\bullet$ In $d=3$: for the ordinary Newtonian model, the
series of equilibria is parameterized by $\alpha$. When $\alpha\rightarrow +\infty$,
the distribution tends to the singular isothermal sphere $\rho_s(r)=1/(2\pi G\beta m r^2)$ and the
inverse temperature $\eta\rightarrow \eta_s=2$. The curve
$\eta(\alpha)$ displays damped oscillations around this value. When
$\alpha\rightarrow 0$, the distribution is homogeneous and the inverse
temperature $\eta\rightarrow 0$.   For the
modified Newtonian model, the series of equilibria is parameterized by
$1/\lambda$.  When $1/\lambda\rightarrow +\infty$, the distribution is
concentrated at the center and we numerically find that
$\eta\rightarrow 3.05...$ (the value is different from the Newtonian result $\eta_s=2$ due to
different boundary conditions). The curve $\eta(\lambda)$ displays damped
oscillations around this value. When $\lambda=1$, the distribution is
homogeneous and $\eta=\eta_c^*=\frac{1}{3} x_{1}^2\simeq
6.7302445$ (see Appendix \ref{sec_bif}). When $1/\lambda\rightarrow
0$, the distribution is concentrated at the boundary and we
numerically find that $\eta\rightarrow +\infty$.

\subsection{The energy}
\label{sec_modifE}

We must also relate the normalized central density  $1/\lambda$ to the energy $E$. The total energy is given by (see Appendix \ref{sec_ext}):
\begin{eqnarray}
\label{mn11}
E=\int f\frac{v^2}{2}\, d{\bf r}d{\bf v}+\frac{1}{2}\int\left (\rho-\overline{\rho}\right ) \Phi \, d\mathbf{r}.
\end{eqnarray}
Using the Maxwell-Boltzmann distribution (\ref{mce6}), the kinetic energy is simply
\begin{eqnarray}
\label{mn12}
K=\frac{d}{2}Nk_BT.
\end{eqnarray}
Using the modified Poisson equation (\ref{mn1}) and an integration by parts, the potential energy can be written
\begin{eqnarray}
\label{mn13}
W=-\frac{1}{2S_dG}\int(\nabla \Phi )^2\, d\bf{r}.
\end{eqnarray}
The total energy $E=K+W$ is therefore given by
\begin{eqnarray}
\label{mn14}
E=\frac{d}{2}Nk_BT-\frac{1}{2S_dG}\int(\nabla
\Phi )^2\, d\bf{r}.
\end{eqnarray}
Introducing the dimensionless variables defined previously, recalling that $r=\xi R/\alpha$, and introducing the normalized energy
\begin{eqnarray}
\label{mn15}
\Lambda\equiv -\frac{ER^{d-2}}{GM^2},
\end{eqnarray}
we obtain
\begin{eqnarray}
\label{mn16}
\Lambda=-\frac{d}{2\eta}+\frac{1}{2\eta^2}\frac{1}{\alpha^{d-2}}\int_{0}^{\alpha}\left (\frac{d\psi}{d\xi}\right )^2\xi^{d-1}d\xi.
\end{eqnarray}
Recalling that $\alpha=\alpha_n(\lambda)$ and $\eta=\eta_n(\lambda)$ for the $n$-th branch, this equation gives the relation between the energy $\Lambda$ and the central density
$1/\lambda$ for the $n$-th branch.

\begin{figure}
\begin{center}
\includegraphics[clip,scale=0.3]{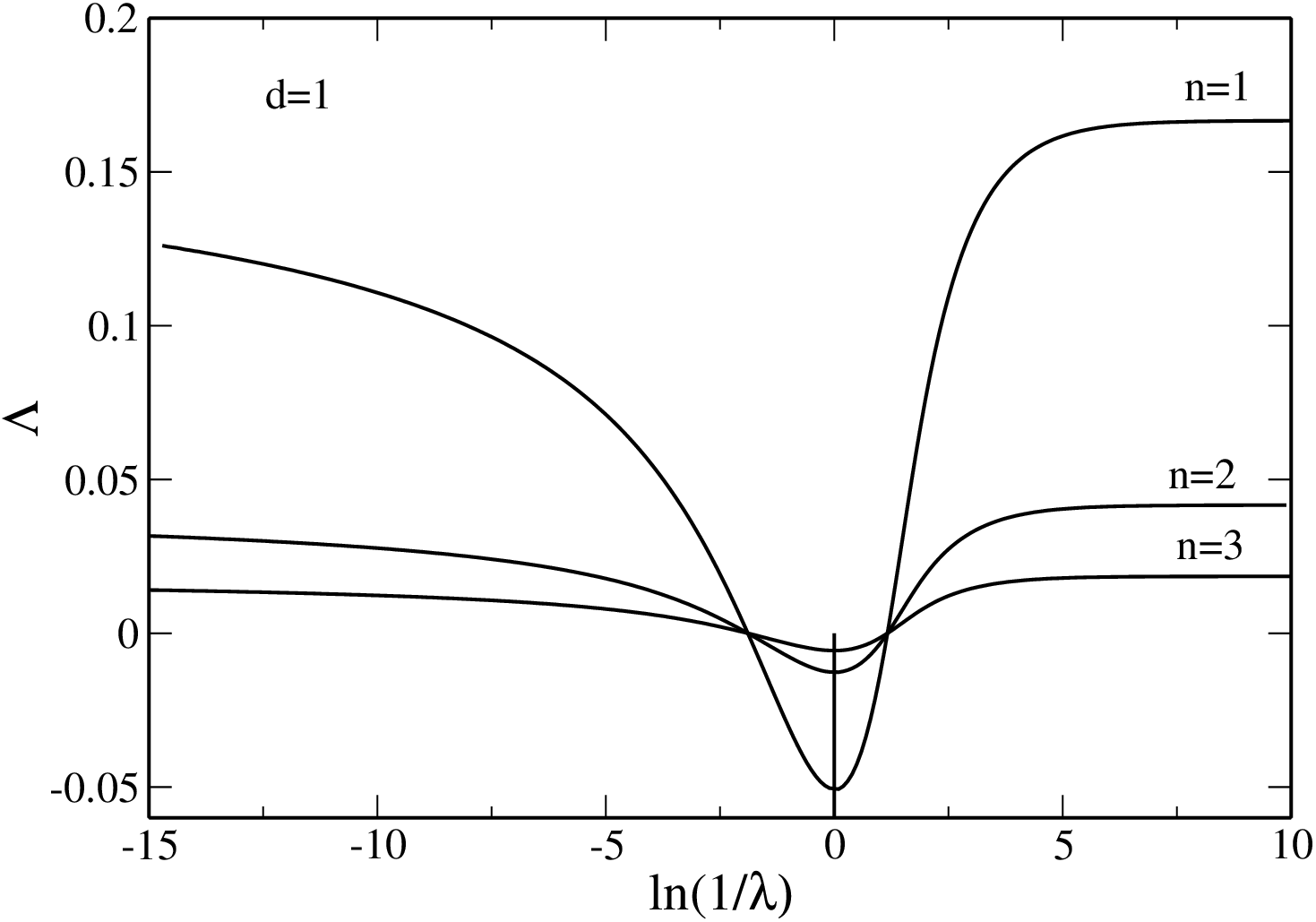}
\caption{Energy $\Lambda$ as a function of the central density $1/\lambda$ for the first three branches in $d=1$. }
\label{ene1}
\end{center}
\end{figure}

\begin{figure}
\begin{center}
\includegraphics[clip,scale=0.3]{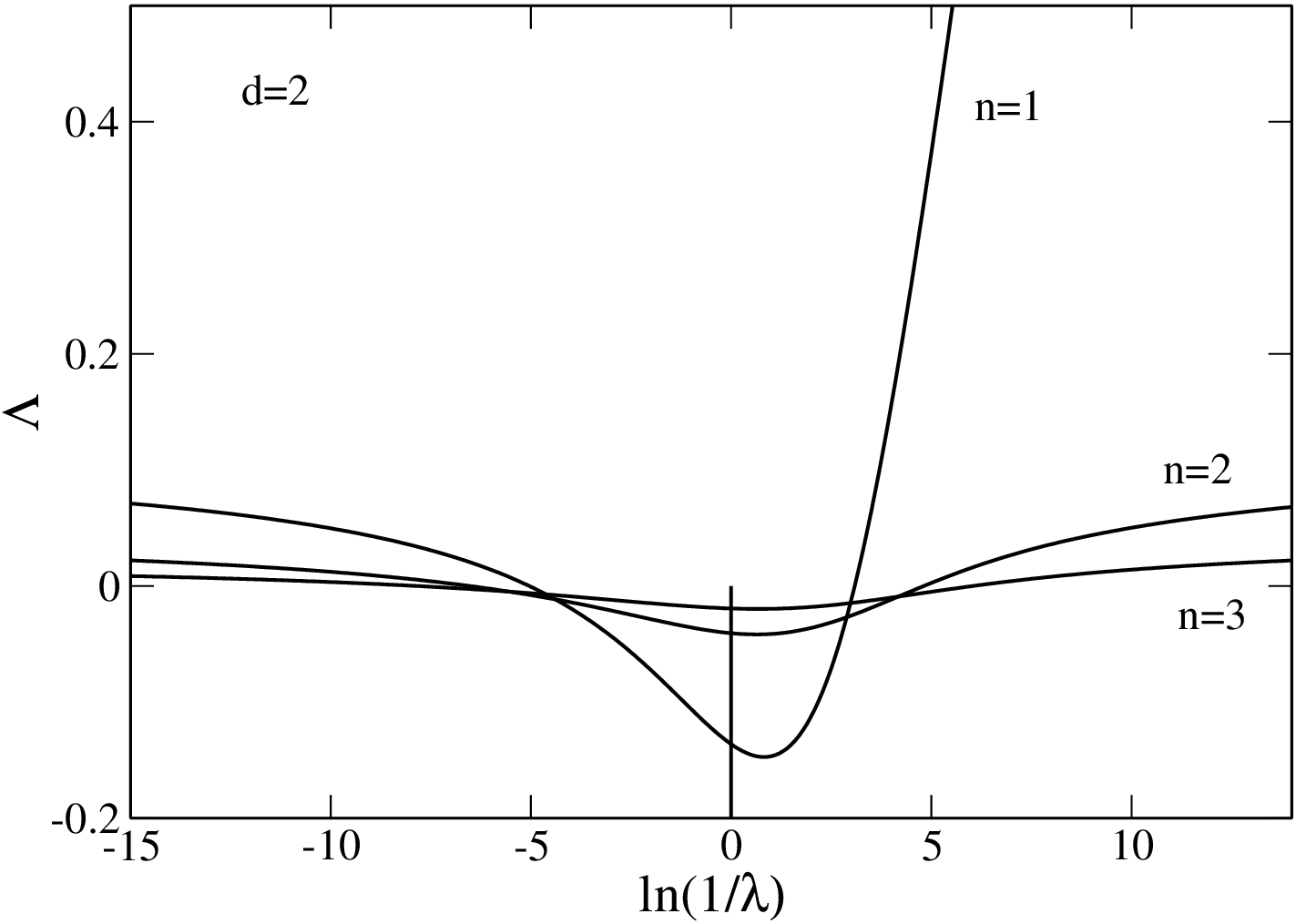}
\caption{Energy $\Lambda$ as a function of the central density $1/\lambda$ for the first three branches in $d=2$.}
\label{ene2}
\end{center}
\end{figure}

\begin{figure}
\begin{center}
\includegraphics[clip=0.3,scale=0.3]{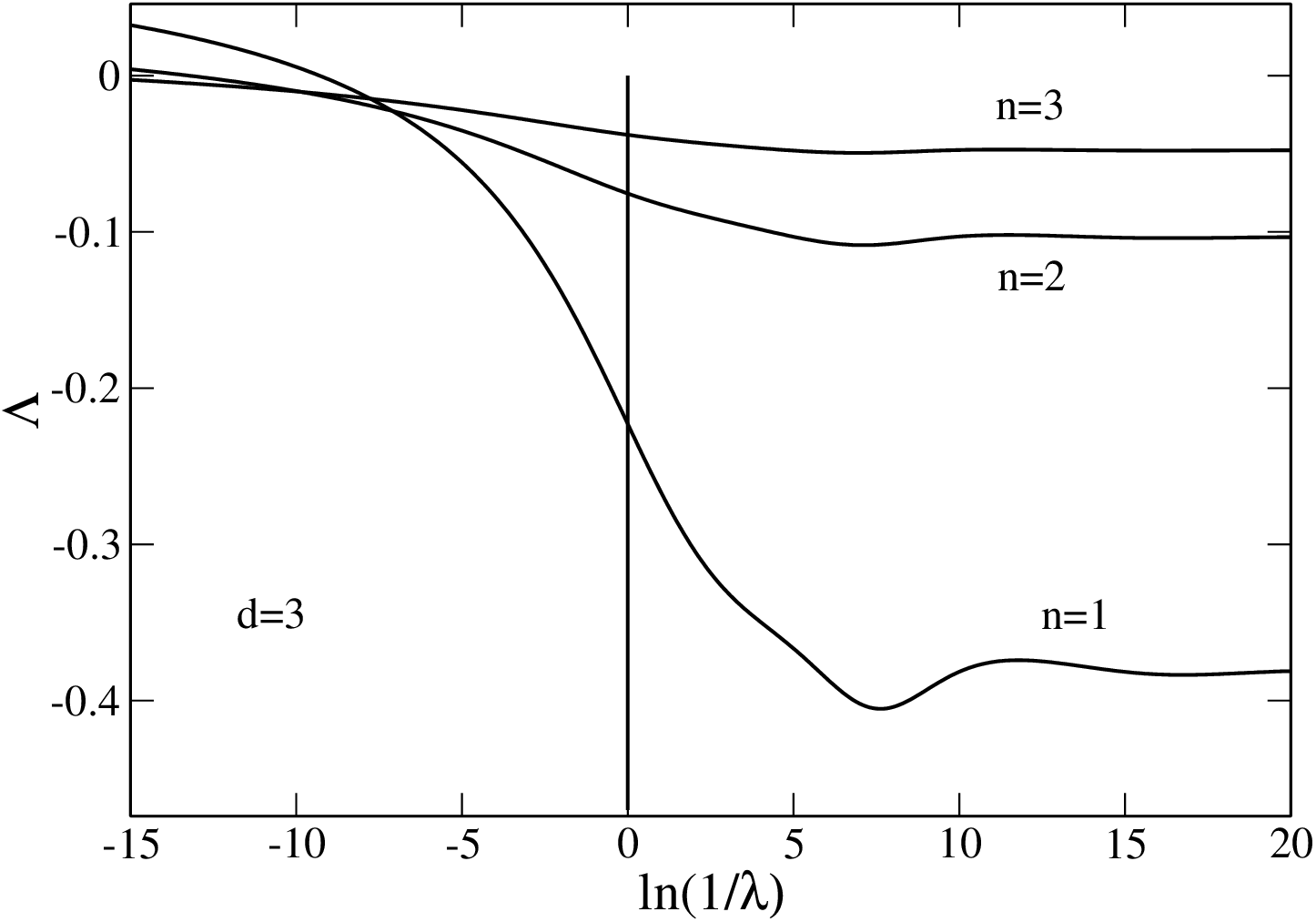}
\caption{Energy $\Lambda$ as a function of the central density $1/\lambda$ for the first three branches in $d=3$.}
\label{ene3}
\end{center}
\end{figure}

In Figs. \ref{ene1}, \ref{ene2} and \ref{ene3}, we plot the normalized energy $\Lambda$ as a function of the central density $1/\lambda$ for the first three branches $n=1,2,3$ in different dimensions of space  $d=1, 2, 3$.

Let us consider the asymptotic behaviors of the energy (we only describe the first branch $n=1$) and compare with the Newtonian model (see, e.g., \cite{sc}):

$\bullet$ In $d=1$: for the ordinary Newtonian model, the series of equilibria is parameterized by $\alpha$, which is a measure of the central density. When $\alpha\rightarrow +\infty$, the distribution tends to a Dirac peak $\rho=M\delta(x)$ and the energy $\Lambda\rightarrow 0$.  When $\alpha\rightarrow 0$, the distribution is homogeneous and the energy $\Lambda\rightarrow -\infty$.  For the modified Newtonian model, the series of equilibria is parameterized by the central density $1/\lambda$. When $1/\lambda\rightarrow +\infty$, the distribution tends to a Dirac peak $\rho=M\delta(x)$ and $\Lambda\rightarrow \Lambda_{max}^{(1)}=1/6$ (see Appendix \ref{sec_me}). When $\lambda=1$ the distribution is homogeneous and $\Lambda=\Lambda_c^*=-1/(2\eta_c^*)\simeq -0.0506606$. When $1/\lambda\rightarrow 0$, the distribution tends to a Dirac peak $\rho=\frac{M}{2}(\delta(x-R)+\delta(x+R))$ concentrated at the box and  $\Lambda\rightarrow \Lambda_{max}^{(1)}$.

$\bullet$ In $d=2$: for the ordinary Newtonian model, the series of equilibria is parameterized by $\alpha$. When $\alpha\rightarrow +\infty$, the distribution tends to a Dirac peak $\rho=M\delta({\bf r})$ and the energy $\Lambda\rightarrow +\infty$. When $\alpha\rightarrow 0$, the distribution is homogeneous and the energy $\Lambda\rightarrow -\infty$.  For the modified Newtonian model, the series of equilibria is parameterized by $1/\lambda$. When $1/\lambda\rightarrow +\infty$, the distribution tends to a Dirac peak $\rho=M\delta({\bf r})$ and $\Lambda\rightarrow +\infty$. When $\lambda=1$ the distribution is homogeneous and $\Lambda=\Lambda_c^*=-1/\eta_c^*\simeq -0.13622121$. When $1/\lambda\rightarrow 0$, the distribution is concentrated at the boundary and we numerically find that $\Lambda\rightarrow 0.1$.

$\bullet$ In $d=3$: for the ordinary Newtonian model, the
series of equilibria is parameterized by $\alpha$. When $\alpha\rightarrow +\infty$, the
distribution tends to the singular isothermal sphere $\rho_s(r)=1/(2\pi G\beta m r^2)$ with energy
$\Lambda_s=1/4$. The curve $\Lambda(\alpha)$ undergoes damped
oscillations around this value. When
$\alpha\rightarrow 0$, the distribution is homogeneous and the energy
$\Lambda\rightarrow -\infty$.   For the modified Newtonian model, the
series of equilibria is parameterized by $1/\lambda$. When
$1/\lambda\rightarrow +\infty$, the distribution is concentrated at
the center and we numerically find that $\Lambda\rightarrow -0.38...$
(the value is different from the Newtonian result $\Lambda_s=1/4$ due to different 
boundary conditions). The
curve $\Lambda(\lambda)$ undergoes damped oscillations around this
value. When $\lambda=1$ the distribution is homogeneous and
$\Lambda=\Lambda_c^*=-3/(2\eta_c^*)\simeq -0.22287452$. When
$\lambda\rightarrow 0$, the distribution is concentrated at the
boundary and we numerically find that $\Lambda\rightarrow 0.05$.

\subsection{The entropy and the free energy}
\label{sec_sfmodif}

Finally, we relate the central density $1/\lambda$ to the entropy $S$ and to the free energy $F$. Using Eqs. (\ref{mce2}), (\ref{mce6}) and (\ref{mce18}), the entropy is given by
\begin{eqnarray}
\label{mn16b}
S=\frac{d}{2}Nk_B \ln T-k_B\int \frac{\rho}{m}\ln\frac{\rho}{m}\, d{\bf r}.
\end{eqnarray}
Substituting Eq. (\ref{mn3}) in Eq. (\ref{mn16b}), and introducing the dimensionless variables defined previously, we get
\begin{eqnarray}
\label{mn17}
\frac{S}{N k_B}=-\frac{d}{2}\ln\beta-\ln\rho_0\nonumber\\
+\frac{\rho_0}{Nm}S_d\left (\frac{R}{\alpha}\right )^d\int_{0}^{\alpha}\psi e^{-\psi}\xi^{d-1}\, d\xi,
\end{eqnarray}
up to some unimportant constants. Using $\alpha= (S_d G\beta m \rho_0)^{1/2} R$ to express $\rho_0$ in terms of $\alpha$ and introducing the normalized  temperature (\ref{mn9}), we finally obtain
\begin{eqnarray}
\label{mn18}
\frac{S}{N k_B}=-\frac{d-2}{2}\ln\eta-2\ln\alpha\nonumber\\
+\frac{1}{\eta}\frac{1}{\alpha^{d-2}}\int_{0}^{\alpha}\psi e^{-\psi}\xi^{d-1}\, d\xi,
\end{eqnarray}
up to some unimportant constants. Using the previous results, this expression relates the entropy $S/N k_B$ to the central density $1/\lambda$. The free energy is $F=E-TS$. In the following, it will be more convenient to work in terms of the Massieu function $J=S-k_B \beta E$ (by an abuse of language, we shall often refer to $J$ as the free energy). We have
\begin{eqnarray}
\label{mn19}
\frac{J}{N k_B}=\frac{S}{N k_B}+\eta\Lambda.
\end{eqnarray}
Using the previous results, this expression relates the free energy  $J/N k_B$ to the central density $1/\lambda$.

\begin{figure}
\begin{center}
\includegraphics[clip,scale=0.3]{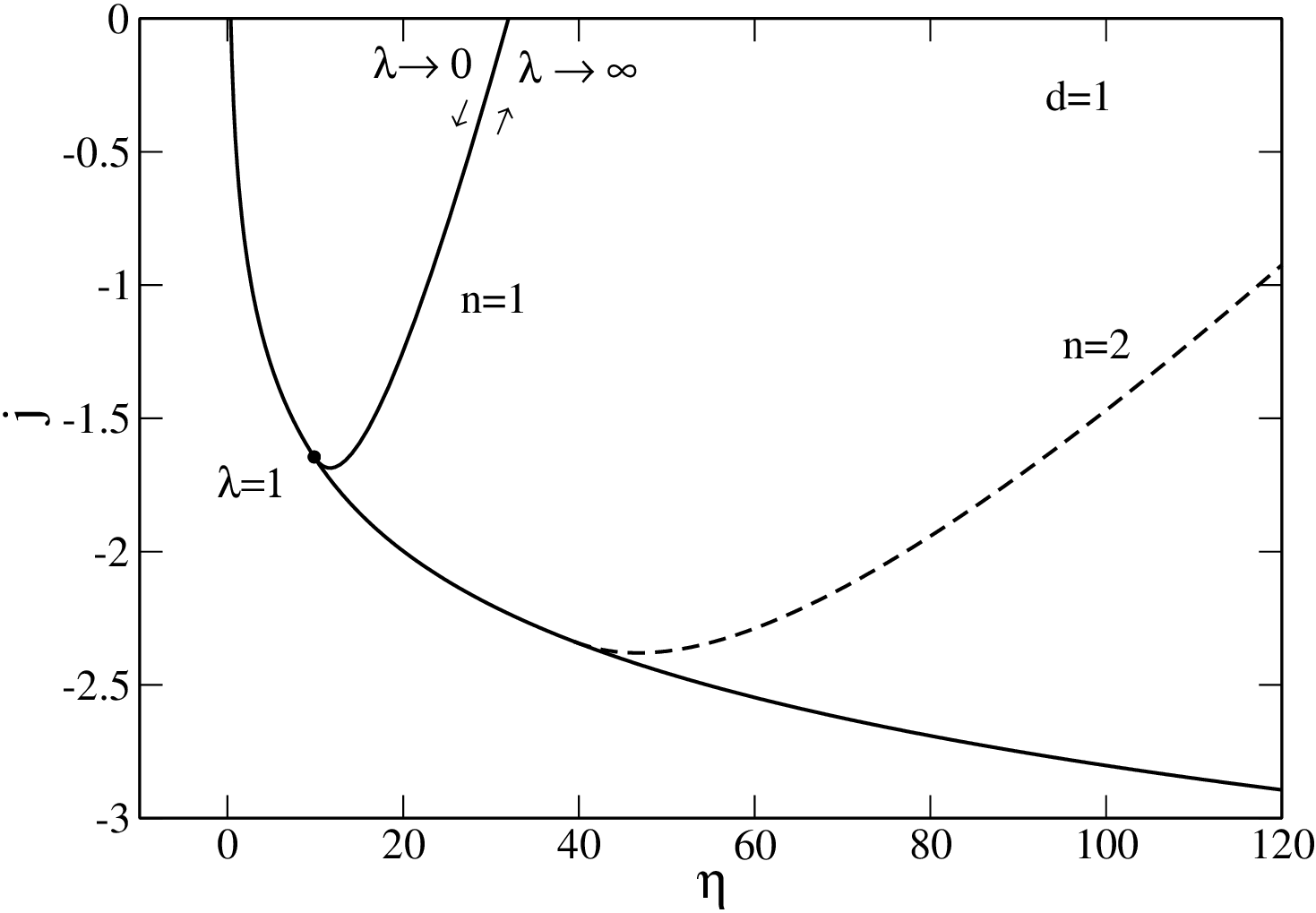}
\caption{Free energy $\frac{J}{Nk_B}$ as a function of the inverse temperature $\eta$ in $d=1$. Note that the branches $\lambda<1$ and $\lambda>1$ coincide. }
\label{free1}
\end{center}
\end{figure}

\begin{figure}
\begin{center}
\includegraphics[clip,scale=0.3]{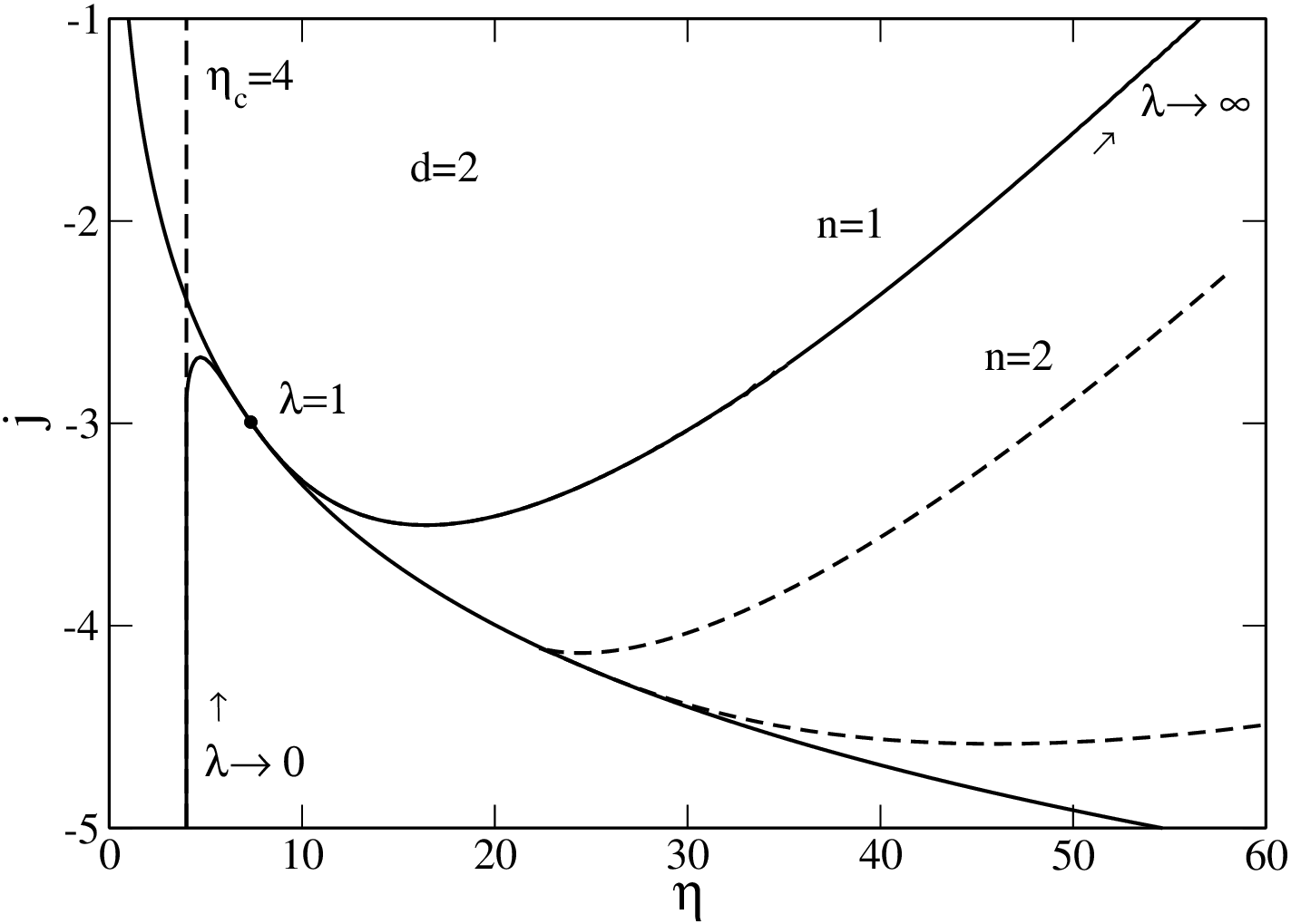}
\caption{Free energy $\frac{J}{Nk_B}$ as a function of the inverse temperature $\eta$ in $d=2$. }
\label{free2}
\end{center}
\end{figure}

\begin{figure}
\begin{center}
\includegraphics[clip,scale=0.3]{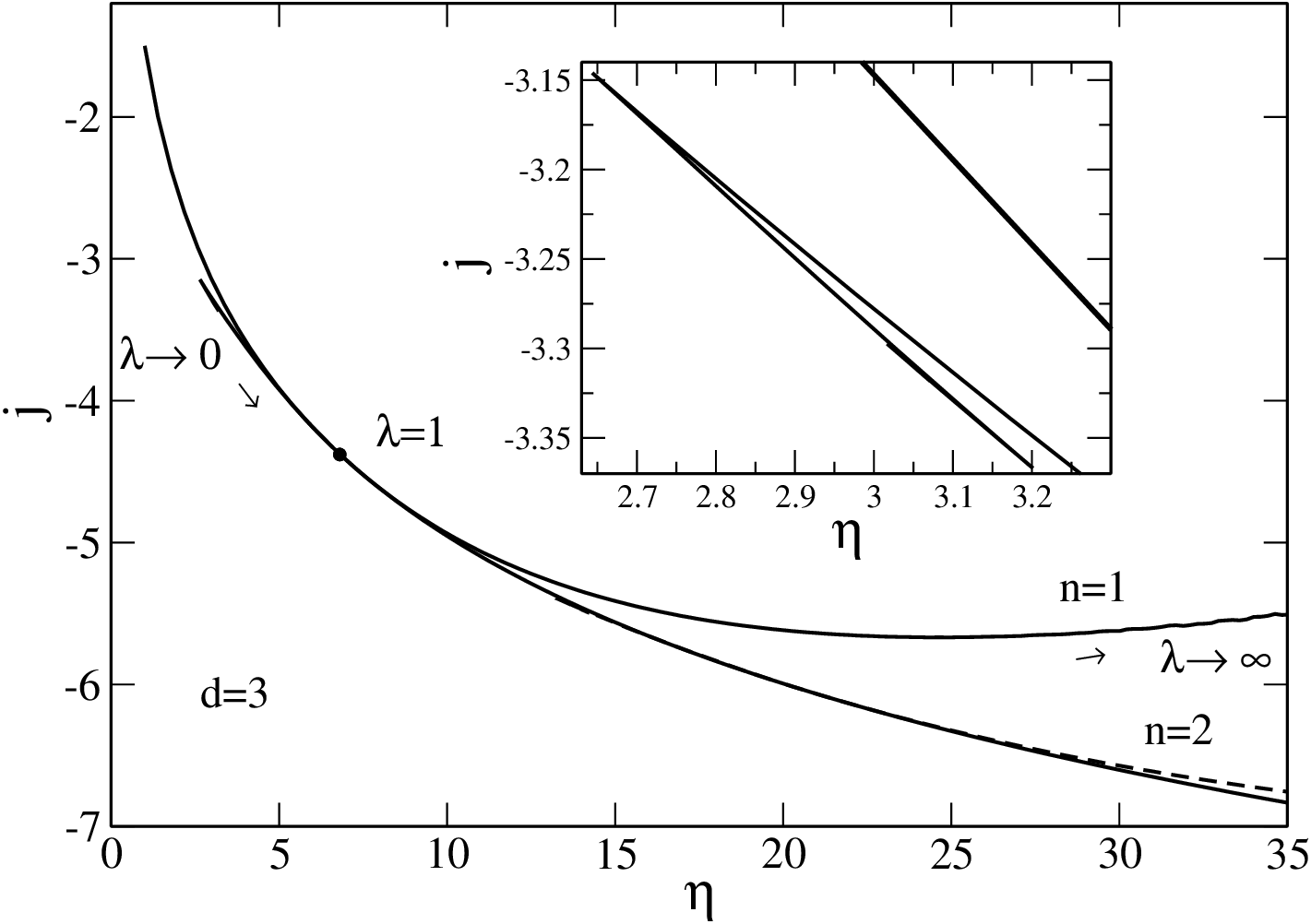}
\caption{Free energy $\frac{J}{Nk_B}$ as a function of the inverse temperature $\eta$ in $d=3$. }
\label{free3}
\end{center}
\end{figure}

\begin{figure}
\begin{center}
\includegraphics[clip,scale=0.3]{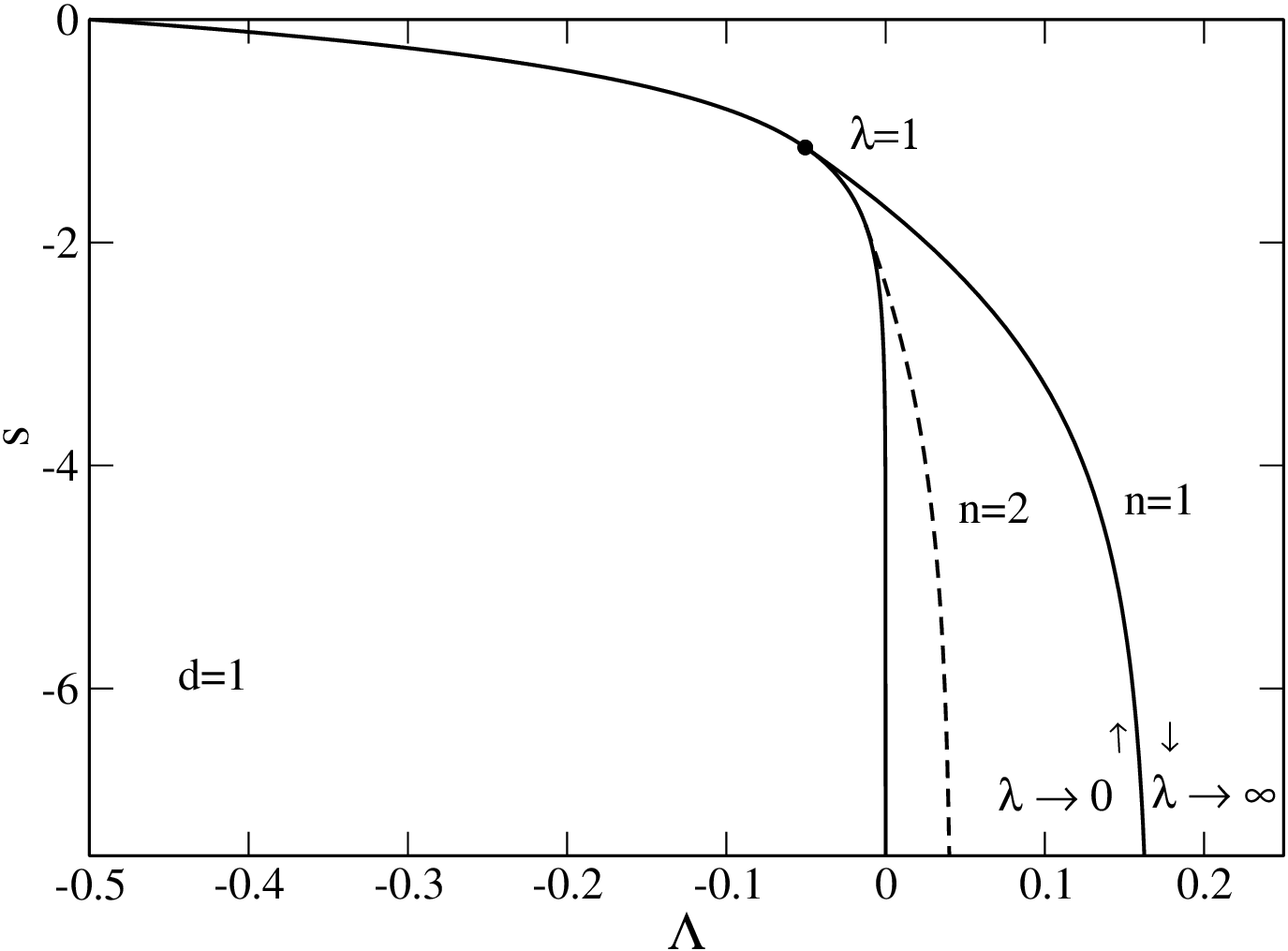}
\caption{Entropy $\frac{S}{Nk_B}$ as a function of energy  $\Lambda$ in $d=1$. Note that the branches $\lambda<1$ and $\lambda>1$ coincide.}
\label{entropy1}
\end{center}
\end{figure}

\begin{figure}
\begin{center}
\includegraphics[clip,scale=0.3]{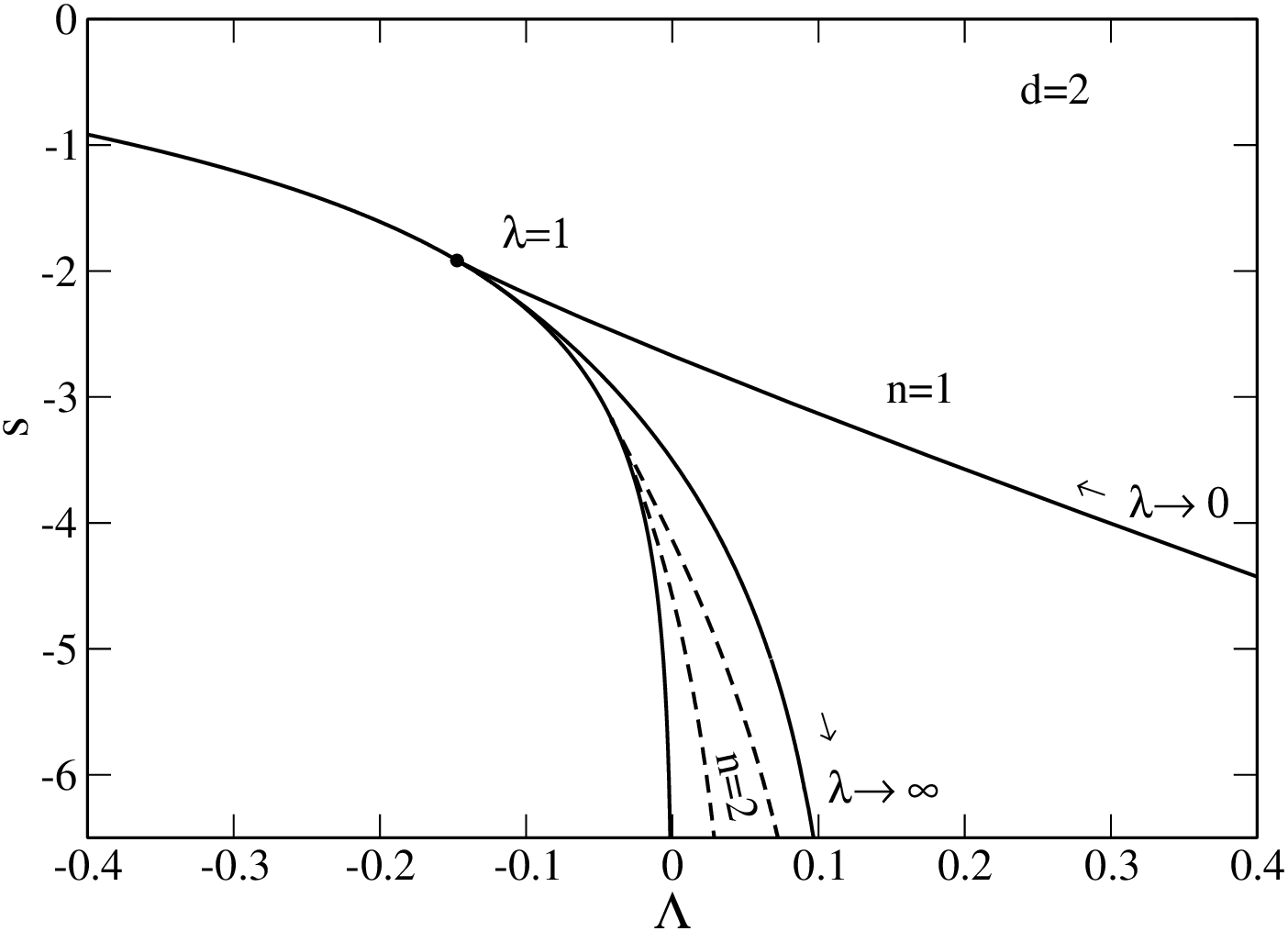}
\caption{Entropy $\frac{S}{Nk_B}$ as a function of energy  $\Lambda$ in $d=2$.}
\label{entropy2}
\end{center}
\end{figure}

\begin{figure}
\begin{center}
\includegraphics[clip,scale=0.3]{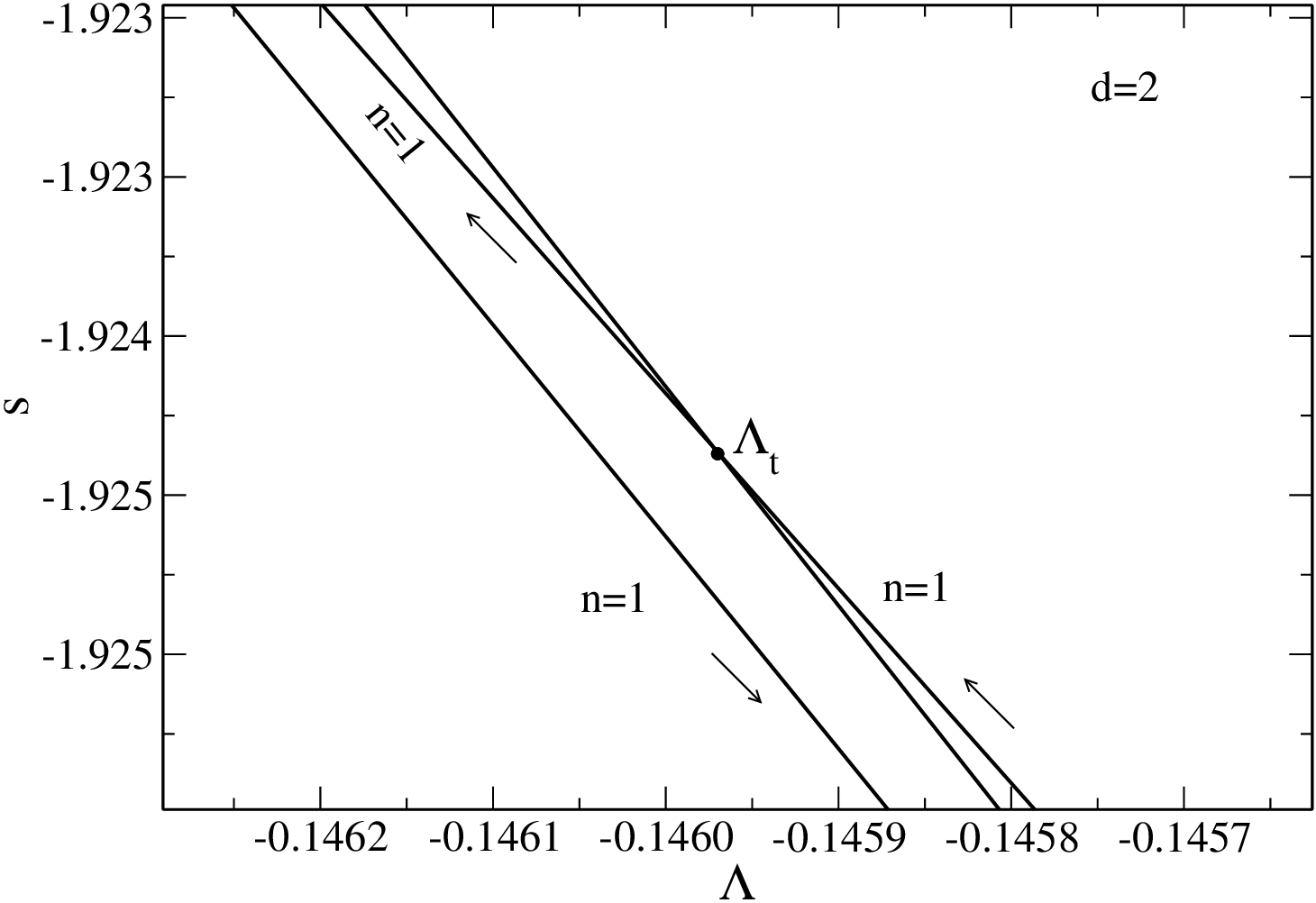}
\caption{Enlargement of Fig. \ref{entropy2}. The entropies of the homogeneous phase and inhomogeneous phase become equal at
$\Lambda=\Lambda_t\simeq -0.146$. This corresponds to a first-order phase transition in the microcanonical ensemble marked by the discontinuity of the slope $S'(E)=1/T$.}
\label{entropy2-zoom}
\end{center}
\end{figure}

\begin{figure}
\begin{center}
\includegraphics[clip,scale=0.3]{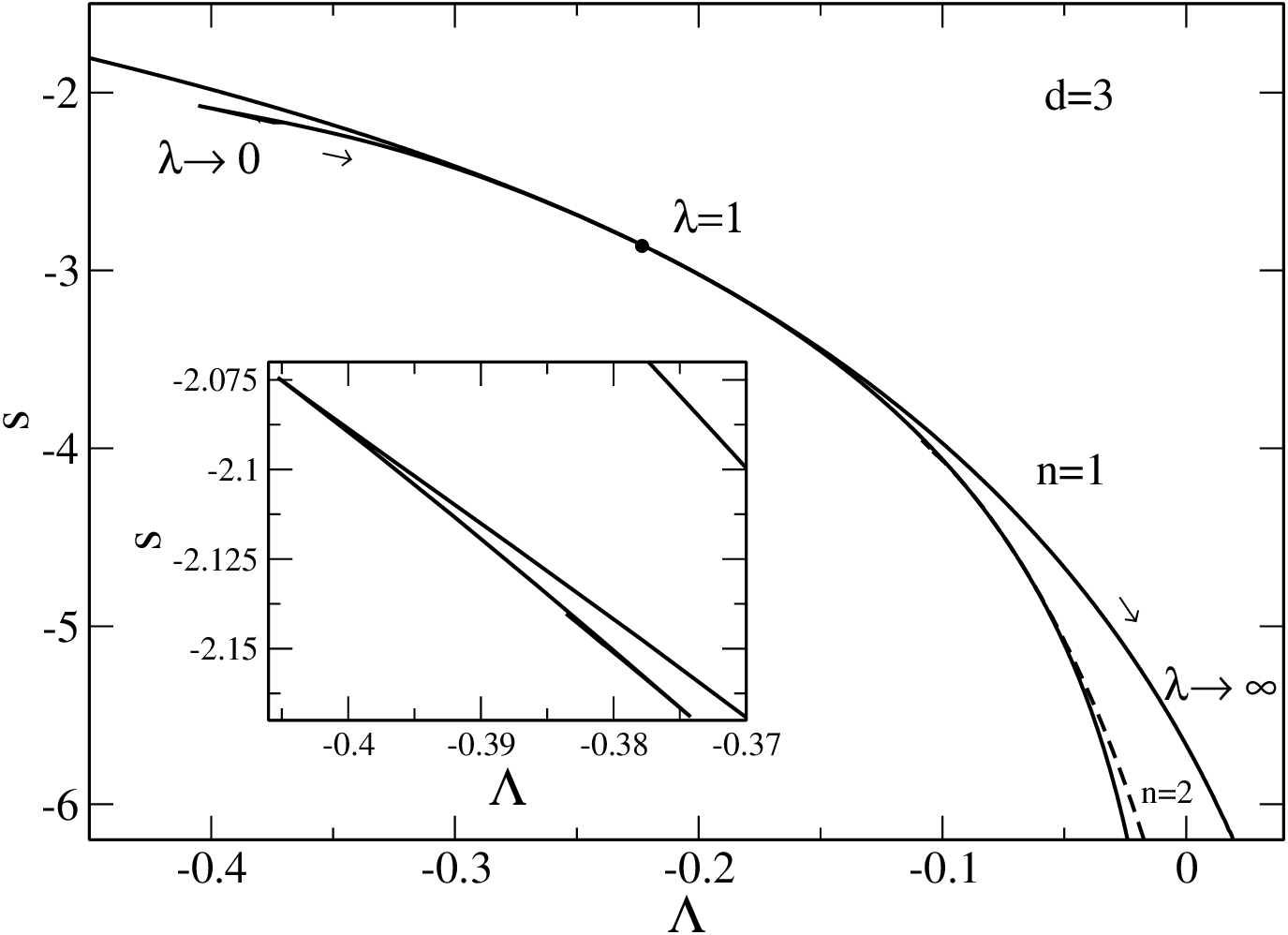}
\caption{Entropy $\frac{S}{Nk_B}$ as a function of energy  $\Lambda$ in $d=3$.}
\label{entropy3}
\end{center}
\end{figure}

In Figs. \ref{free1}, \ref{free2} and \ref{free3}, we have plotted the free energy  $J/Nk_B$ as a function of the inverse temperature $\eta$ (parameterized by the central density $1/\lambda$) in $d=1,2,3$.
In Figs. \ref{entropy1}, \ref{entropy2}, \ref{entropy2-zoom} and \ref{entropy3}, we have plotted the entropy $S/Nk_B$ as a function of the energy  $\Lambda$ (parameterized by the central density $1/\lambda$) in $d=1,2,3$. In these figures, the solid lines without label refer to the homogeneous phase. The solid lines with label $n=1$ refer to the first inhomogeneous  branch. The dashed lines with label $n=2$ refer to the second inhomogeneous branch.  These curves will be helpful in the next section to analyze the phase transitions in the canonical and microcanonical ensembles respectively.

{\it Remark:} Since $\delta S=k_B\beta\delta E$, the extrema of
entropy $S(\lambda)$ and energy $E(\lambda)$ coincide.  Since the
series of equilibria $E(\lambda)$ exhibits damped oscillations for
$1/\lambda\rightarrow +\infty$ in $d=3$ (see Fig. \ref{ene3}), this
implies that the curve $S(\lambda)$ will also exhibit damped
oscillations at the same locations. Correspondingly, $S(E)$ will
present some ``spikes'' for $1/\lambda\rightarrow +\infty$ in $d=3$
(see inset of Fig. \ref{entropy3}). Similarly, since $\delta J=-E
k_B\delta \beta$, the extrema of free energy $J(\lambda)$ and
temperature $\beta(\lambda)$ coincide. Since the series of equilibria
$\beta(\lambda)$ undergoes damped oscillations for
$1/\lambda\rightarrow +\infty$ in $d=3$ (see Fig. \ref{eta3}), this
implies that the curve $J(\lambda)$ will also exhibit damped
oscillations at the same location, and that the curve $J(\beta)$ will
present some ``spikes'' for $1/\lambda\rightarrow +\infty$ in $d=3$
(see inset of Fig. \ref{free3}). In addition, the curve $J(\beta)$
presents a minimum for $\eta\simeq 24.7$ corresponding to
$E=0$. Similar behaviors were previously observed in the model of
self-gravitating fermions \cite{prefermi,ijmpb}.

\subsection{Caloric curves and phase transitions}

We shall now determine the caloric curve $\beta(E)$ corresponding to the modified Newtonian model. First of all, we note that for the homogeneous phase, the potential energy $W=0$ so that the energy reduces to the kinetic energy. Therefore, the series of equilibria of the homogeneous phase is simply
\begin{eqnarray}
\label{mn20}
\eta=-\frac{d}{2\Lambda}.
\end{eqnarray}
On the other hand, eliminating $\lambda$ between $\eta_n(\lambda)$ and
$\Lambda_n(\lambda)$ given by Eqs. (\ref{mn10}) and (\ref{mn16}), we
get the series of equilibria $\eta_n(\Lambda)$ for the $n$-th
inhomogeneous branch. The series of equilibria contain all the
critical points of the optimization problems (\ref{mce1}) and
(\ref{ce1}). The series of equilibria are the same in the canonical
and microcanonical ensembles because the critical points are the
same. They contain fully stable states (global maxima of $S$ or $J$),
metastable states (local maxima of $S$ or $J$) and unstable states
(saddle points of $S$ or $J$). The stable parts of the series of
equilibria form the caloric curves in the canonical and microcanonical
ensembles. We shall distinguish the strict caloric curves formed by
fully stable states and the physical caloric curves containing fully
stable and metastable states \footnote{In many papers, only fully stable states forming the strict caloric curve are indicated. We think that clarity is gained when the full series of equilibria is shown. Then, we can see where the stable, metastable and unstable branches are located and how they are connected to each other. This also allows us to use the Poincar\'e theory of linear series of equilibria to settle their stability without being required to study an eigenvalue equation associated with the second order variations of the thermodynamical potential \cite{katz,ijmpb}.}. Metastable states are important because
they can be long-lived in systems with long-range interaction
\cite{ruffo,lifetime}. The caloric curves may differ in CE and MCE
in case of ensembles inequivalence. They are described below in
different dimensions of space.

{\it Remark:} In order to determine the stable branch, we shall compare the entropy (in MCE)  or the free energy (in CE) of the different solutions in competition (with the same values of energy or temperature). However, this is not sufficient because a distribution could have a high entropy and be an unstable saddle point. A more rigorous study should therefore investigate the sign of the second order variations of entropy or free energy for each critical point. But this is a difficult task that is left for future works. In order to find the stable states, we shall use physical considerations and exploit results obtained in related studies.

\subsubsection{The dimension $d=1$}

\begin{figure}[!ht]
\begin{center}
\includegraphics[clip,scale=0.3]{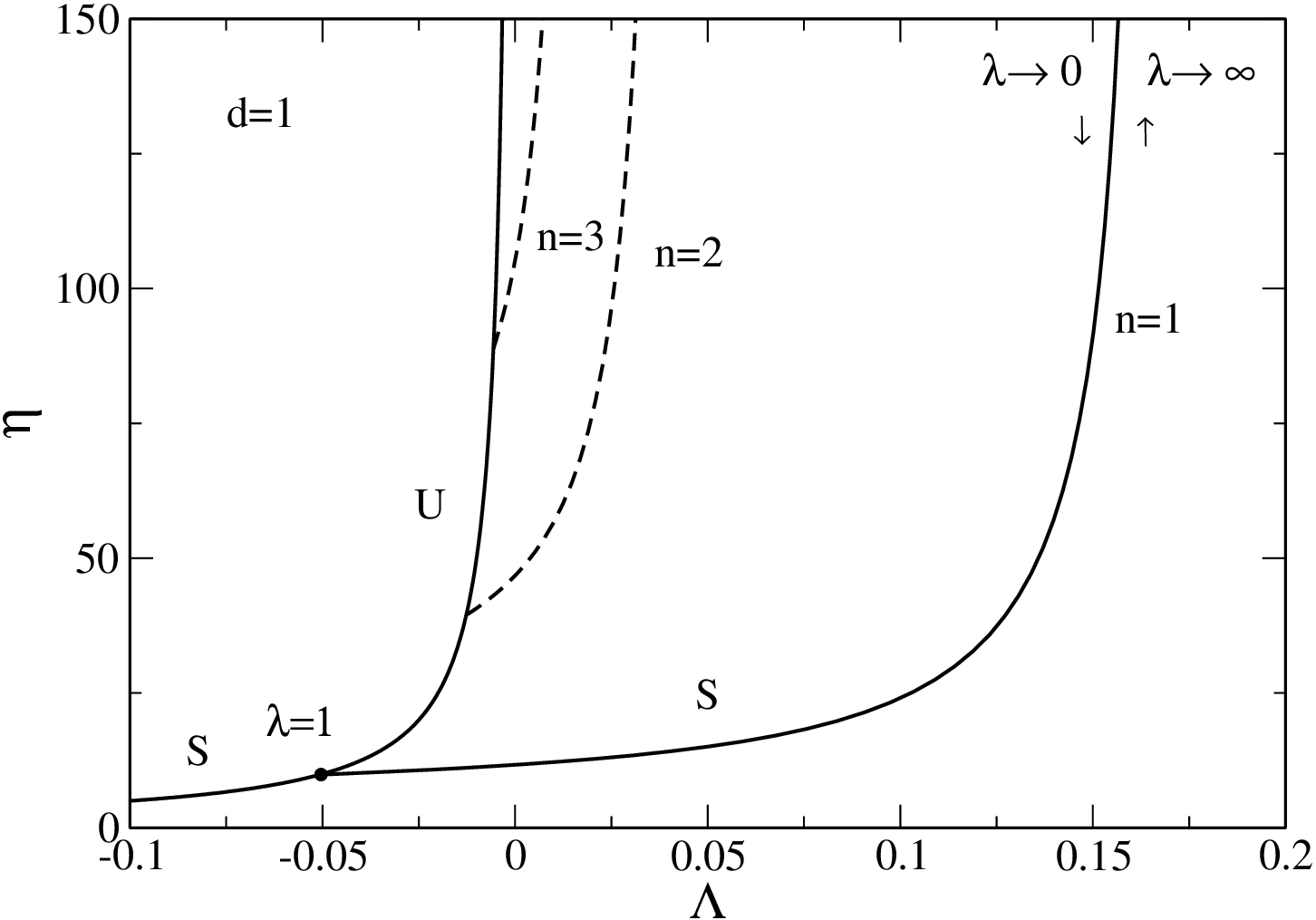}
\caption{Series of equilibria in $d=1$.
The caloric curve displays a second order phase transition in CE and
MCE taking place at $\eta=\eta_c^*$ and $\Lambda=\Lambda_c^*$ (corresponding to $\lambda=1$).  It is
marked by the discontinuity of $\partial\beta/\partial E$ in MCE or
$\partial E/\partial \beta$ in CE. Note that the branches $\lambda<1$
and $\lambda>1$ coincide. The corresponding density profiles are
plotted in Fig. \ref{prof}.}
\label{fig1}
\end{center}
\end{figure}

In Fig. \ref{fig1} we plot the series of equilibria in $d=1$.

\begin{figure}
\begin{center}
\includegraphics[clip,scale=0.3]{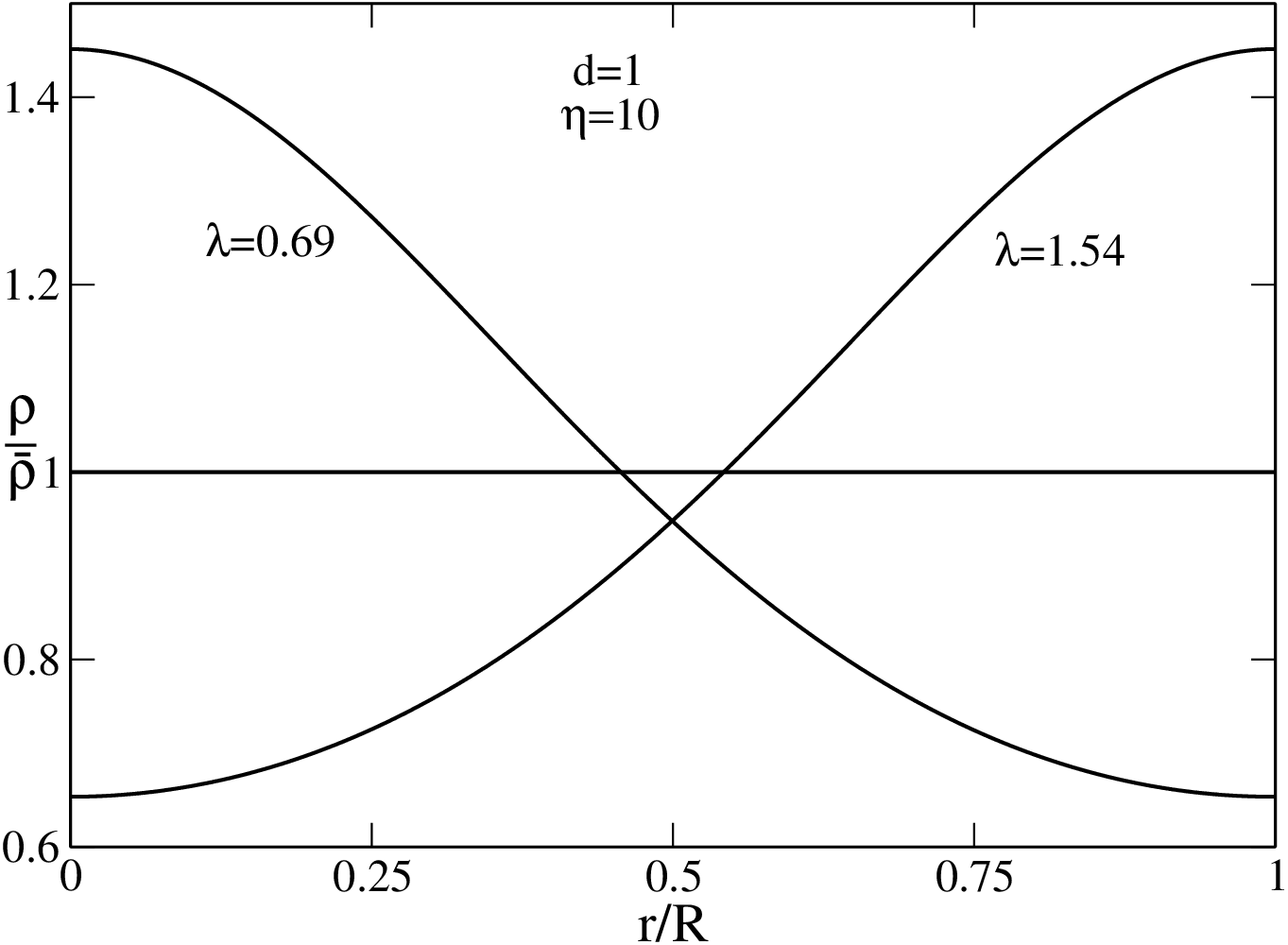}
\caption{Density profiles of the two stable inhomogeneous solutions $\lambda_1=0.69<1$ and $\lambda_2=1.54>1$  corresponding to $\eta=10$ in $d=1$. We have also represented the unstable homogeneous solution.}
\label{prof}
\end{center}
\end{figure}

Let us first describe the canonical ensemble (CE). The control
parameter is the inverse temperature $\eta$ and the stable states are
maxima of free energy $J$ at fixed mass $M$. The homogeneous phase
exists for any value of $\eta$. It is fully stable for $\eta<\eta_c^*$
and unstable for $\eta>\eta_c^*$ (see Sec. \ref{sec_stab}). The first
branch $n=1$ of inhomogeneous states exists only for
$\eta>\eta_c^*$. It has a higher free energy $J$ than the homogeneous
phase (see Fig. \ref{free1}) and it is fully stable. Secondary
branches of inhomogeneous states appear for smaller values of the
temperature but they have smaller values of free energy $J$ (see
Fig. \ref{free1}) and they are unstable (saddle points of free
energy). Therefore, the canonical caloric curve displays a second
order phase transition between homogeneous and inhomogeneous states
marked by the discontinuity of $\frac{\partial E}{\partial\beta}$ at
$\beta=\beta_c^*$. For the inhomogeneous states, there
exists two solutions with the same temperature and the same free
energy but with different density profiles corresponding to
$\lambda_1<1$ and $\lambda_2>1$ (see Fig. \ref{prof}). Thus, the
inhomogeneous branch is degenerate. These two states can be
distinguished by their central density $1/\lambda$. In conclusion: (i)
for $\eta<\eta_c^*$, there is only one stable state $\lambda=1$
(homogeneous); (ii) for $\eta>\eta_c^*$, there are two stable states
$\lambda_1<1$ and $\lambda_2>1$ (inhomogeneous) with the same free
energy and one unstable state $\lambda=1$ (homogeneous).  Therefore,
the central density $1/\lambda$ plays the role of an order parameter
(see Fig. \ref{eta1}). In $d=1$, there exists a fully stable
equilibrium state for any temperature. This is consistent with the
usual Newtonian model in $d=1$ \cite{kl,sc}. This is also consistent
with results of chemotaxis since it has been rigorously proven that
the Keller-Segel model does not blow up in $d=1$ \cite{nagai}.

Let us now describe the microcanonical ensemble (MCE). The control parameter is the energy $\Lambda$ and the stable states are maxima of entropy $S$ at fixed mass $M$ and energy $E$. The homogeneous phase exists for any value of energy $\Lambda<0$. It is fully stable for $\Lambda<\Lambda_c^*$  and unstable for $\Lambda>\Lambda_c^*$ (see Sec. \ref{sec_stab}). The first branch $n=1$ of inhomogeneous states exists only for $\Lambda_c^*<\Lambda<\Lambda_{max}$. It has a higher entropy $S$ than the homogeneous phase (see Fig. \ref{entropy1}) and it is fully  stable. Secondary branches of inhomogeneous states appear for smaller values of the energy but they have smaller values of  entropy $S$ (see Fig. \ref{entropy1}) and they are unstable (saddle points of entropy). Therefore, the microcanonical caloric curve displays a second order phase transition marked by the discontinuity of $\frac{\partial \beta}{\partial E}$ at $E=E_c^*$. For the inhomogeneous states, there exists two solutions with the same energy  and the same entropy but with different density profiles corresponding to $\lambda_1<1$ and $\lambda_2>1$. Thus, the inhomogeneous branch is degenerate. These two states can be distinguished by their central density  $1/\lambda$. In conclusion: (i) for $\Lambda<\Lambda_c^*$, there is only one stable state $\lambda=1$ (homogeneous); (ii) for $\Lambda_c^*<\Lambda<0$, there are two stable states $\lambda_1<1$ and $\lambda_2>1$ (inhomogeneous) with the same entropy and one unstable state $\lambda=1$ (homogeneous). (iii) for $0<\Lambda<\Lambda_{max}$, there are two stable states $\lambda_1<1$ and $\lambda_2>1$ (inhomogeneous) with the same entropy. Therefore, the central density $1/\lambda$ plays the role of an order parameter (see Fig. \ref{ene1}). In $d=1$, there exists a fully stable  equilibrium state for any accessible energy. This is consistent with the usual Newtonian model in $d=1$ \cite{kl,sc}.

The caloric curve, corresponding to the fully stable states in the series of equilibria, is denoted by (S) in Fig. \ref{fig1}. The branch (U) corresponds to unstable states. There exists a fully stable equilibrium state for any accessible values of energy in MCE and temperature in CE. The microcanonical and canonical ensembles are equivalent (like in the Newtonian case).

In conclusion, the system displays second order phase transition in CE and MCE. This is similar to the HMF model \cite{inagaki,ar,cvb}.

\subsubsection{The dimension $d=2$}
\label{sec_modif2}

\begin{figure}[!ht]
\begin{center}
\includegraphics[clip,scale=0.3]{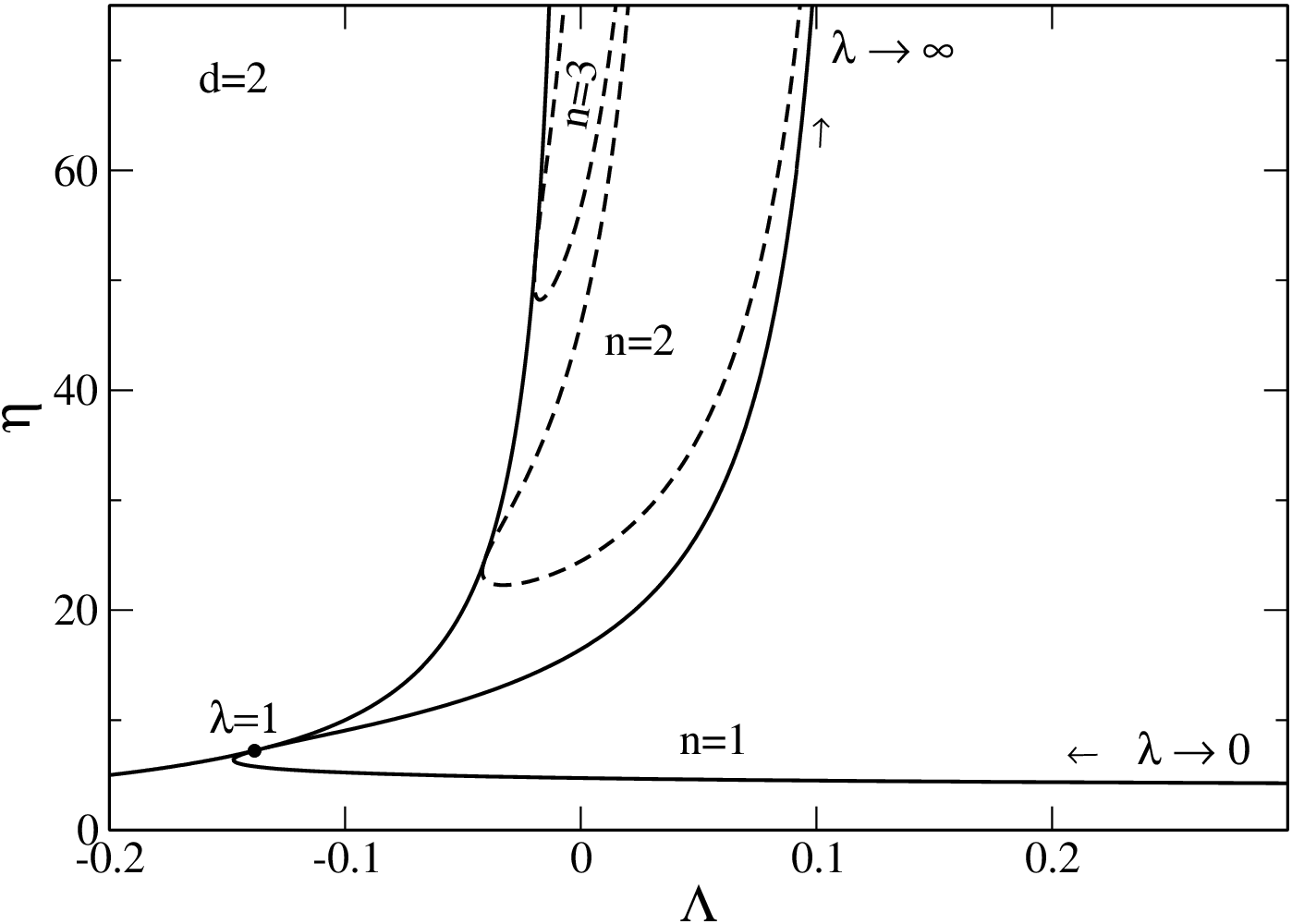}
\caption{Series of equilibria in $d=2$. The first inhomogeneous branch $n=1$  tends to a plateau $\eta_c=4$ for large central densities $1/\lambda\rightarrow +\infty$ due to the formation of a Dirac peak. This is similar to the plateau appearing in the caloric curve of the classical self-gravitating gas \cite{sc}.}
\label{fig2}
\end{center}
\end{figure}

In Fig. \ref{fig2} we plot the series of equilibria in $d=2$.

\begin{figure}
\begin{center}
\includegraphics[clip,scale=0.3]{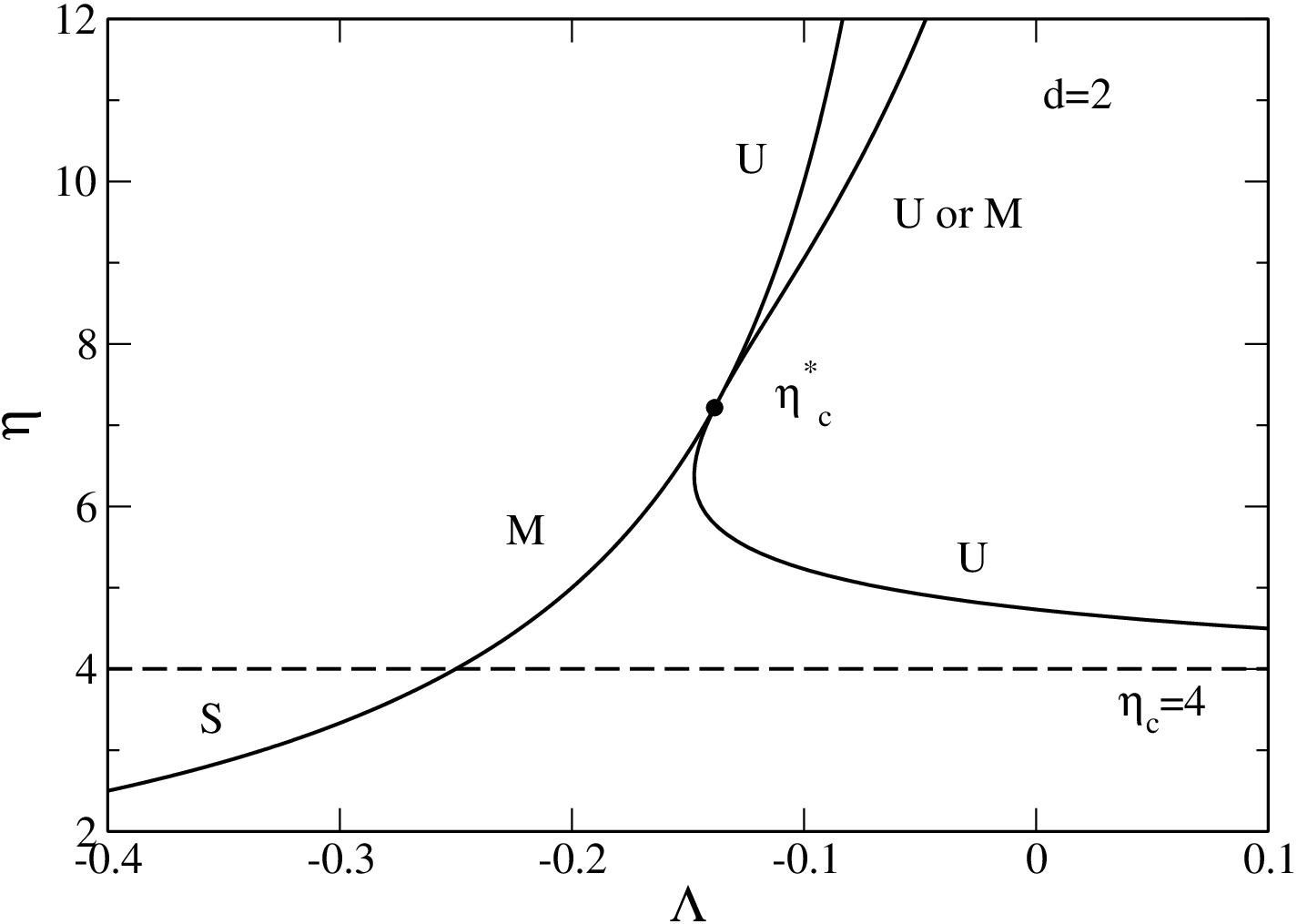}
\caption{Caloric curve in the canonical ensemble in $d=2$. The homogeneous branch is fully stable for $\eta<\eta_c=4$, metastable for $\eta_c<\eta<\eta_c^*$ and unstable for $\eta>\eta_c^*$. The inhomogeneous branch is always unstable (or, possibly,  metastable for $\eta>\eta_c^*$). For sufficiently low temperatures, the system can experience an isothermal collapse. }
\label{canonic}
\end{center}
\end{figure}

Let us first describe the canonical ensemble (CE). The control
parameter is the inverse temperature $\eta$.  The homogeneous phase
exists for any $\eta$. It is stable for $\eta<\eta_c^*$ and unstable
for $\eta>\eta_c^*$ (see Sec. \ref{sec_stab}). The first branch $n=1$
of inhomogeneous states exists for $\eta>\eta_c=4$ and it connects the
homogeneous branch at $\eta_c^*$. For $\eta<\eta_c^*$, it has a lower
free energy $J$ than the homogeneous phase (see Fig. \ref{free2}) and
it is unstable. For $\eta>\eta_c^*$, it has a higher free energy $J$
than the homogeneous phase (see Fig. \ref{free2}). However, it is
expected to be unstable or, possibly, metastable (to settle this issue
we have to study the sign of the second order variations of free
energy as explained above). Secondary inhomogeneous branches appear
for smaller values of the temperature but they have smaller values of
the free energy (see Fig. \ref{free2}) and they are unstable. The
homogeneous branch is expected to be fully stable for $\eta<\eta_c=4$
and metastable for $\eta_c=4<\eta<\eta_c^*$ (see Fig. \ref{canonic}).
These conclusions are motivated by two arguments: (i) in the Newtonian
model in $d=2$, we know that there exists a fully stable equilibrium
state for $\eta<\eta_c=4$ and no equilibrium state for
$\eta>\eta_c=4$. In that case, the system undergoes an {\it isothermal
collapse} \cite{klb,sc}. For $\eta>\eta_c=4$, there is no global
maximum of free energy $J$ because we can make it diverge by creating
a Dirac peak containing all the particles. In the modified Newtonian
model, the same argument applies since it is independent of boundary
conditions.  Since we know that the homogeneous branch is stable for
$\eta<\eta_c^*$, we conclude that it must be metastable in the range
$\eta_c=4<\eta<\eta_c^*$. There is therefore a zeroth
order phase transition at $\eta_c=4$ marked by the discontinuity of
the free energy. (ii) In the chemotactic literature, it has been
rigorously established that the Keller-Segel model in $d=2$ does not
blow up for $\eta<\eta_c=4$ while it can blow up for $\eta>\eta_c=4$
\cite{nagai}. This is consistent with our stability results.

\begin{figure}
\begin{center}
\includegraphics[clip,scale=0.3]{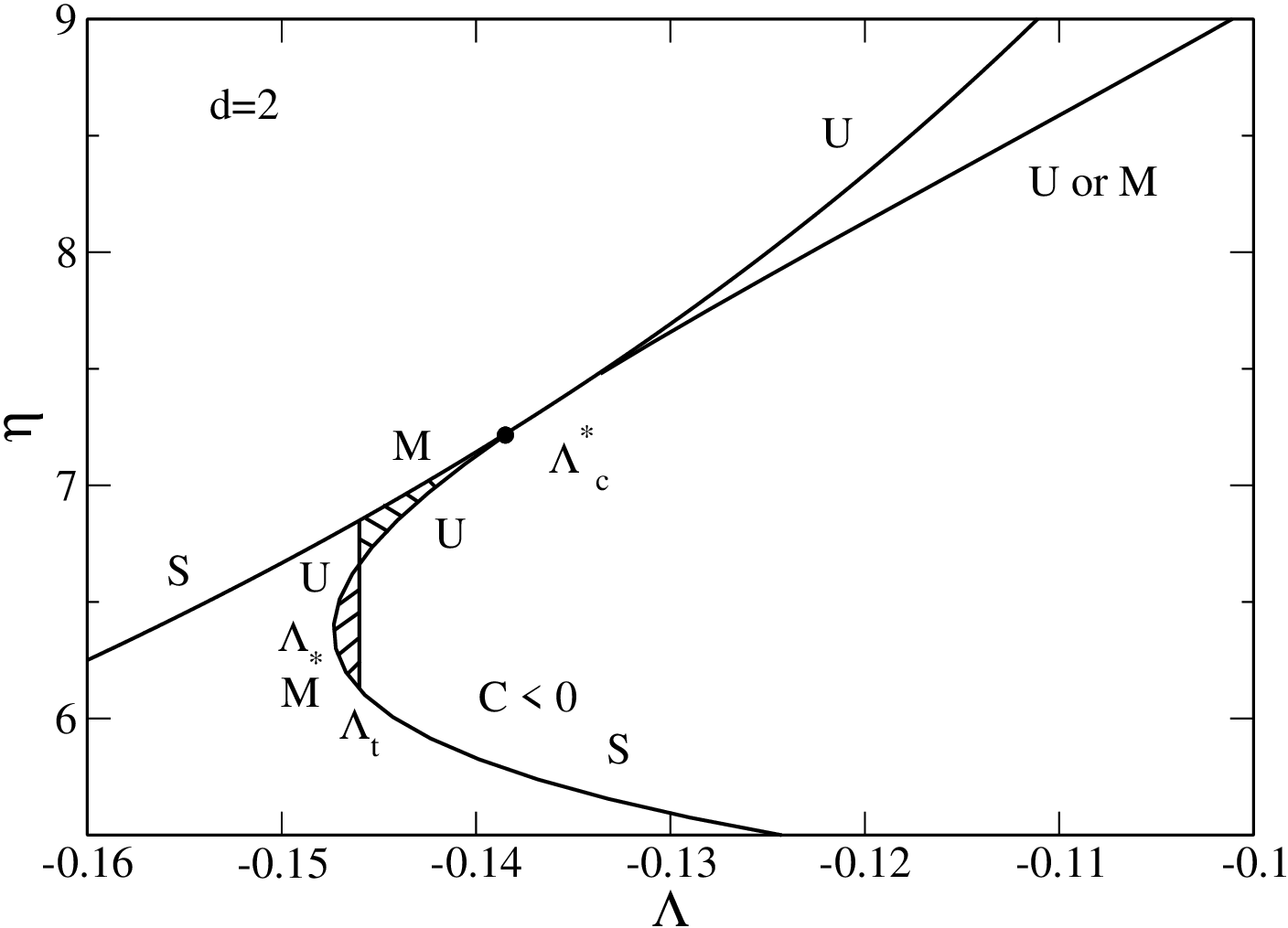}
\caption{Caloric curve in the microcanonical ensemble in $d=2$. A first-order phase transition is expected to take place in the microcanonical ensemble at $\Lambda=\Lambda_t$. Note that the lower branch has negative specific heats.  $\Lambda_*$ and  possibly $\Lambda_c^*$ represent microcanonical spinodal points marking the end of the metastable phase.}
\label{micro}
\end{center}
\end{figure}

Let us now describe the microcanonical ensemble (MCE). The control parameter is the energy  $\Lambda$. The homogeneous phase exists for all values of $\Lambda<0$. It is stable for $\Lambda<\Lambda_c^*$ and unstable for $\Lambda>\Lambda_c^*$  (see Sec. \ref{sec_stab}). The first branch $n=1$ of inhomogeneous states exists for $\Lambda>\Lambda_*$ and it connects the homogeneous branch at $\Lambda_c^*$. We see that the inhomogeneous branch  $\beta(E)$ is multi-valued. Considering the value of the entropy in the different phases (see Figs. \ref{entropy2} and \ref{entropy2-zoom}), the caloric curve is expected to display a microcanonical first order phase transition at $\Lambda = \Lambda_t\simeq -0.146$ marked by the discontinuity of the temperature (see Fig. \ref{micro}). The energy of transition has been determined by comparing the entropy of the homogeneous and inhomogeneous phases and looking at which point the curves $S(E)$ intersect (see Fig. \ref{entropy2-zoom}). Equivalently, it can be obtained by performing a vertical Maxwell construction \cite{ijmpb}.  The homogeneous phase is fully stable for $\Lambda<\Lambda_t$, metastable for $\Lambda_t<\Lambda<\Lambda_c^*$ and unstable for $\Lambda>\Lambda_c^*$. The lower part of the first inhomogeneous branch is fully stable for $\Lambda>\Lambda_t$ and  metastable for $\Lambda_*<\Lambda<\Lambda_t$. The upper part of the  first inhomogeneous branch is unstable for $\Lambda_*<\Lambda<\Lambda_c^*$. For $\Lambda>\Lambda_c^*$, it is unstable or, possibly, metastable. Secondary inhomogeneous branches appear for smaller values of the energy but they have smaller values of the entropy (see Fig. \ref{entropy2}) and they are unstable.  The stable states of the inhomogeneous branch have $1/\lambda>1$ indicating that the density is concentrated at the center. The possibly metastable states for $\Lambda>\Lambda_c^*$ have $1/\lambda<1$ indicating that the density is concentrated near the box. In conclusion, there exists a fully stable equilibrium state for any value of energy. This is similar to the Newtonian model in $d=2$ \cite{ap,sc}. However, in the modified Newtonian model, we expect a first order phase transition at $\Lambda_t$ that is not present in the Newtonian model.

The strict caloric curve, corresponding to the fully stable states (global maxima) in the series of equilibria, is denoted (S) in Figs. \ref{canonic} and \ref{micro}. The unstable states (saddle points) are denoted (U) and the metastable states (local maxima) are denoted (M). There exists a fully stable equilibrium state for any accessible value of energy in MCE and for sufficiently high values of the temperature in CE ($\eta<\eta_c=4$). Here, the microcanonical and canonical ensembles are inequivalent (unlike in the Newtonian case). In particular, the lower part of the  first inhomogeneous branch is stable in MCE while it is unstable in CE. This branch  has negative specific heats $C<0$ (see Fig. \ref{micro}) which is not possible in the canonical ensemble.

In conclusion, the system displays a zeroth order phase transition in CE (associated with an isothermal collapse) and a first order phase transition in MCE. Note also that the energy $E(\beta)$ and its first derivative $E'(\beta)$ are continuous at the critical point $\beta_c^*$ but its second derivative $E''(\beta)$ is discontinuous. Provided that the inhomogeneous branch for $\eta>\eta_c^*$ is metastable, this would correspond to a third order canonical phase transition between a homogeneous metastable state and an inhomogeneous metastable state.

\subsubsection{The dimension $d=3$}
\label{sec_modif3}

\begin{figure}[!ht]
\begin{center}
\includegraphics[clip,scale=0.3]{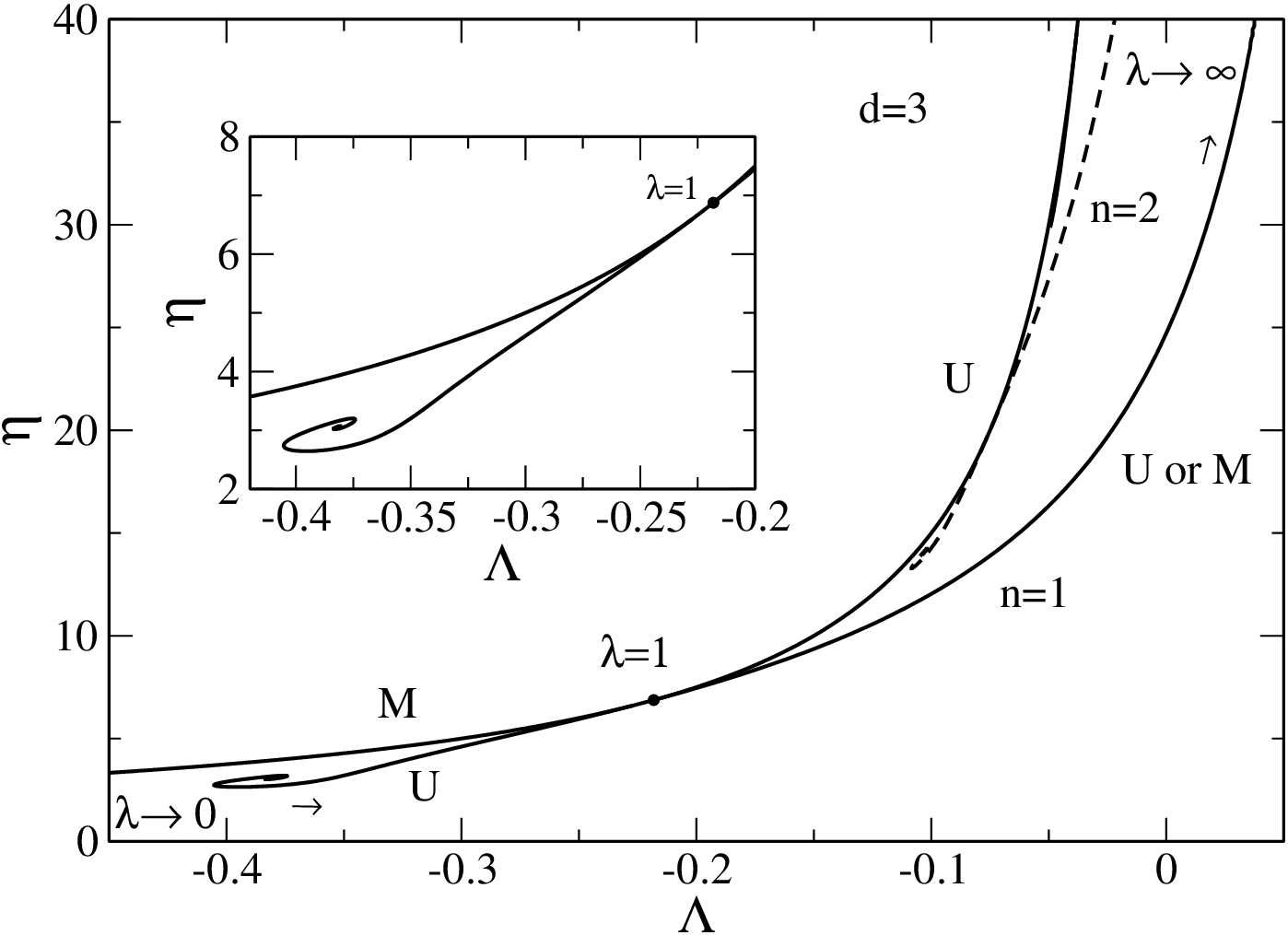}
\caption{Series of equilibria in $d=3$. The inhomogeneous branch forms a spiral for large central densities $1/\lambda\rightarrow +\infty$ due to the damped oscillations of the inverse temperature $\eta(\lambda)$ and energy $\Lambda(\lambda)$. This is similar to the spiral appearing in the series of equilibria of the classical self-gravitating gas as we approach the singular isothermal sphere \cite{sc}.}
\label{fig3}
\end{center}
\end{figure}

In Fig. \ref{fig3} we plot the series of equilibria in $d=3$.

Let us first describe the canonical ensemble (CE). The control
parameter is $\eta$. The homogeneous phase exists for all $\eta$.  It
is stable for $\eta<\eta_c^*$ and unstable for $\eta>\eta_c^*$ (see
Sec. \ref{sec_stab}).  The first branch $n=1$ of inhomogeneous states
exists for $\eta>2.64$ and it connects the homogeneous branch at
$\eta=\eta_c^*$. For large central densities $1/\lambda$, it forms a
spiral towards a singular solution. For $\eta<\eta_c^*$, it has a
lower free energy $J$ than the homogeneous phase (see
Fig. \ref{free3}) and it is unstable. For $\eta>\eta_c^*$, it has a
higher free energy $J$ than the homogeneous phase (see
Fig. \ref{free3}). However, it is expected to be unstable or,
possibly, metastable. Secondary inhomogeneous branches appear for
smaller values of the temperature but they have a higher value of free
energy $J$ and they are unstable. The homogeneous branch is metastable
for $\eta<\eta_c^*$. These conclusions are motivated by two arguments:
(i) in the Newtonian model in $d=3$, we know that there is no fully
stable equilibrium state in CE. The system can undergo an {\it
isothermal collapse} \cite{sc}. There is no global maximum of free
energy $J$ because we can make it diverge by creating a Dirac peak
containing all the particles \cite{kiessling}. In the modified
Newtonian model, the same argument applies since it is independent on
boundary conditions. Since we know that the homogeneous branch is stable
for $\eta<\eta_c^*$, then it can only be metastable. (ii) In the
chemotactic literature, it has been rigorously established that the
Keller-Segel model in $d=3$ can blow up for any $\eta$
\cite{nagai}. This is consistent with our stability results.

Let us now describe the microcanonical ensemble (MCE). The control
parameter is the energy $\Lambda$. The homogeneous phase exists for
all $\Lambda<0$. It is stable for $\Lambda<\Lambda_c^*$ and unstable
for $\Lambda>\Lambda_c^*$ (see Sec. \ref{sec_stab}). The first branch
$n=1$ of inhomogeneous states exists for $\Lambda>-0.405$ and it
connects the homogeneous branch at $\Lambda=\Lambda_c^*$.  For large
central densities $1/\lambda$, it forms a spiral towards a singular
solution. For $\Lambda<\Lambda_c^*$, it has a lower entropy $S$ than
the homogeneous phase (see Fig. \ref{entropy3}) and it is
unstable. For $\Lambda>\Lambda_c^*$, it has a higher entropy than the
homogeneous phase (see Fig. \ref{entropy3}). However, it is expected
to be unstable or, possibly, metastable. Secondary inhomogeneous branches appear for
smaller values of the energy but they have a lower value of entropy
$S$ and they are unstable.  The homogeneous branch is metastable for
$\Lambda<\Lambda_c^*$. These conclusions are motivated by two
arguments: (i) in the Newtonian model in $d=3$, we know that there is
no fully stable equilibrium state in MCE. The system undergoes a {\it
gravothermal catastrophe} \cite{antonov,lbw}. There is no global
maximum of entropy $S$ at fixed mass and energy because we can make it
diverge by creating a binary star surrounded by a hot halo
\cite{paddy,ijmpb}. In the modified Newtonian model, the same argument
applies. Since we know that the homogeneous branch is stable for
$\Lambda<\Lambda_c^*$, then it can only be metastable.

There is no strict  caloric curve since there is no fully stable states (global maxima). But there is a physical caloric curve made of metastable states (local maxima) denoted (M) in Fig. \ref{fig3}. The unstable states (saddle points) are denoted (U). Here, the microcanonical and canonical ensembles, regarding the metastable states, are equivalent unlike in the Newtonian case. This is because the homogeneous branch and the inhomogeneous branch connect each other at a {\it single} point at $\lambda=1$ by making a cusp (see inset in Fig. \ref{fig3}) while the Newtonian series of equilibria  is smooth and presents {\it two} distinct turning points of temperature and energy (denoted CE and MCE in Fig. 8 of \cite{ijmpb}) separated by a region of negative specific heats.

In conclusion, if we take  metastable states into account, the system displays a zeroth order phase transition in CE and MCE corresponding to a discontinuity of entropy or free energy. They are associated with an  isothermal collapse or a  gravothermal catastrophe respectively.

{\it Remark:} There is no natural external parameter in the modified
Newtonian model. However, the dimension of space $d$ could play the
role of an effective external parameter. The preceding results predict
the existence of a critical dimension $d_c$ between $1$ and $2$ at
which the phase transition passes from second order ($d<d_c$) to first
order ($d>d_c$). However, this transition turns out to occur in a very
small range of parameters since we find that the critical dimension
$d_c$ is between $1$ and $d=1.00001$ and the concerned range of
energies and temperatures is extremely narrow. We have not
investigated this transition in detail since the dimension of space is
not a physical (tunable) parameter. Furthermore, in the next model, we
have an external parameter $\mu$ played by screening length that is
more relevant.

\section{The screened Newtonian model}
\label{sec_yuk}

In this section, we discuss phase transitions that appear in the
screened Newtonian model corresponding to an attractive Yukawa
potential.

\subsection{Physical motivation of the model}

We consider a system of particles interacting via the potential $\Phi({\bf r},t)$ that is solution of the screened Poisson equation
\begin{eqnarray}
\label{ay1}
\Delta\Phi-k_0^2\Phi=S_d G\rho,
\end{eqnarray}
where $k_0$ is the inverse of the screening length. At statistical equilibrium, the density is given by the Boltzmann distribution
\begin{eqnarray}
\label{ay2}
\rho=A e^{-\beta m\Phi}.
\end{eqnarray}
We assume that the system is confined in a finite domain (box) and we impose the Neumann boundary conditions
\begin{eqnarray}
\nabla \Phi\cdot {\bf n}=0, \qquad  \nabla \rho\cdot {\bf n}=0,
\label{ksbcbbis}
\end{eqnarray}
where ${\bf n}$ is a unit vector normal to the boundary of the box (the explicit expression of the potential in $d=1$ is given in Appendix \ref{sec_exp}).  This model admits spatially homogeneous solutions ($\rho=\rho_0$ and $\Phi=\Phi_0$ with $-k_0^2\Phi_0=S_d G\rho_0$) at any temperature. It also admits spatially inhomogeneous solutions at sufficiently low temperatures. We shall study this model in arbitrary dimensions of space $d$ with explicit computations for $d=1,2,3$.  This model has different physical applications:

(i) It describes a system of particles interacting via  a screened attractive (Newtonian) potential.

(ii) By  a proper reinterpretation of the parameters, the field equation (\ref{ay1}) describes the relation between the concentration of the chemical and the density of bacteria in the Keller-Segel model (\ref{ks5}) where the degradation of the chemical reduces the range of the interaction. In that case, the boundary conditions are of the form (\ref{ksbcbbis}). Furthermore, the relevant ensemble is the CE since the KS model has a canonical structure. This model has been studied by Childress \& Percus \cite{cp} in $d=1$ using an approach different from the one we are going to develop.

For the sake of generality, we shall study this model in the microcanonical and canonical ensembles in any dimension of space.

\subsection{The screened Emden equation}

In the screened Newtonian model, the equilibrium density profile is given by the Boltzmann distribution (\ref{ay2})
coupled to the screened Poisson equation (\ref{ay1}). As in Sec. \ref{sec_modifemden}, we look for spherically symmetric solutions. Introducing the central density $\rho_0=\rho(0)$, the central potential $\Phi_0=\Phi(0)$, the new field $\psi = \beta m (\Phi - \Phi_0)$ and the scaled distance $\xi = (S_dG\beta m \rho_0)^{1/2} r$ the Boltzmann distribution can be rewritten
\begin{eqnarray}
\label{ay3}
\rho=\rho_0 e^{-\psi(\xi)}.
\end{eqnarray}
Substituting this relation in the screened Poisson equation (\ref{ay1}), we obtain the screened Emden equation
\begin{eqnarray}
\label{ay4}
\frac{1}{\xi^{d-1}}\frac{d}{d\xi}\left (\xi^{d-1}\frac{d\psi}{d\xi}\right )-\kappa^2\psi=e^{-\psi} - \lambda,
\end{eqnarray}
where $\kappa={k_0}/{(S_dG\beta m \rho_0)^{1/2}}$ and $\lambda=-{k_0^2\Phi_0}/{S_d G\rho_0}$. The boundary conditions at the origin are
\begin{eqnarray}
\label{ay5}
\psi (0)=\psi^{\prime}(0)=0.
\end{eqnarray}
The normalized box radius is $\alpha= (S_dG\beta m \rho_0)^{1/2} R$ and the boundary condition $\Phi^{\prime} (R)=0$ becomes
\begin{eqnarray}
\label{ay6}
\psi^{\prime}(\alpha)=0.
\end{eqnarray}
Introducing the normalized screening length
\begin{eqnarray}
\label{ay7}
\mu=k_0 R,
\end{eqnarray}
the parameter $\kappa$ can be rewritten $\kappa={\mu}/{\alpha}$.
For given $\mu$, we solve the problem as follows: (i) We fix $\alpha$. (ii) $\kappa=\mu/\alpha$ is then given. (iii) We determine $\lambda$ by an iterative method such that $\psi'(\alpha)=0$. (iv) We obtain different solutions $\lambda_n(\alpha)$ determining different branches $n=1$, $n=2$ etc. This procedure determines for each value of $\alpha$, and for each branch, the normalized density profile $e^{-\psi(\xi)}$. The homogeneous solution corresponds to $\psi=0$ and $\lambda = 1$. This solution is degenerate because the boundary condition (\ref{ay6}) is satisfied for any value of $\alpha$.

\subsection{The temperature}

\begin{figure}
\begin{center}
\includegraphics[clip,scale=0.3]{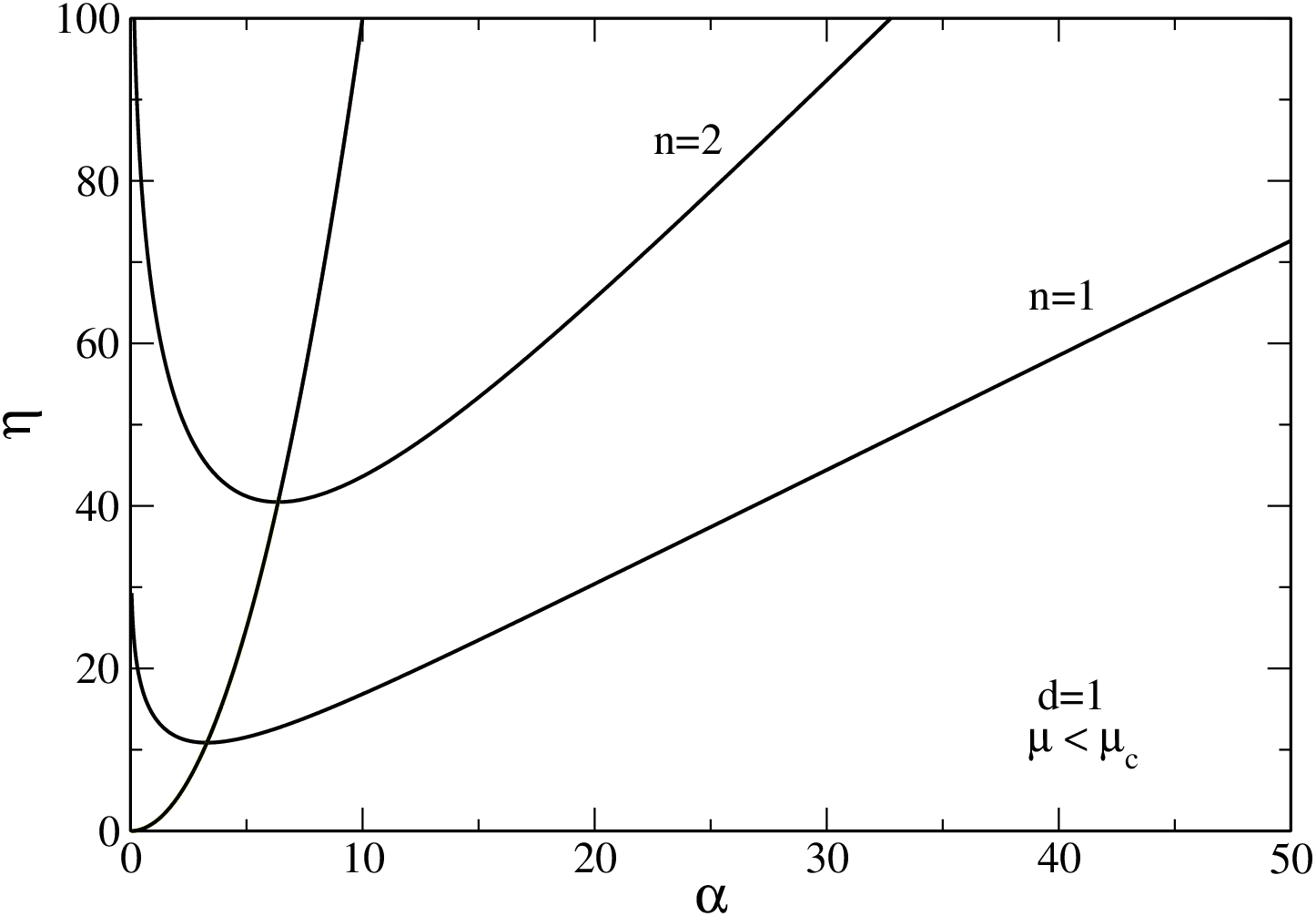}
\caption{Inverse temperature $\eta$ as a function of $\alpha$ for the first two branches in $d=1$. We have taken $\mu=1<\mu_c$. }
\label{eta1-mu1-mod2}
\end{center}
\end{figure}

\begin{figure}
\begin{center}
\includegraphics[clip,scale=0.3]{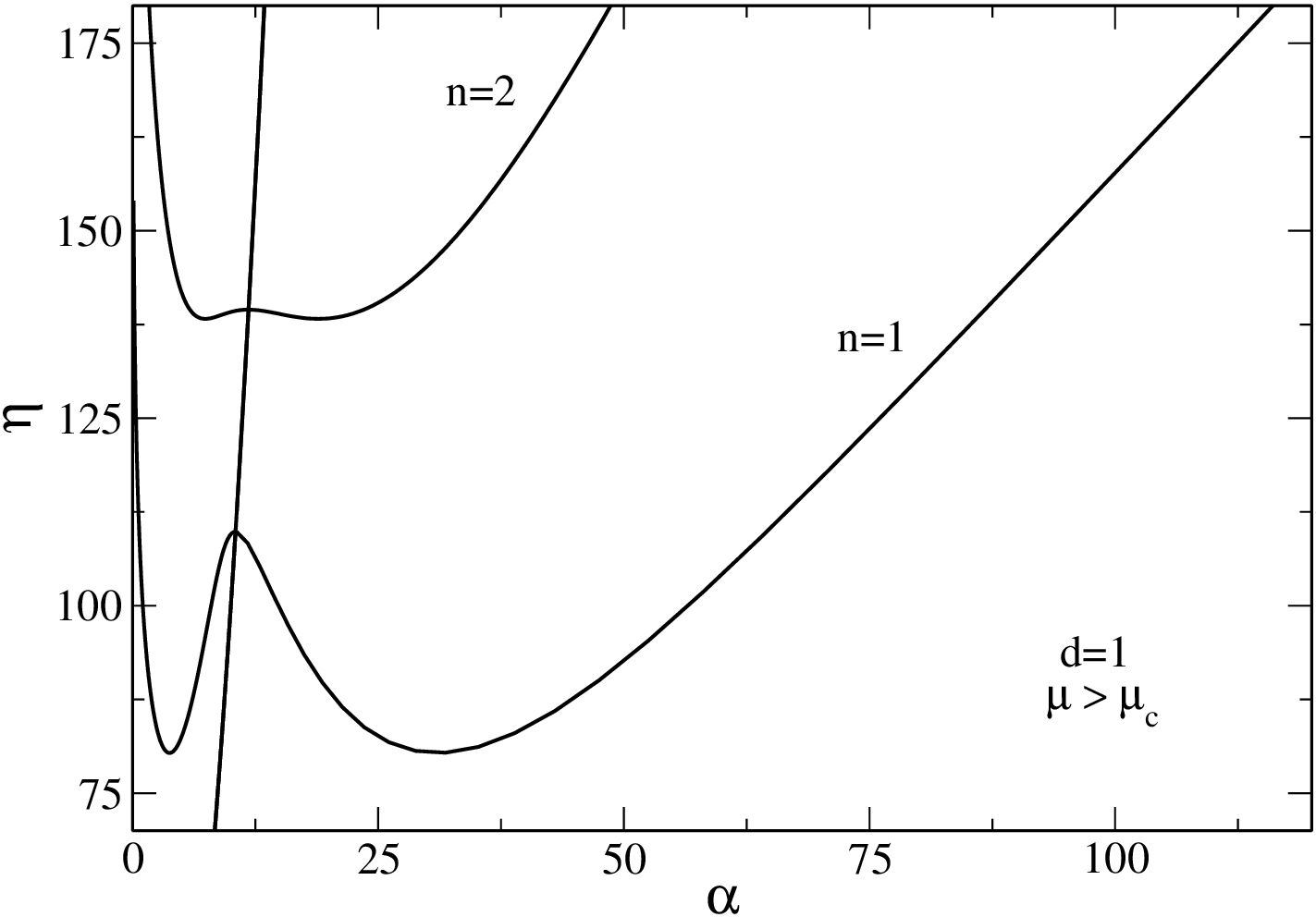}
\caption{Inverse temperature $\eta$ as a function of $\alpha$ for the first two branches in $d=1$. We have taken $\mu=10>\mu_c$.}
\label{eta1-mu10-mod2}
\end{center}
\end{figure}

\begin{figure}
\begin{center}
\includegraphics[clip,scale=0.3]{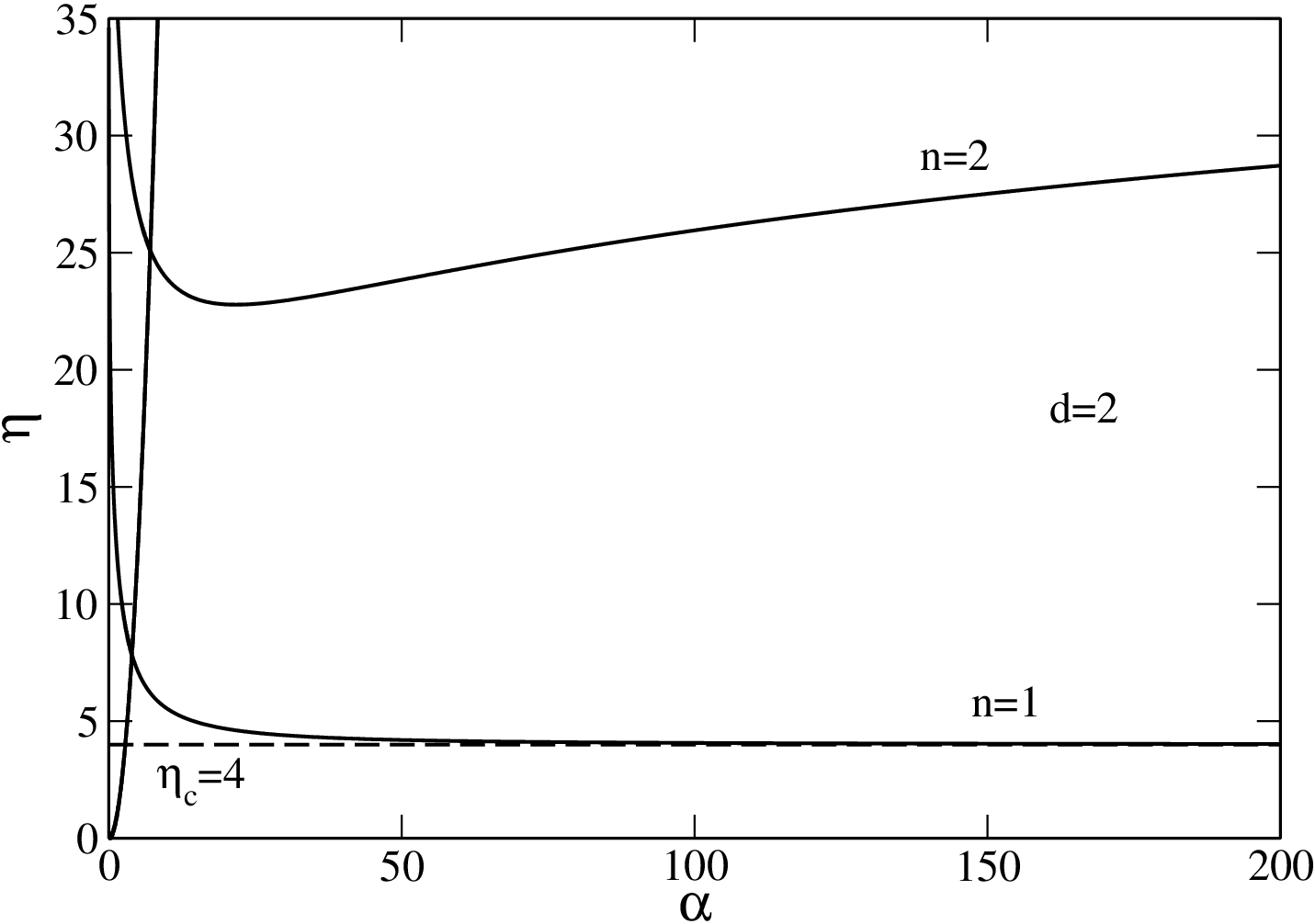}
\caption{Inverse temperature $\eta$ as a function of $\alpha$ for the first two branches in $d=2$. For $\alpha\rightarrow +\infty$, the inverse temperature of the first branch tends to $\eta_c=4$. We have taken $\mu=1$. }
\label{eta2-mod2}
\end{center}
\end{figure}

\begin{figure}
\begin{center}
\includegraphics[clip,scale=0.3]{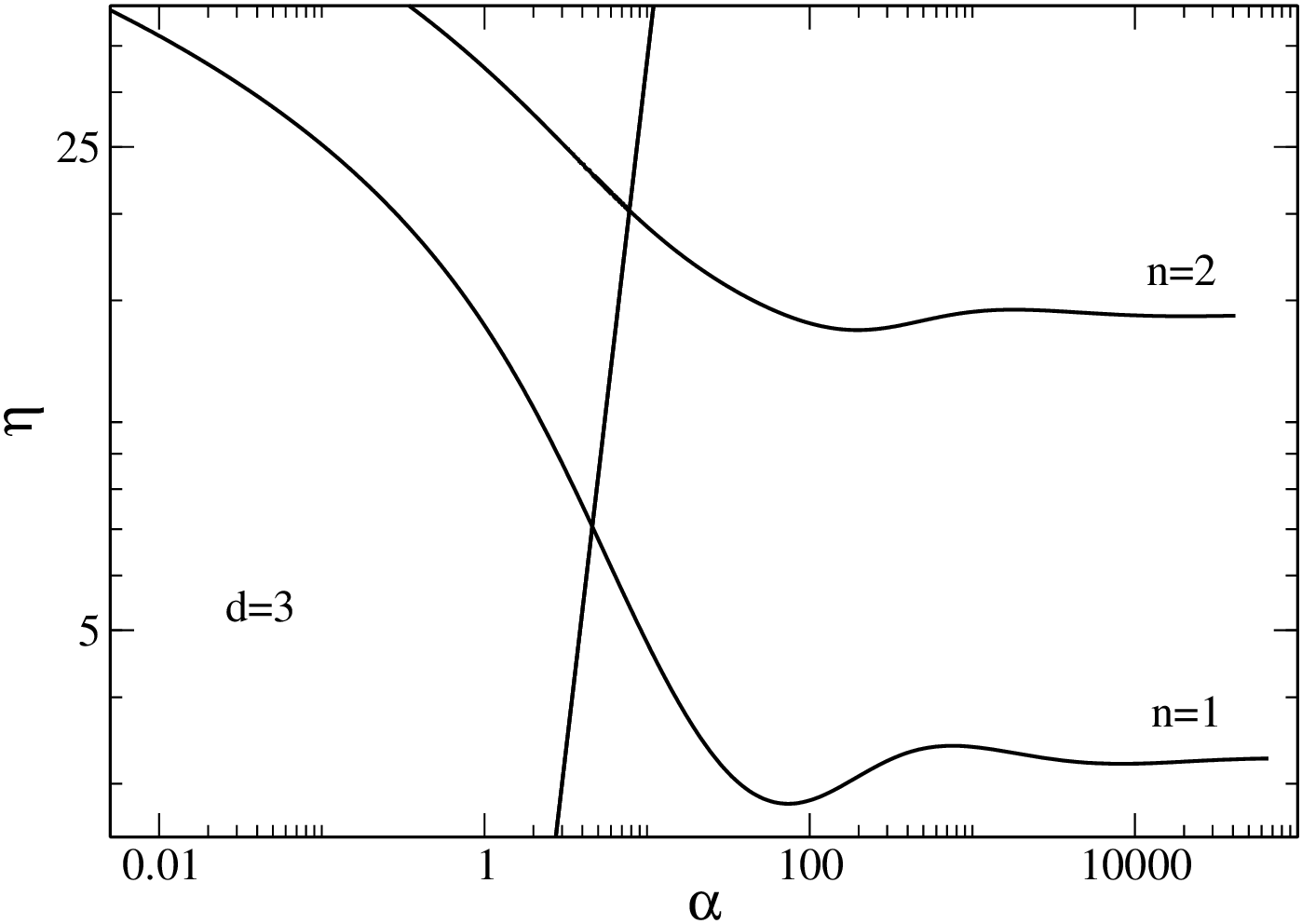}
\caption{Inverse temperature $\eta$ as a function of $\alpha$ for the first two branches in $d=3$.  For $\alpha\rightarrow +\infty$, the inverse temperature of the first branch undergoes damped oscillations around the value $\eta_s\simeq 3.25$. We have taken $\mu=1$.}
\label{eta3-mu1-mod2}
\end{center}
\end{figure}

We must now relate the parameter $\alpha$ to the temperature $T$. Introducing the dimensionless variables defined previously and recalling that $r=R\xi/\alpha$, the mass can be written
\begin{eqnarray}
\label{ay9}
M=\rho_0 S_d \left (\frac{R}{\alpha}\right )^d\int_0^{\alpha}e^{-\psi}\xi^{d-1}\, d\xi.
\end{eqnarray}
Using $\alpha= (S_dG\beta m \rho_0)^{1/2} R$  and introducing the dimensionless temperature (\ref{mn9}), we obtain
\begin{eqnarray}
\label{ay10}
\eta=\frac{1}{\alpha^{d-2}}\int_{0}^{\alpha}e^{-\psi}\xi^{d-1}d\xi.
\end{eqnarray}
This equation gives the relation between the inverse temperature $\eta$ and $\alpha$ for the $n$-th branch. In Figs. \ref{eta1-mu1-mod2}, \ref{eta1-mu10-mod2}, \ref{eta2-mod2} and \ref{eta3-mu1-mod2}, we plot the inverse temperature $\eta$ as a function of $\alpha$ for the first two branches $n=1,2$ in different dimensions of space $d=1,2,3$. The discussion is similar to the one given in Sec. \ref{sec_modifT}.
We have also represented the branch corresponding to the homogeneous solution. Its equation is given by
$\eta=\alpha^2/d$. The branch $n=1$ of inhomogeneous solutions connects the branch of homogeneous solutions at $\alpha_c^2=d\eta_c^*$ (see Appendix \ref{sec_bif}).

\subsection{The energy}

\begin{figure}
\begin{center}
\includegraphics[clip,scale=0.3]{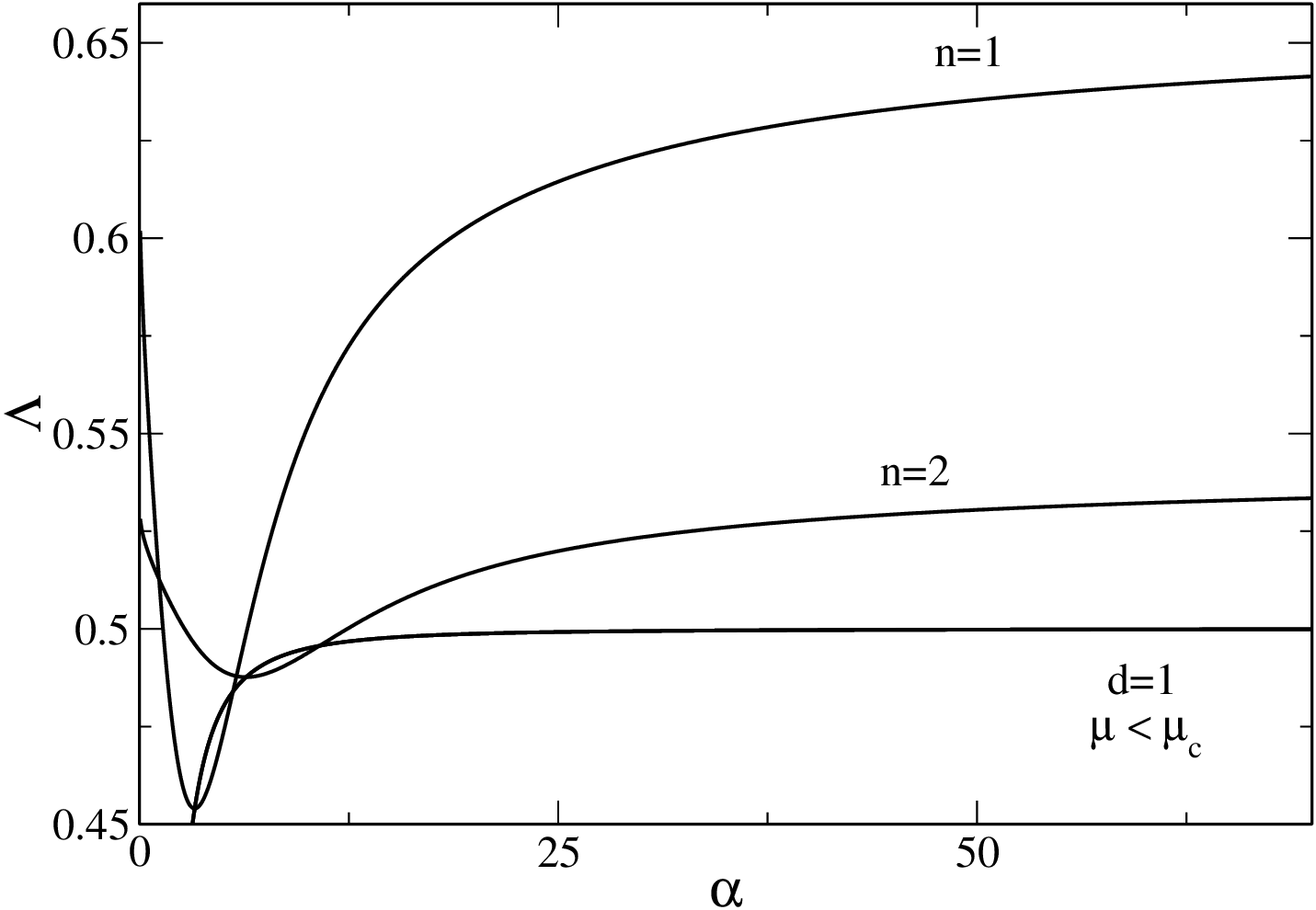}
\caption{Energy $\Lambda$ as a function of $\alpha$ for the first two branches in $d=1$. We have taken $\mu=1<\mu_c$.}
\label{ene1-mu1-mod2}
\end{center}
\end{figure}

\begin{figure}
\begin{center}
\includegraphics[clip,scale=0.3]{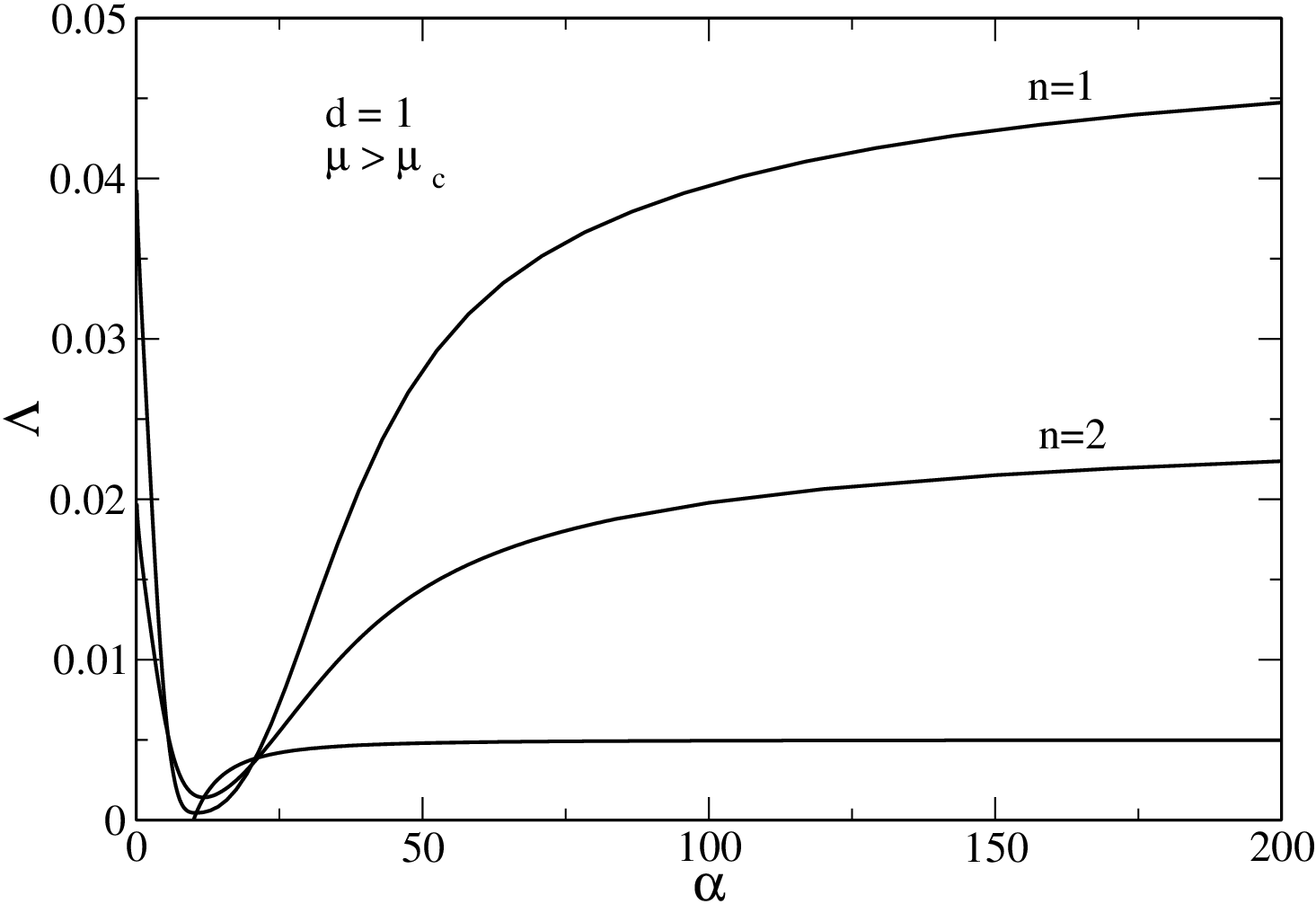}
\caption{Energy $\Lambda$ as a function of $\alpha$ for the first two branches in $d=1$. We have taken $\mu=10>\mu_c$.}
\label{ene1-mu10-mod2}
\end{center}
\end{figure}

\begin{figure}[!ht]
\begin{center}
\includegraphics[clip,scale=0.3]{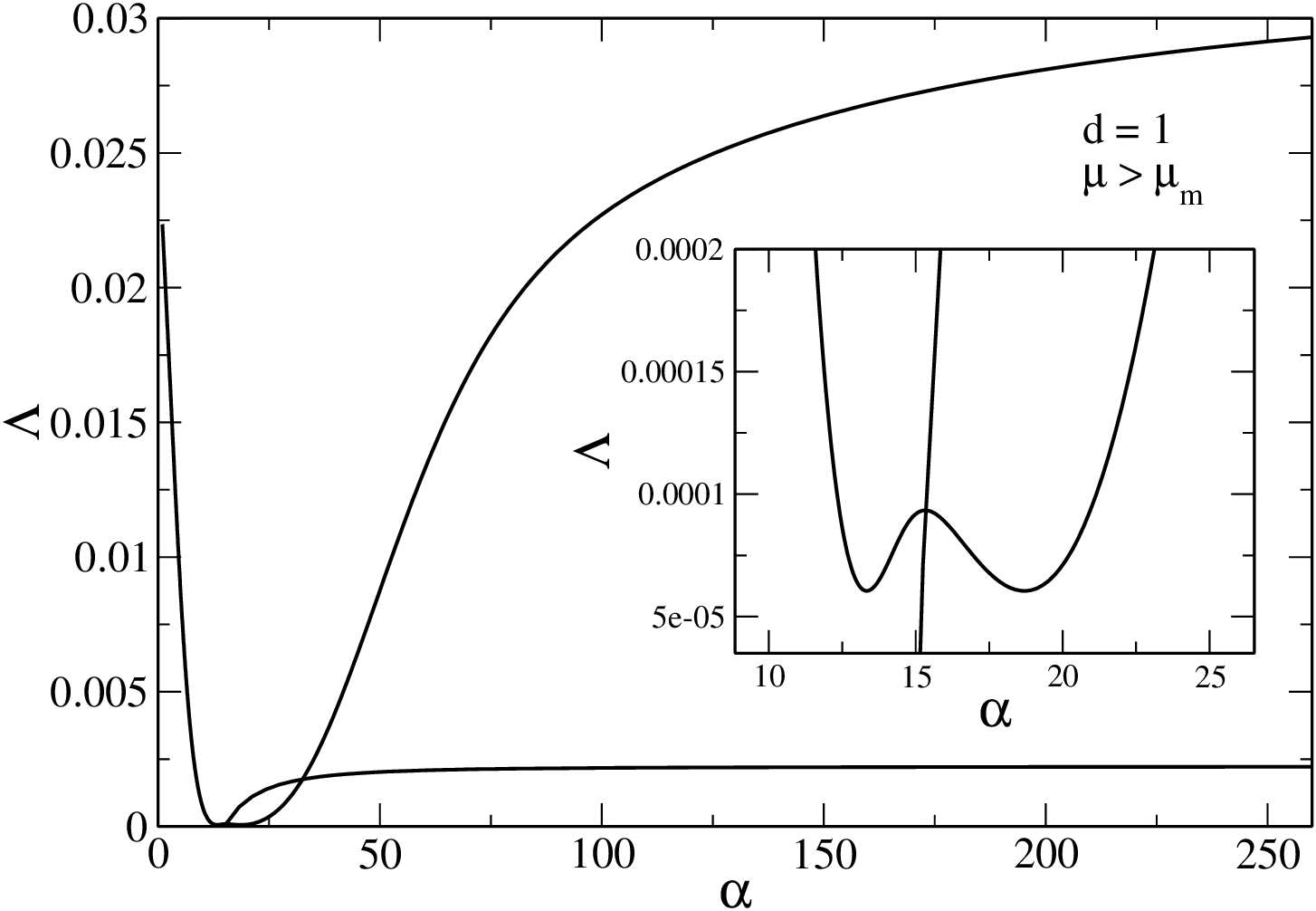}
\caption{Energy $\Lambda$ as a function of $\alpha$ for the first two branches in $d=1$. We have taken $\mu=15>\mu_m$.}
\label{ene-alfa-mu15}
\end{center}
\end{figure}

\begin{figure}
\begin{center}
\includegraphics[clip,scale=0.3]{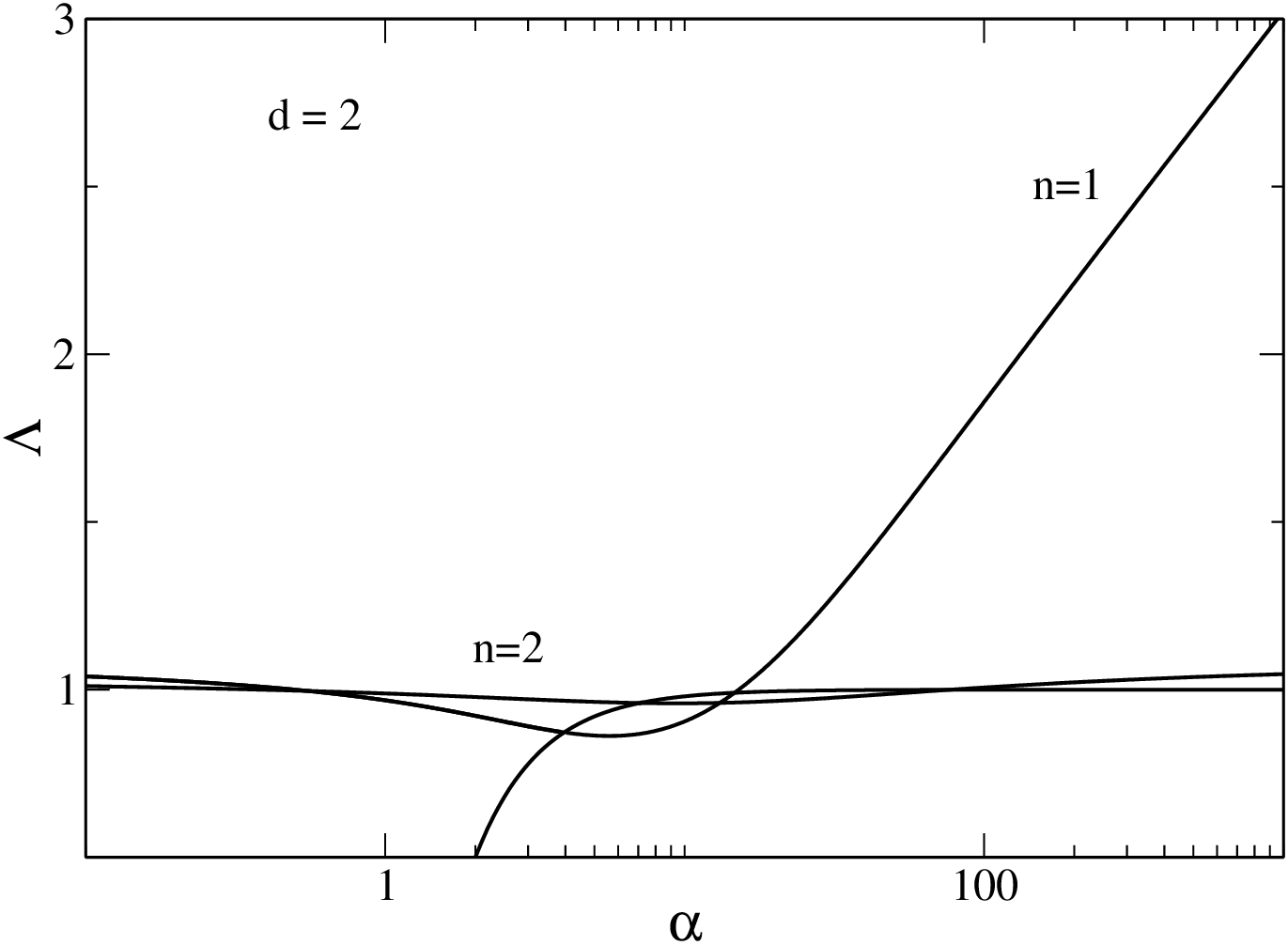}
\caption{Energy $\Lambda$ as a function of $\alpha$ for the first two branches in $d=2$. We have taken $\mu=1$.}
\label{ene2-mod2}
\end{center}
\end{figure}

\begin{figure}
\begin{center}
\includegraphics[clip,scale=0.3]{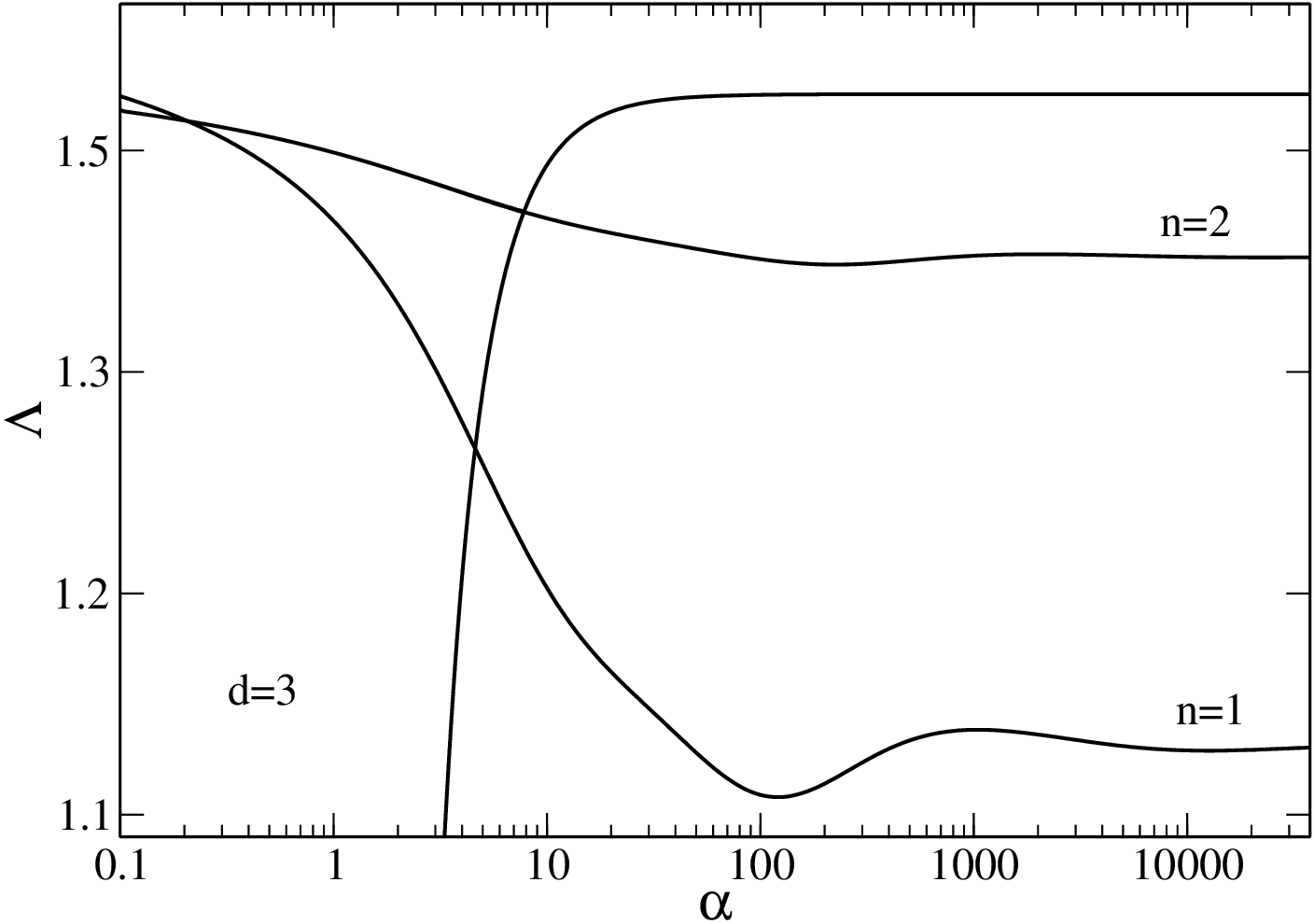}
\caption{Energy $\Lambda$ as a function of $\alpha$ for the first two branches in $d=3$.  For $\alpha\rightarrow +\infty$, the energy of the first branch undergoes damped oscillations around the value $\Lambda_s\simeq 1.13$. We have taken $\mu=1$.}
\label{ene3-mu1-mod2}
\end{center}
\end{figure}

We must also relate $\alpha$ to the energy $E$. The total energy is given by
\begin{eqnarray}
\label{ay11}
E=\int f\frac{v^2}{2}\, d{\bf r}d{\bf v}+\frac{1}{2}\int \rho \Phi \, d\mathbf{r}.
\end{eqnarray}
Using the Maxwell-Boltzmann distribution (\ref{mce6}), the kinetic energy is simply
\begin{eqnarray}
\label{ay12}
K=\frac{d}{2}Nk_BT.
\end{eqnarray}
Using the screened Poisson equation (\ref{ay1}) and integrating by parts, the potential energy can be written
\begin{eqnarray}
\label{ay13}
W=-\frac{1}{2S_dG}\int \left\lbrack (\nabla
\Phi )^2+k_0^2\Phi^2\right\rbrack \, d\bf{r}.
\end{eqnarray}
The total energy $E=K+W$ is therefore given by
\begin{eqnarray}
\label{ay14}
E=\frac{d}{2}Nk_BT-\frac{1}{2S_dG}\int \left\lbrack (\nabla
\Phi )^2+k_0^2\Phi^2\right\rbrack \, d\bf{r}.
\end{eqnarray}
Introducing the dimensionless variables defined previously, recalling that $r=\xi R/\alpha$ and $\mu=k_0 R$, and introducing the normalized energy (\ref{mn15}), we obtain
\begin{eqnarray}
\label{ay15}
\Lambda=-\frac{d}{2\eta}+\frac{1}{2\eta^2}\frac{1}{\alpha^{d-2}}\int_{0}^{\alpha}\left (\frac{d\psi}{d\xi}\right )^2\xi^{d-1}d\xi\nonumber\\
+\frac{\mu^2}{2\eta^2}\frac{1}{\alpha^{d}}\int_{0}^{\alpha} (\psi+\beta m\Phi_0)^2\xi^{d-1}d\xi.
\end{eqnarray}
Using the expressions of $\kappa$ and $\lambda$ following Eq. (\ref{ay4}), we find that
\begin{eqnarray}
\label{ay16}
\beta m\Phi_0=-\frac{\lambda}{\kappa^2},
\end{eqnarray}
so that, finally,
\begin{eqnarray}
\label{ay17}
\Lambda=-\frac{d}{2\eta}+\frac{1}{2\eta^2}\frac{1}{\alpha^{d-2}}\int_{0}^{\alpha}\left (\frac{d\psi}{d\xi}\right )^2\xi^{d-1}d\xi\nonumber\\
+\frac{\mu^2}{2\eta^2}\frac{1}{\alpha^{d}}\int_{0}^{\alpha} \left (\psi-\frac{\lambda\alpha^2}{\mu^2}\right )^2\xi^{d-1}d\xi.
\end{eqnarray}
This equation gives the relation between the energy $\Lambda$ and
$\alpha$ for the $n$-th branch.  In Figs. \ref{ene1-mu1-mod2},
\ref{ene1-mu10-mod2}, \ref{ene2-mod2} and \ref{ene3-mu1-mod2}, we plot
the energy $\Lambda$ as a function of $\alpha$ for the first two
branches $n=1$ and $n=2$ in different dimensions of space $d=1,2,3$.
The discussion is similar to the one given in Sec. \ref{sec_modifE}.
We have also represented the branch corresponding to the homogeneous
solution. Using Eq. (\ref{ay24}) and $\eta=\alpha^2/d$, its equation
is given by $\Lambda=-d^2/(2\alpha^2)+d/(2\mu^2)$.

\subsection{The entropy and the free energy}

\begin{figure}
\begin{center}
\includegraphics[clip,scale=0.3]{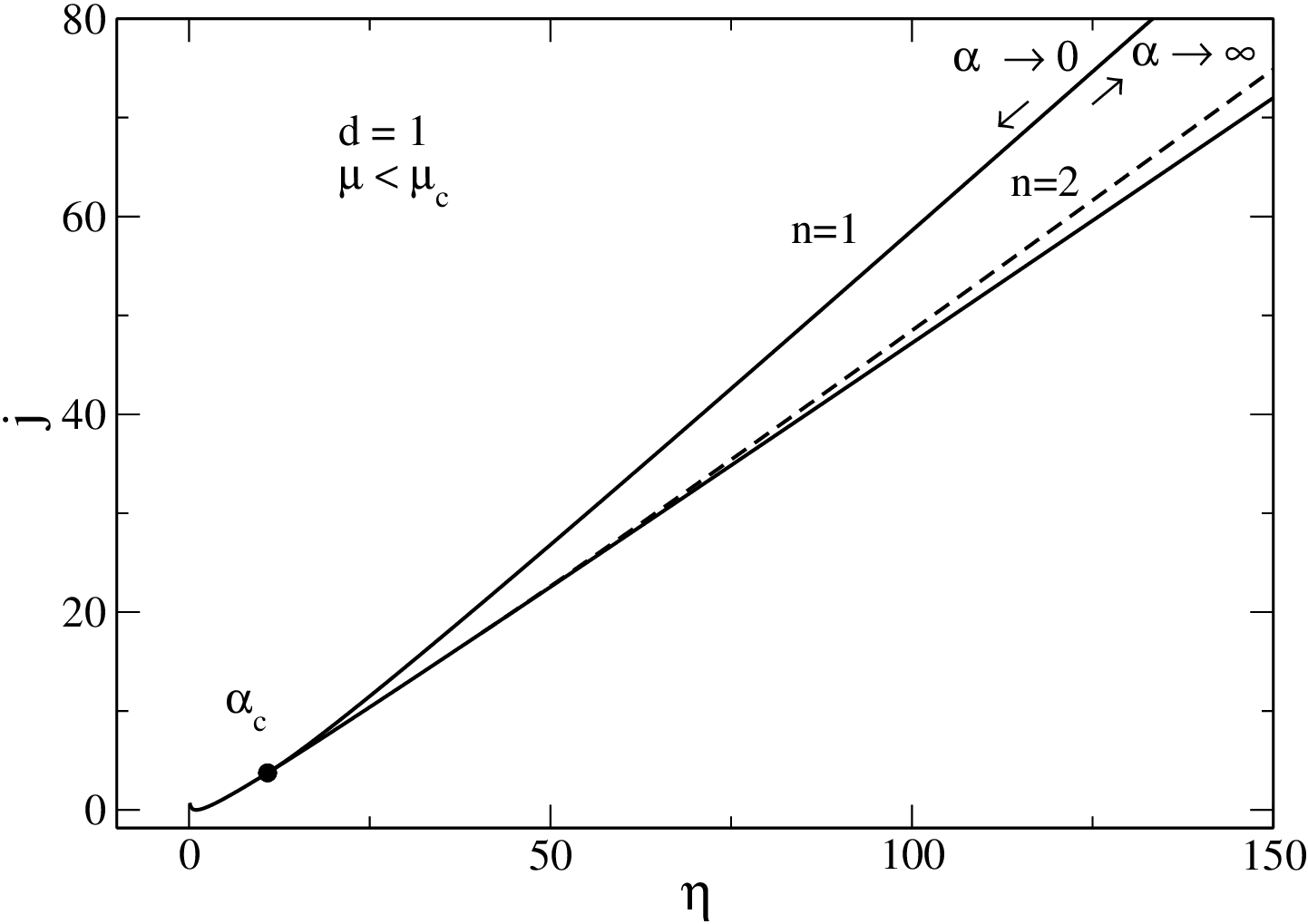}
\caption{Free energy $\frac{J}{Nk_B}$ as a function of the inverse temperature $\eta$ for $d=1$. We have taken $\mu=1<\mu_c$.}
\label{free1-mu3-mod2}
\end{center}
\end{figure}

\begin{figure}
\begin{center}
\includegraphics[clip,scale=0.3]{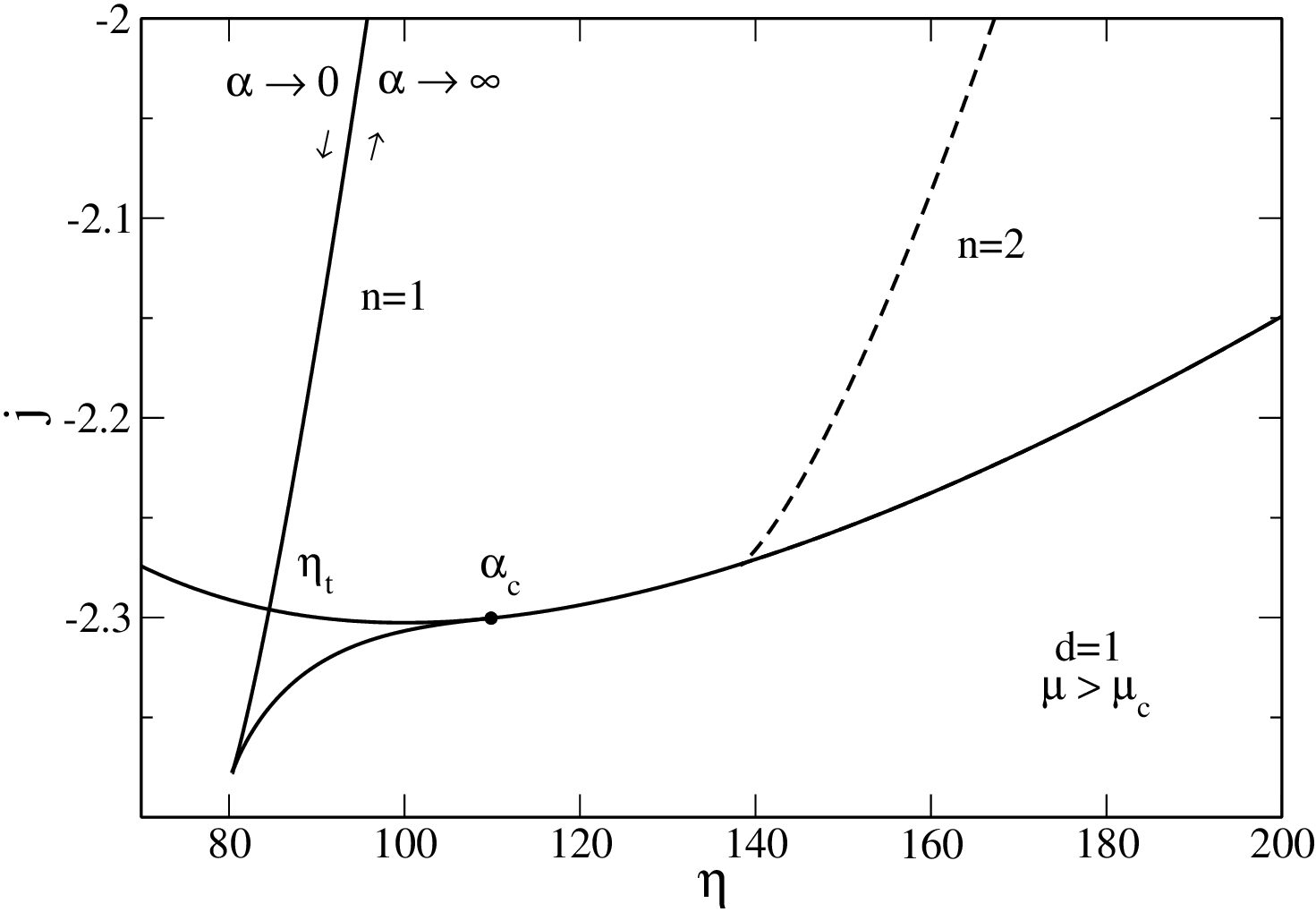}
\caption{Free energy $\frac{J}{Nk_B}$ as a function of the inverse temperature $\eta$ for $d=1$. We have taken $\mu=10>\mu_c$. The free energies of the homogeneous phase and inhomogeneous phase become equal at $\eta=\eta_t(\mu)$. This corresponds to a first order phase transition in the canonical ensemble marked by the discontinuity of the slope $J'(\beta)=-E$.}
\label{free1-mu10-mod2}
\end{center}
\end{figure}

\begin{figure}
\begin{center}
\includegraphics[clip,scale=0.3]{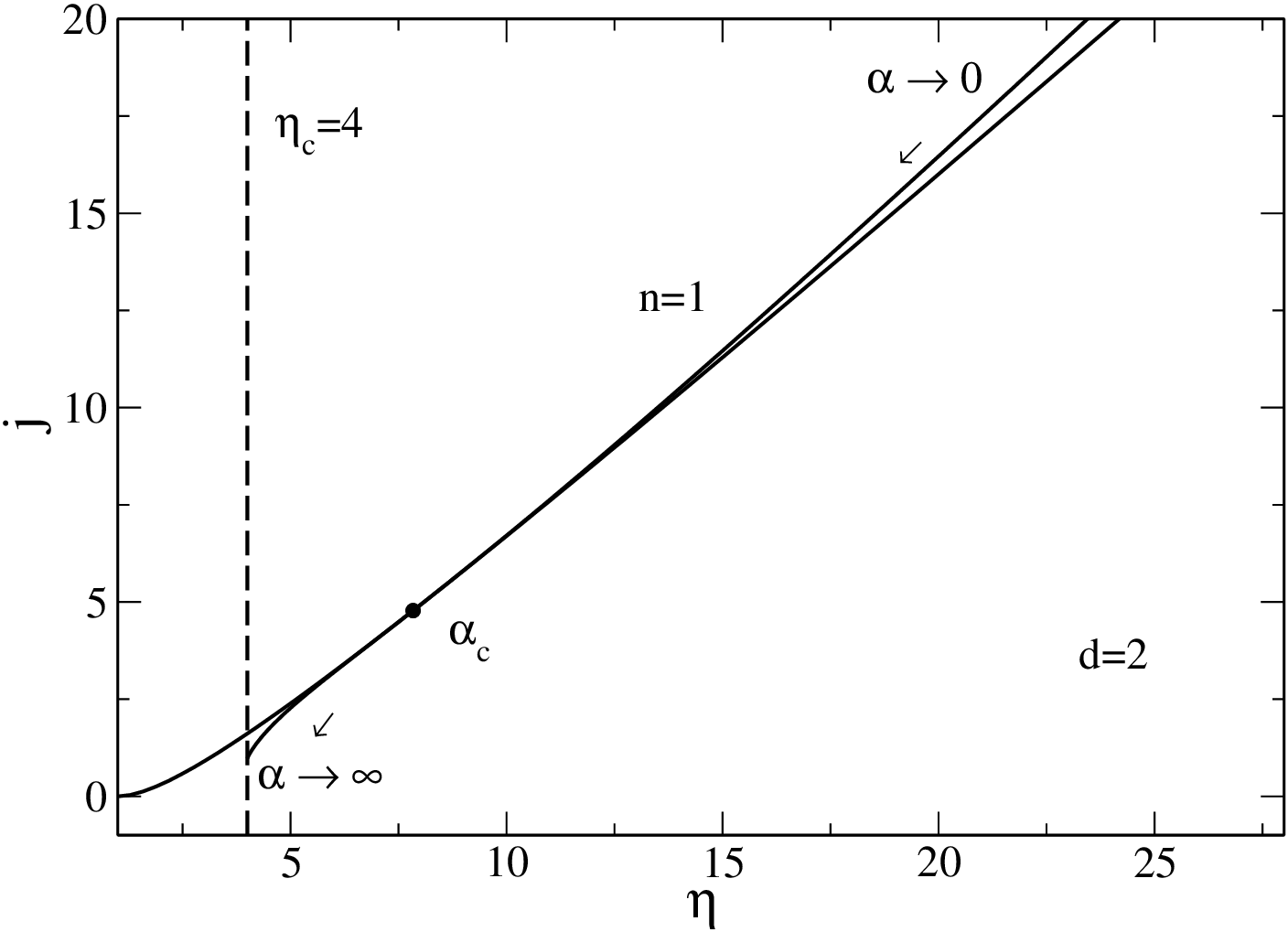}
\caption{Free energy $\frac{J}{Nk_B}$ as a function of the inverse temperature $\eta$ for $d=2$. We have taken $\mu=1$.}
\label{free2-mu1-mod2}
\end{center}
\end{figure}

\begin{figure}
\begin{center}
\includegraphics[clip,scale=0.3]{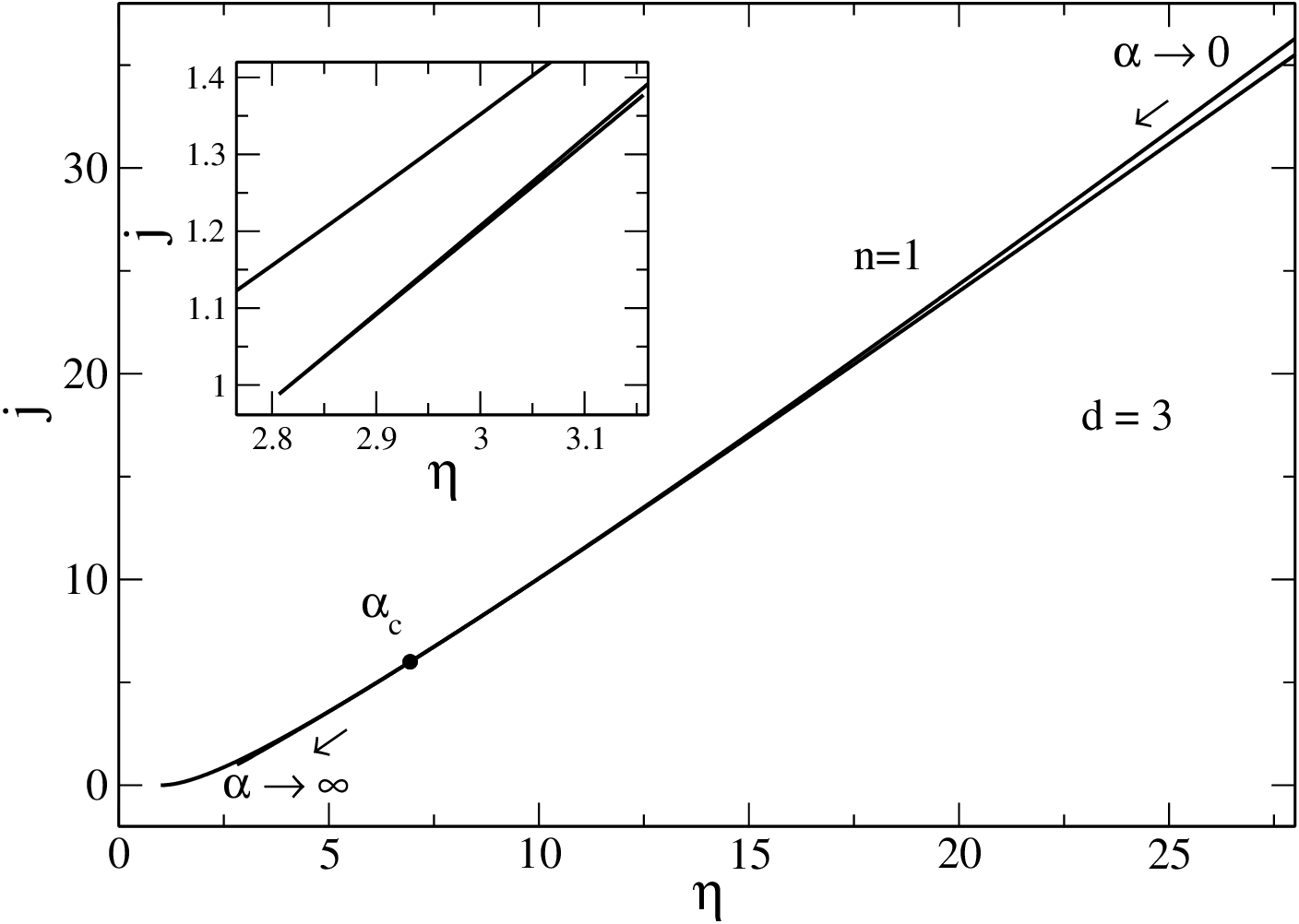}
\caption{Free energy $\frac{J}{Nk_B}$ as a function of the inverse temperature $\eta$ for $d=3$. We have taken  $\mu=1$.}
\label{free3-mod2}
\end{center}
\end{figure}

\begin{figure}
\begin{center}
\includegraphics[clip,scale=0.3]{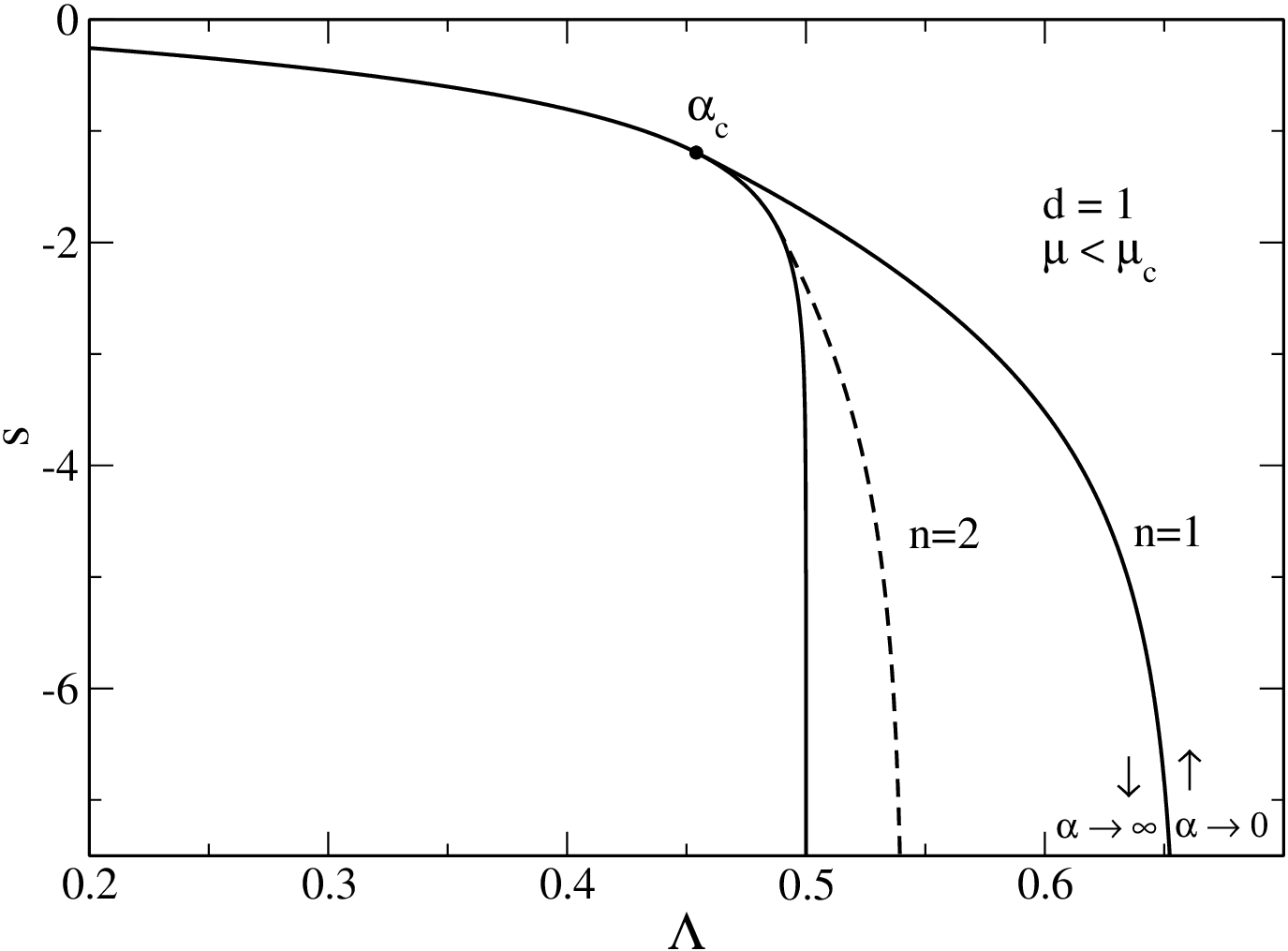}
\caption{Entropy $\frac{S}{Nk_B}$ as a function of the energy $\Lambda$ for $d=1$. We have taken $\mu=1<\mu_c$.}
\label{entropy1-mu3-mod2}
\end{center}
\end{figure}

\begin{figure}
\begin{center}
\includegraphics[clip,scale=0.3]{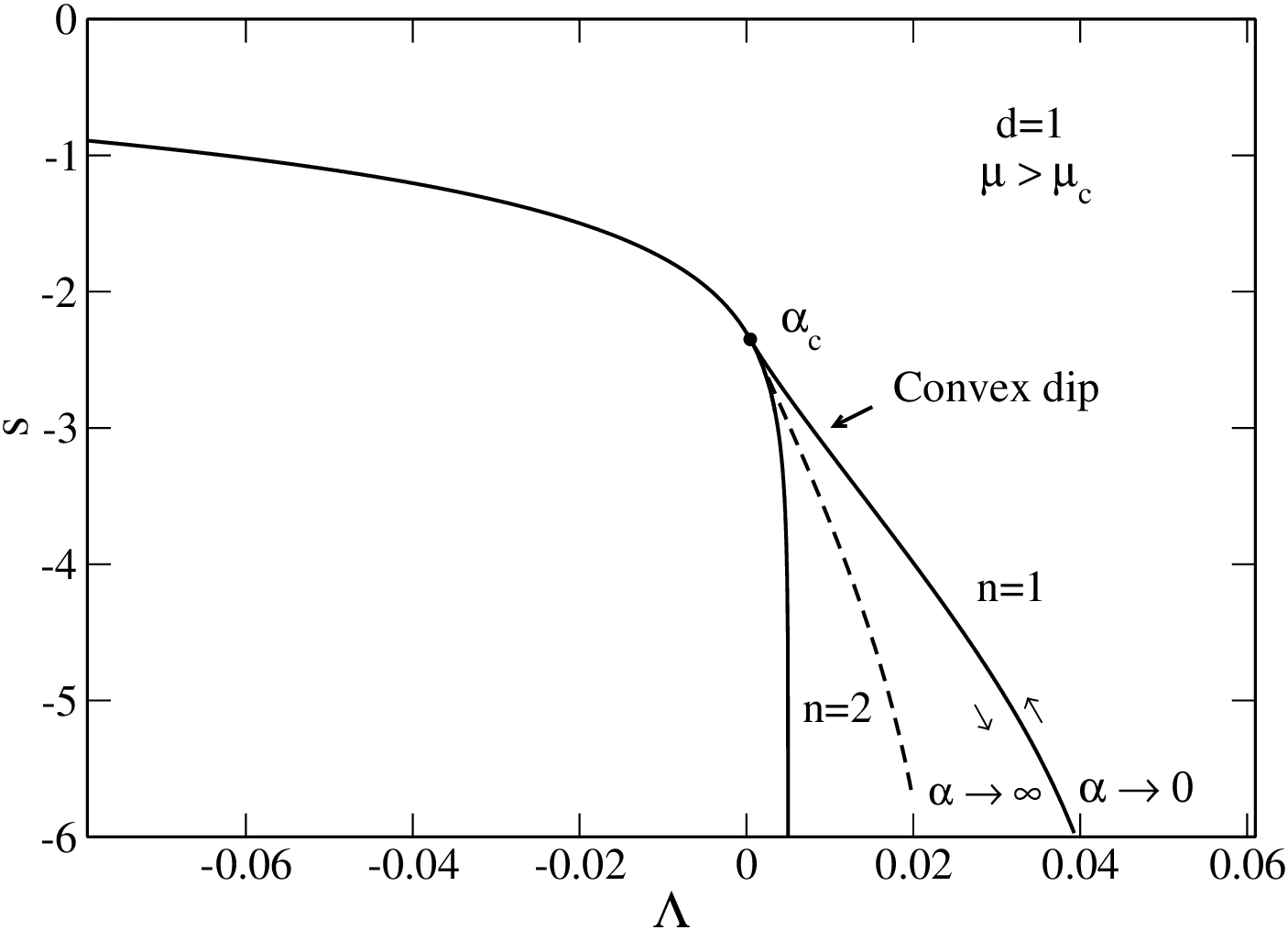}
\caption{Entropy $\frac{S}{Nk_B}$ as a function of the energy $\Lambda$ for $d=1$. We have taken $\mu=10>\mu_c$. There is a (small)  convex dip associated with the region of negative specific heats in the microcanonical ensemble. }
\label{entropy1-mu10-mod2}
\end{center}
\end{figure}

\begin{figure}
\begin{center}
\includegraphics[clip,scale=0.3]{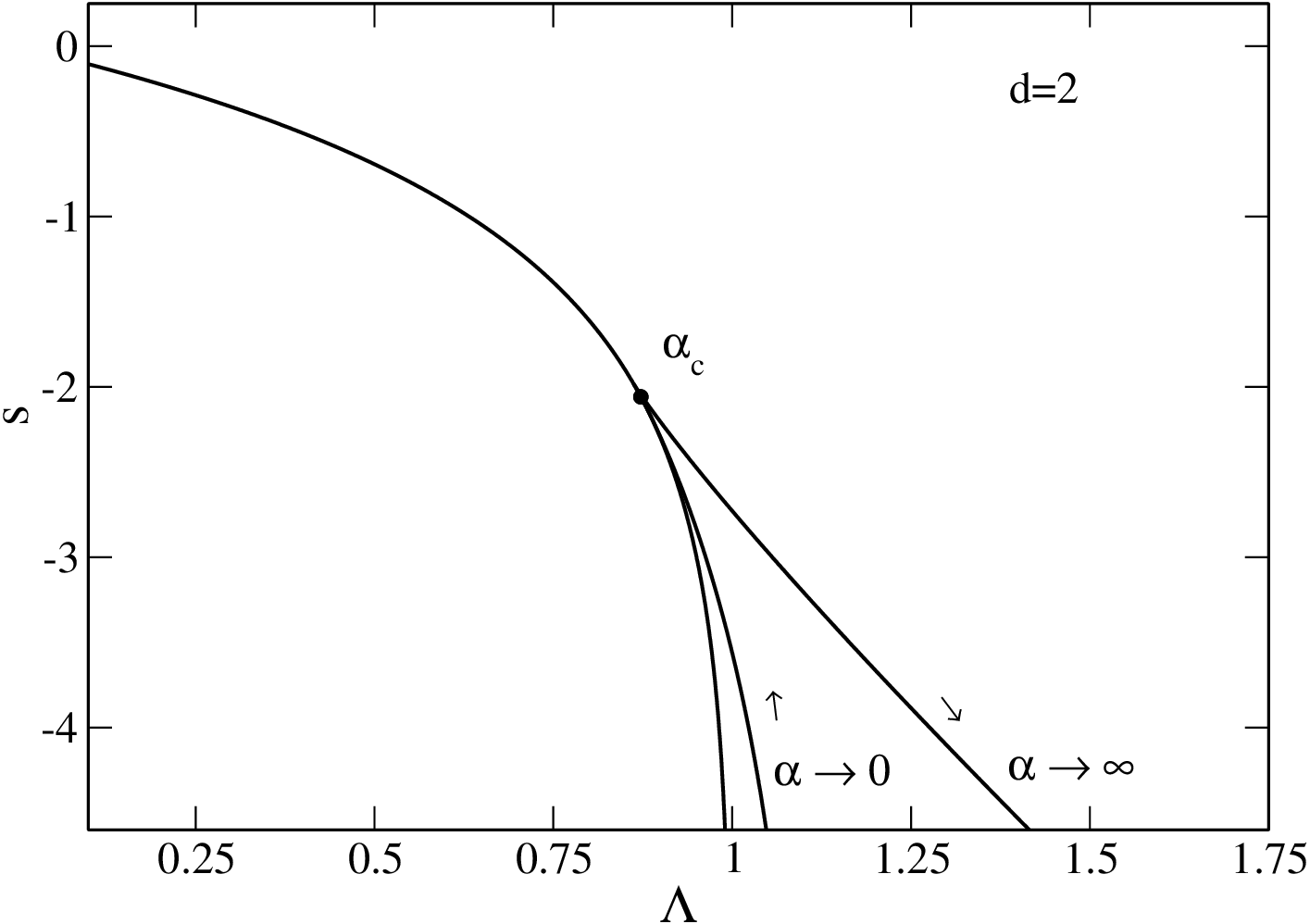}
\caption{Entropy $\frac{S}{Nk_B}$ as a function of the energy $\Lambda$ for $d=2$. We have taken $\mu=1$.}
\label{entropy2-mu1-mod2}
\end{center}
\end{figure}

\begin{figure}
\begin{center}
\includegraphics[clip,scale=0.3]{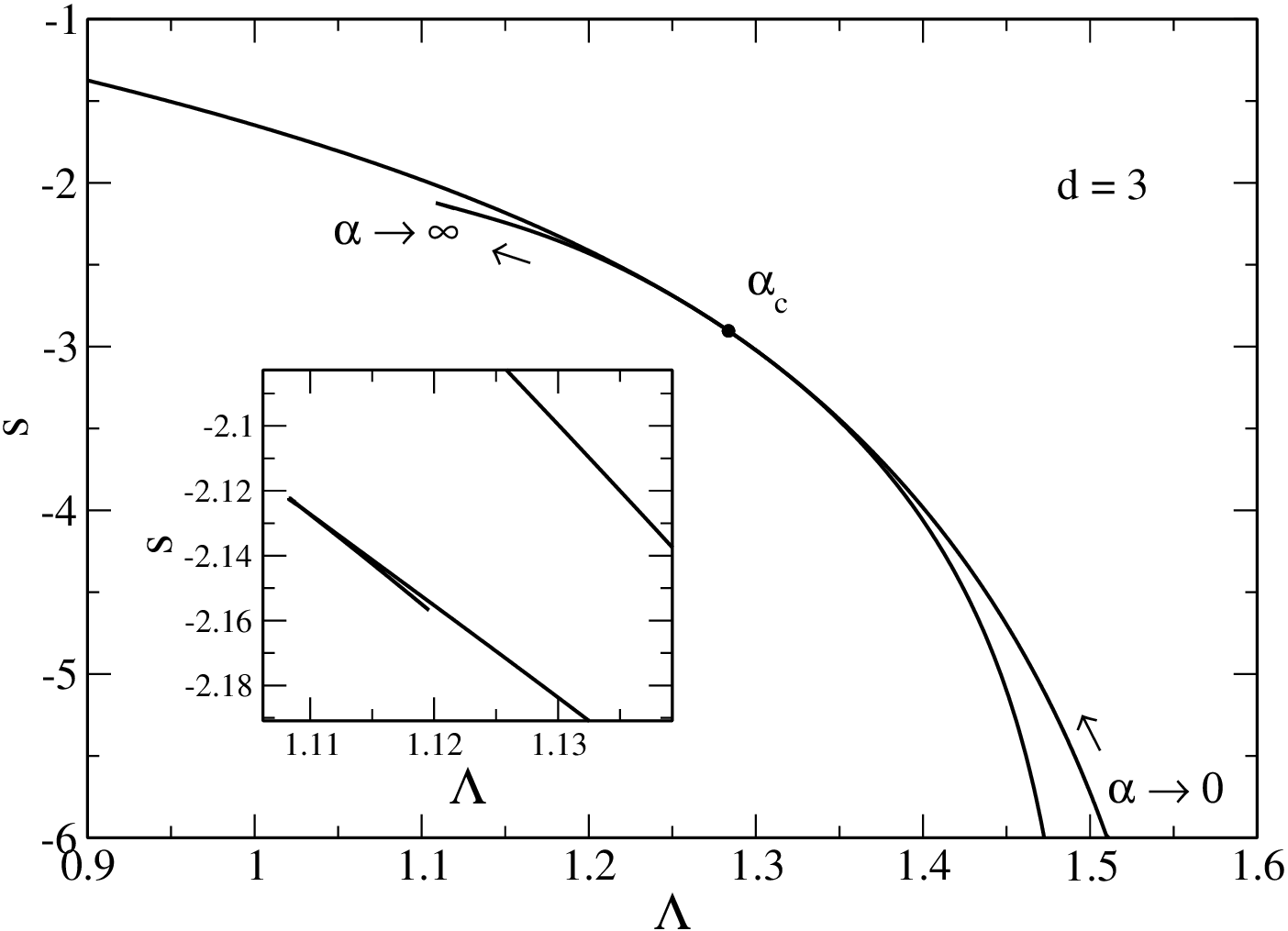}
\caption{Entropy $\frac{S}{Nk_B}$ as a function of the energy $\Lambda$ for $d=3$. We have taken $\mu=1$.}
\label{entropy3-mod2}
\end{center}
\end{figure}

Finally, we relate $\alpha$ to the entropy $S$ and to the free energy $F$.  The entropy is given by
\begin{eqnarray}
\label{ay18}
S=\frac{d}{2}Nk_B \ln T-k_B\int \frac{\rho}{m}\ln\frac{\rho}{m}\, d{\bf r}.
\end{eqnarray}
We can proceed exactly as in Sec. \ref{sec_sfmodif} and obtain
\begin{eqnarray}
\label{ay19}
\frac{S}{N k_B}=-\frac{d-2}{2}\ln\eta-2\ln\alpha\nonumber\\
+\frac{1}{\eta}\frac{1}{\alpha^{d-2}}\int_{0}^{\alpha}\psi e^{-\psi}\xi^{d-1}\, d\xi,
\end{eqnarray}
up to unimportant constants. However, we can also obtain a simpler expression. Substituting $\rho=\rho_0 e^{-\beta m(\Phi-\Phi_0)}$ in Eq. (\ref{ay18}), we obtain
\begin{eqnarray}
\label{ay20}
S=\frac{d}{2}Nk_B\ln T-k_B\int \frac{\rho}{m}\ln\frac{\rho_0}{m}\, d{\bf r}\nonumber\\
+k_B\beta\int\rho (\Phi-\Phi_0)\, d{\bf r}.
\end{eqnarray}
This can be rewritten
\begin{eqnarray}
\label{ay21}
\frac{S}{N k_B}=-\frac{d}{2}\ln\beta-\ln\rho_0+\frac{2\beta E}{N}-\beta m\Phi_0,
\end{eqnarray}
up to unimportant constants. Finally, using Eqs. (\ref{ay16}), (\ref{mn9}),  (\ref{mn15}) and the relations $\kappa=\mu/\alpha$ and $\alpha= (S_dG\beta m \rho_0)^{1/2} R$, we obtain
\begin{eqnarray}
\label{ay22}
\frac{S}{N k_B}=-\frac{d-2}{2}\ln\eta-2\ln\alpha-2\Lambda\eta+\frac{\lambda\alpha^2}{\mu^2},
\end{eqnarray}
which does not involve new integrals. Using the previous results, this
expression relates the entropy $S/N k_B$ to $\alpha$. The free energy
is $F=E-TS$. In the following, it will be more convenient to work in
terms of the Massieu function $J=S-k_B \beta E$ (by an abuse of
language, we shall often refer to $J$ as the free energy). We have
\begin{eqnarray}
\label{ay23}
\frac{J}{N k_B}=\frac{S}{N k_B}+\eta\Lambda.
\end{eqnarray}
Using the previous results, this expression relates the free energy  $J/N k_B$ to $\alpha$.

In Figs. \ref{free1-mu3-mod2}, \ref{free1-mu10-mod2}, \ref{free2-mu1-mod2} and \ref{free3-mod2}, we have plotted the free energy  $J/Nk_B$ as a function of the inverse temperature $\eta$ (parameterized by $\alpha$) in $d=1,2,3$. In Figs. \ref{entropy1-mu3-mod2}, \ref{entropy1-mu10-mod2}, \ref{entropy2-mu1-mod2} and \ref{entropy3-mod2}), we have plotted the entropy $S/Nk_B$ as a function of the energy  $\Lambda$ (parameterized by $\alpha$) in $d=1,2,3$.

\subsection{Caloric curve}

We shall now determine the caloric curve $\beta(E)$ corresponding to the screened Newtonian model. First of all, we note that, for the homogeneous phase, we have $\rho=\rho_0$ and $\Phi=\Phi_0$ with $-k_0^2\Phi_0=S_d G\rho_0$ (or equivalently  $\psi=0$, $\lambda=1$ and $\alpha^2=d\eta$). Therefore, the relationship between the energy and the temperature can be written
\begin{eqnarray}
\label{ay24}
\Lambda=-\frac{d}{2\eta}+\frac{d}{2\mu^2}.
\end{eqnarray}
This shows that $\eta\rightarrow +\infty$ for $\Lambda\rightarrow \Lambda_{max}=d/(2\mu^2)$.
On the other hand, eliminating $\alpha$ between  $\eta_n(\alpha)$ and $\Lambda_n(\alpha)$ given by Eqs. (\ref{ay10}) and (\ref{ay17}), we get the series of equilibria for the $n$-th inhomogeneous branch.
The series of equilibria (critical points) and the caloric curves (stable states) in CE and MCE are described below for different dimensions of space.

\subsubsection{The dimension $d=1$}

\begin{figure}[!ht]
\begin{center}
\includegraphics[clip,scale=0.3]{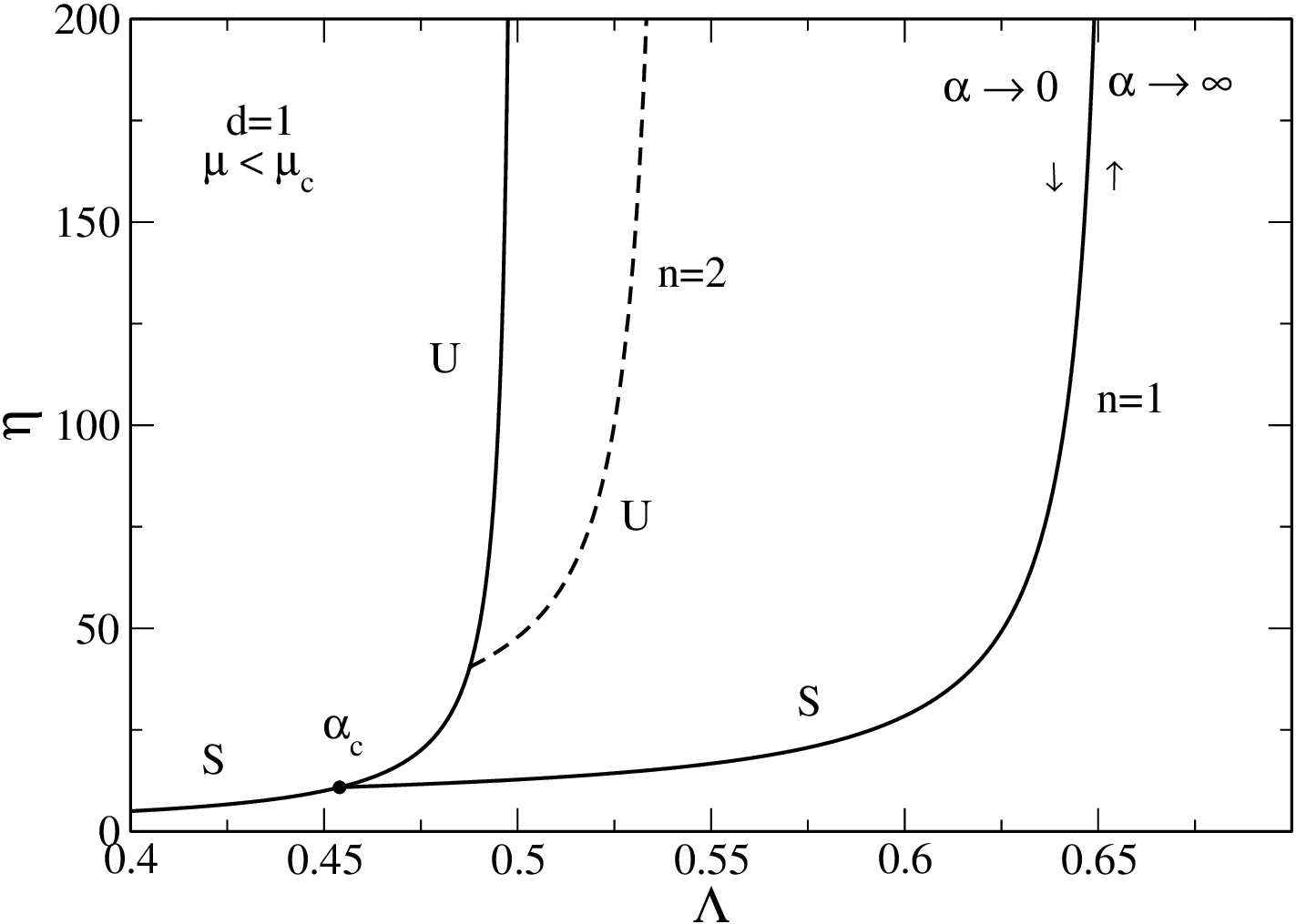}
\caption{Series of equilibria in $d=1$ for $\mu=1<\mu_c$. The caloric curve displays a second order phase transition in CE and MCE taking place at $\eta=\eta_c^*$ and $\Lambda=\Lambda_c^*$. It is marked by the discontinuity of  $\partial E/\partial \beta$ in CE or $\partial\beta/\partial E$ in MCE. Note that the branches $\alpha<\alpha_c$ and $\alpha>\alpha_c$ coincide.   For $\mu<\mu_c$, the CE and MCE ensembles are equivalent.}
\label{fig4}
\end{center}
\end{figure}

\begin{figure}[!ht]
\begin{center}
\includegraphics[clip,scale=0.3]{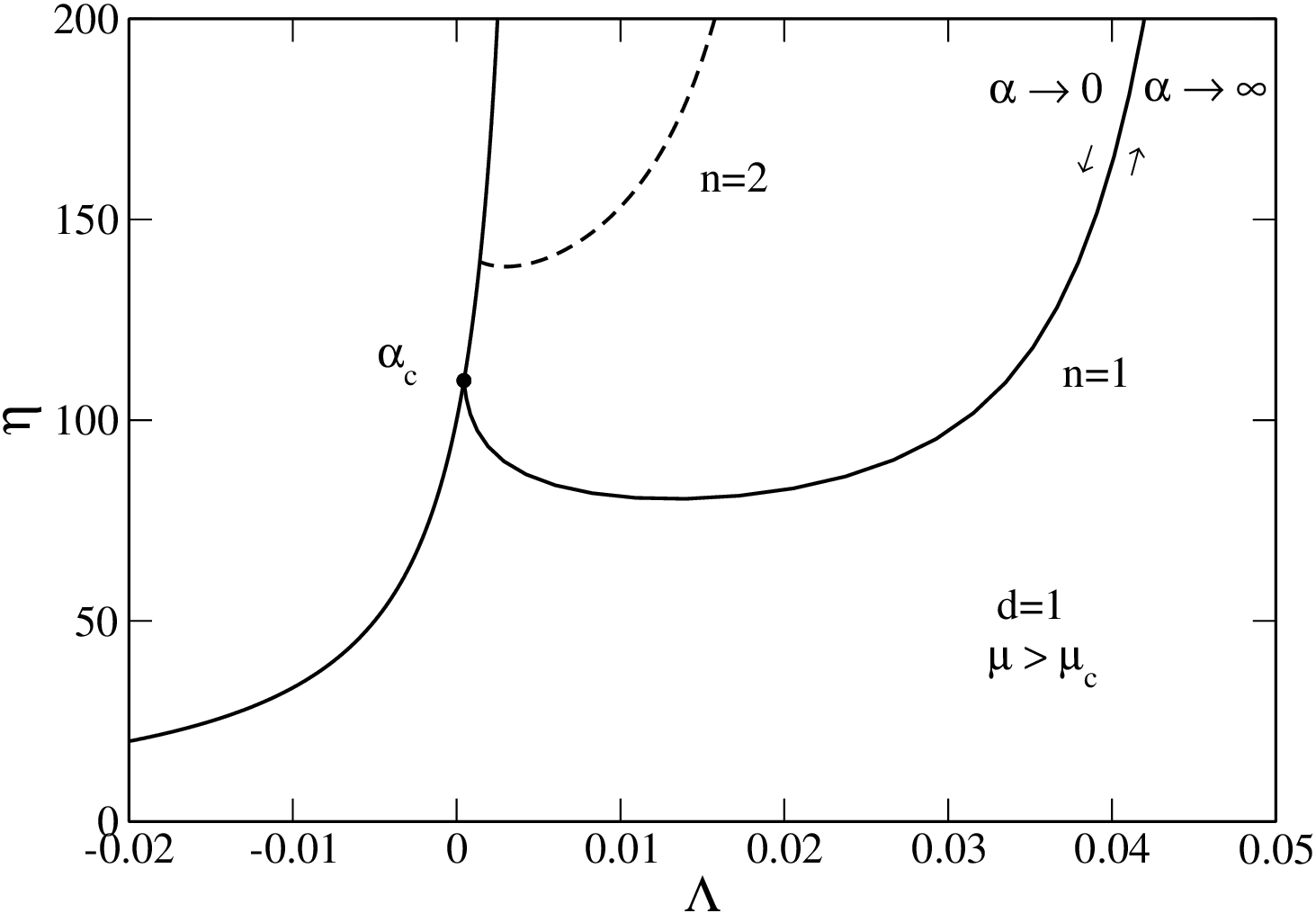}
\caption{Series of equilibria in $d=1$ for $\mu=10 > \mu_c$.}
\label{fig5}
\end{center}
\end{figure}

In Figs. \ref{fig4} and \ref{fig5}, we plot the series of equilibria in $d=1$ for different values of the screening parameter $\mu$.

Let us first describe the canonical ensemble (CE). The control
parameter is the inverse temperature $\eta$ and the stable states are
maxima of free energy $J$ at fixed mass $M$.  The homogeneous phase
exists for any value of $\eta$. It is stable for $\eta<\eta_c^*$
and unstable for $\eta>\eta_c^*$ (see Sec. \ref{sec_stab}). Comparing
Figs. \ref{fig4} and \ref{fig5}, we see that the screened Newtonian
model is characterized by a pitchfork bifurcation at
$\eta=\eta_c^*$. The pitchfork bifurcation is super-critical if
$\mu<\mu_c=\sqrt{2}\pi\simeq 4.4428829$ and sub-critical if $\mu>\mu_c$. This
interesting transition was first evidenced by Childress \& Percus
\cite{cp} using a different approach. In our thermodynamical approach,
this implies the existence of a canonical tricritical point at
$\mu_c=\sqrt{2}\pi$. For $\mu<\mu_c$ the phase transition is second
order and for $\mu>\mu_c$ the phase transition is first order.

Let us first consider $\mu<\mu_c$ (see Fig. \ref{fig4}). The
discussion is similar to that given for the modified Newtonian
model. The first branch $n=1$ of inhomogeneous states exists only for
$\eta>\eta_c^*$. It has a higher free energy $J$ than the homogeneous
phase (see Fig. \ref{free1-mu3-mod2}) and it is fully
stable. Secondary branches appear for smaller values of the
temperature but they have smaller values of free energy $J$ (see
Fig. \ref{free1-mu3-mod2}) and they are unstable (saddle points of
free energy). Therefore, the canonical caloric curve displays a second
order phase transition between homogeneous and inhomogeneous states
marked by the discontinuity of $\frac{\partial E}{\partial\beta}$ at
$\beta=\beta_c^*$. We note that, for the inhomogeneous states, there
exists two solutions with the same temperature and the same free
energy but with different density profiles corresponding to
$\alpha_1<\alpha_c$ and $\alpha_2>\alpha_c$, where $\alpha_c$ is the
value of $\alpha$ at the point of contact with in the homogeneous
branch. Thus, the inhomogeneous branch is degenerate. These two states
can be distinguished by their central density $\alpha$. Since
$\overline{\rho}/\rho_0=d\eta/\alpha^2$, the solution
$\alpha_1<\alpha_c$ corresponds to $\rho_0<\overline{\rho}$ and the
solution $\alpha_2>\alpha_c$ corresponds to
$\rho_0>\overline{\rho}$. The density profiles are similar to those
represented in Fig. \ref{prof} for the modified Newtonian model. In
conclusion: (i) for $\eta<\eta_c^*$, there is only one stable state
(homogeneous); (ii) for $\eta>\eta_c^*$, there are two stable states
$\alpha_1<\alpha_c$ and $\alpha_2>\alpha_c$ (inhomogeneous) with the
same free energy and one unstable state (homogeneous). Therefore, the
central density $\alpha$ plays the role of an order parameter (see
Fig. \ref{eta1-mu1-mod2}).

\begin{figure}
\begin{center}
\includegraphics[clip,scale=0.3]{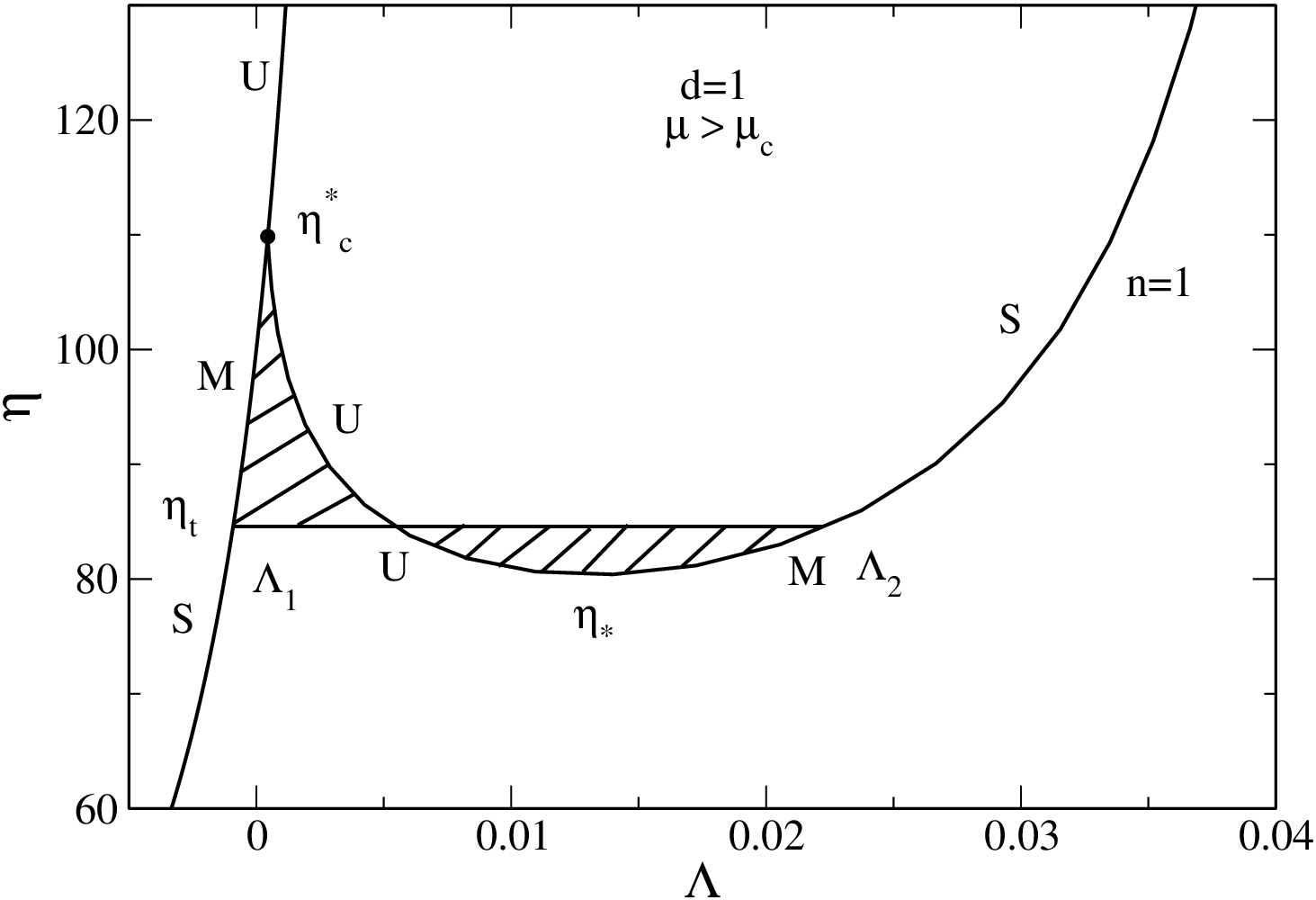}
\caption{Canonical caloric curve in $d=1$ for $\mu=10 > \mu_c$. It displays a canonical first-order phase transition marked by the discontinuity of the energy at $\eta=\eta_t(\mu)$. The region of negative specific heats is unstable in the canonical ensemble and replaced by a phase transition (Maxwell plateau). The temperatures $\eta_c^*$ and $\eta_*$ represent canonical spinodal points marking the end of the metastable phase.
}
\label{canonic-mod2}
\end{center}
\end{figure}

\begin{figure}
\begin{center}
\includegraphics[clip,scale=0.3]{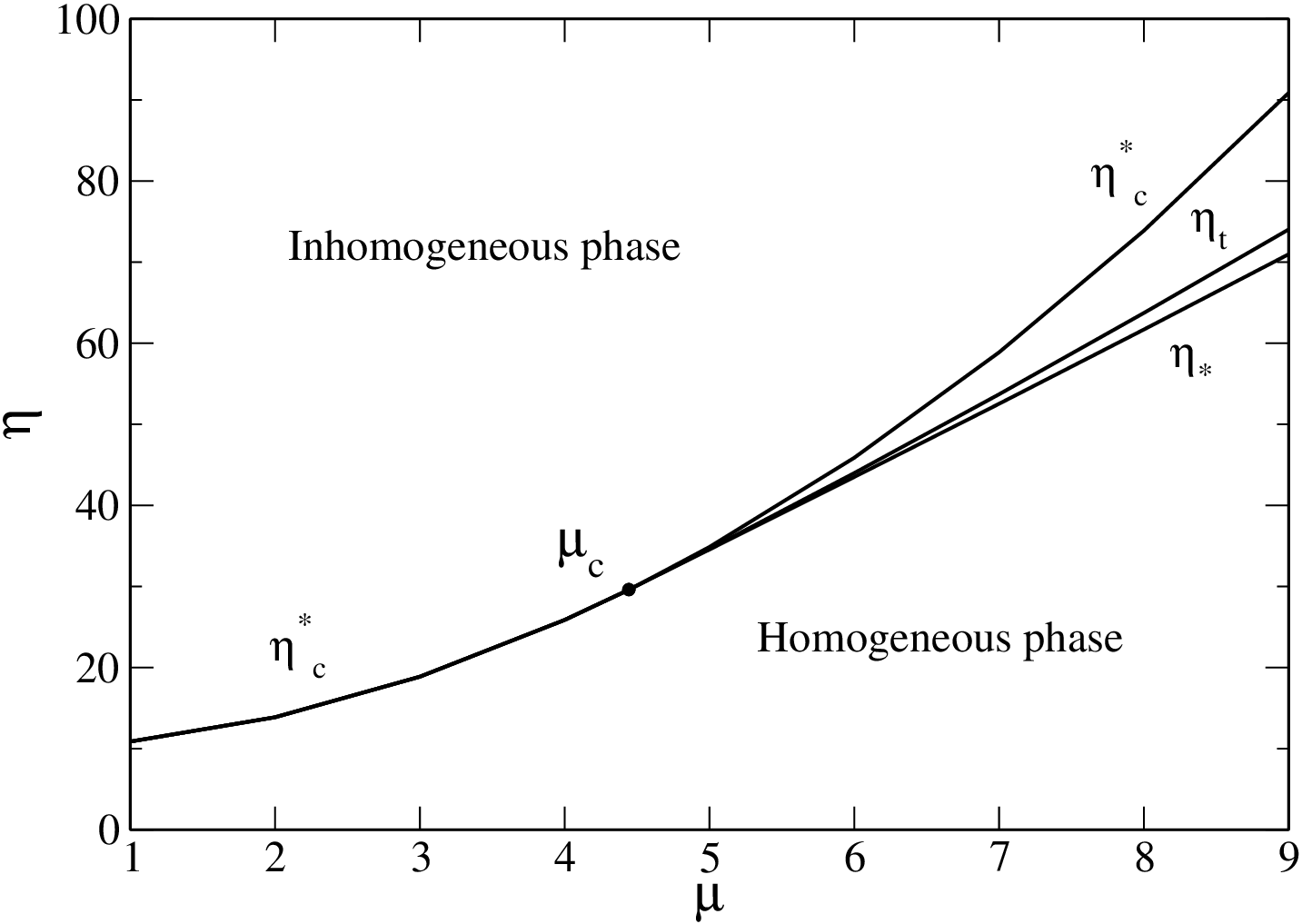}
\caption{Canonical phase diagram in $d=1$ exhibiting  a tricritical point at $\mu_c=\sqrt{2}\pi\simeq 4.44$ and $\eta\simeq 29.6$. We have represented $\eta_c$, $\eta_{*}$ and $\eta_t$ as a function of $\mu$. The region between $\eta_{*}$ and $\eta_c^*$ contains stable and metastable states.}
\label{tricrit}
\end{center}
\end{figure}

\begin{figure}
\begin{center}
\includegraphics[clip,scale=0.3]{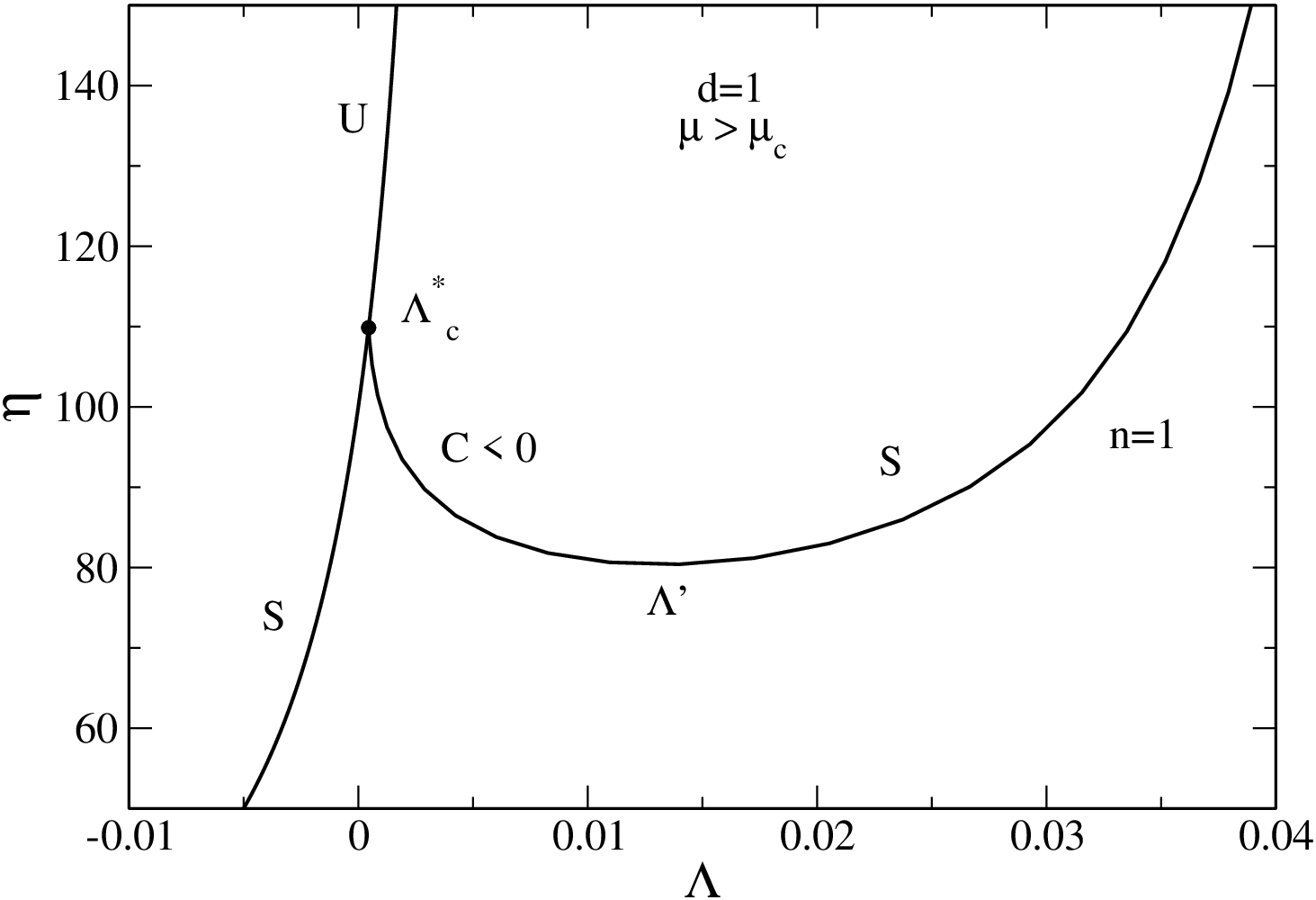}
\caption{Microcanonical caloric curve in $d=1$ for  $\mu=10 > \mu_c$. It displays a microcanonical second order phase transition marked by the discontinuity of $\frac{\partial\beta}{\partial E}$ at $E=E_c^*$. For $\mu > \mu_c$, there exists a region of negative specific heats that is stable in the microcanonical ensemble.}
\label{micro-mod2}
\end{center}
\end{figure}

\begin{figure}[!ht]
\begin{center}
\includegraphics[clip,scale=0.3]{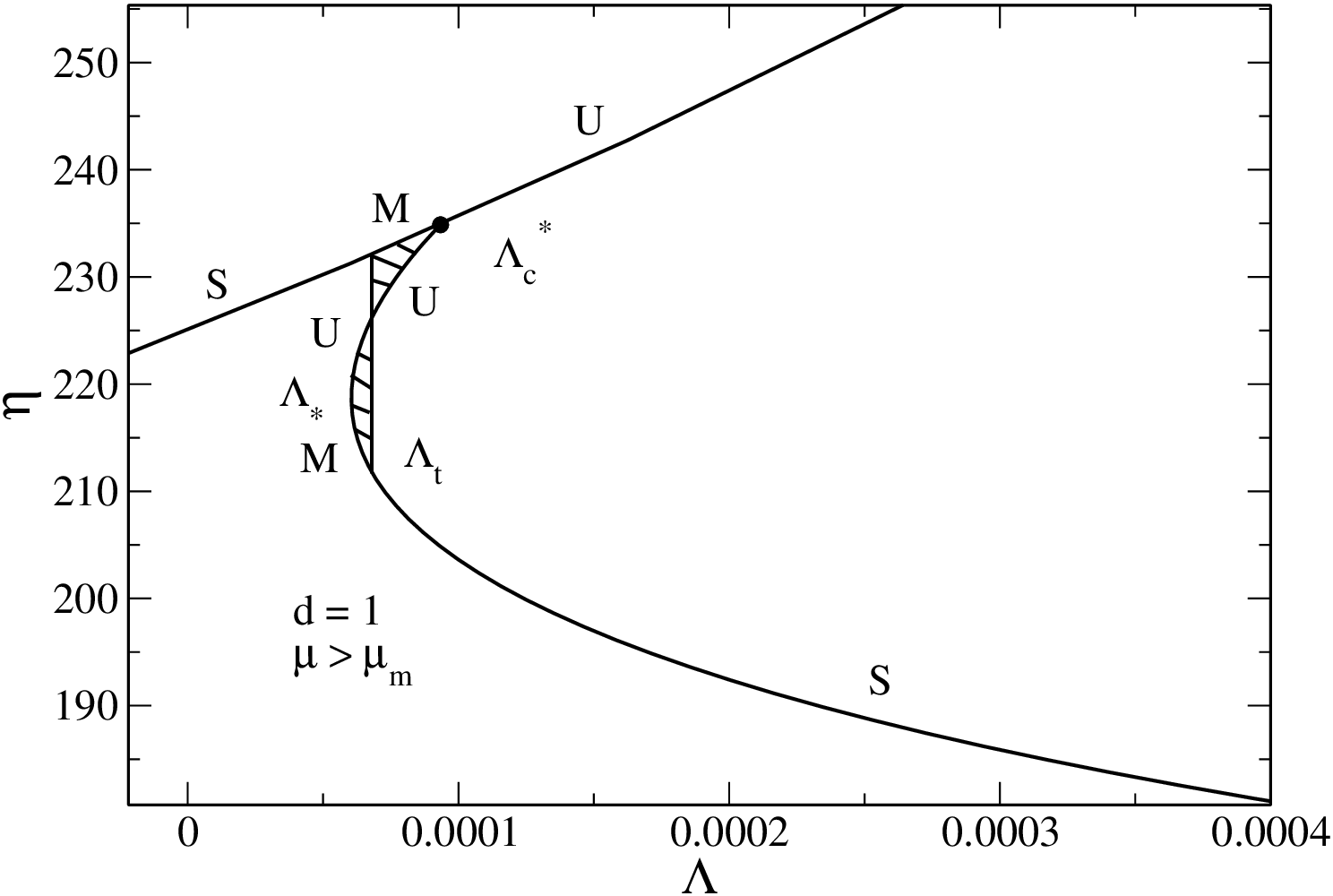}
\caption{Microcanonical caloric curve in $d=1$ for  $\mu=15 > \mu_m$. It displays a microcanonical first order phase transition marked by the discontinuity of $T$ at $E=E_t$. The energies $E_c^*$ and $E_*$ are spinodal points marking the end of the
metastable branches. Note that this first order phase transition
occurs in an extremely small range of energies.}
\label{fig-mu15}
\end{center}
\end{figure}

\begin{figure}
\begin{center}
\includegraphics[clip,scale=0.3]{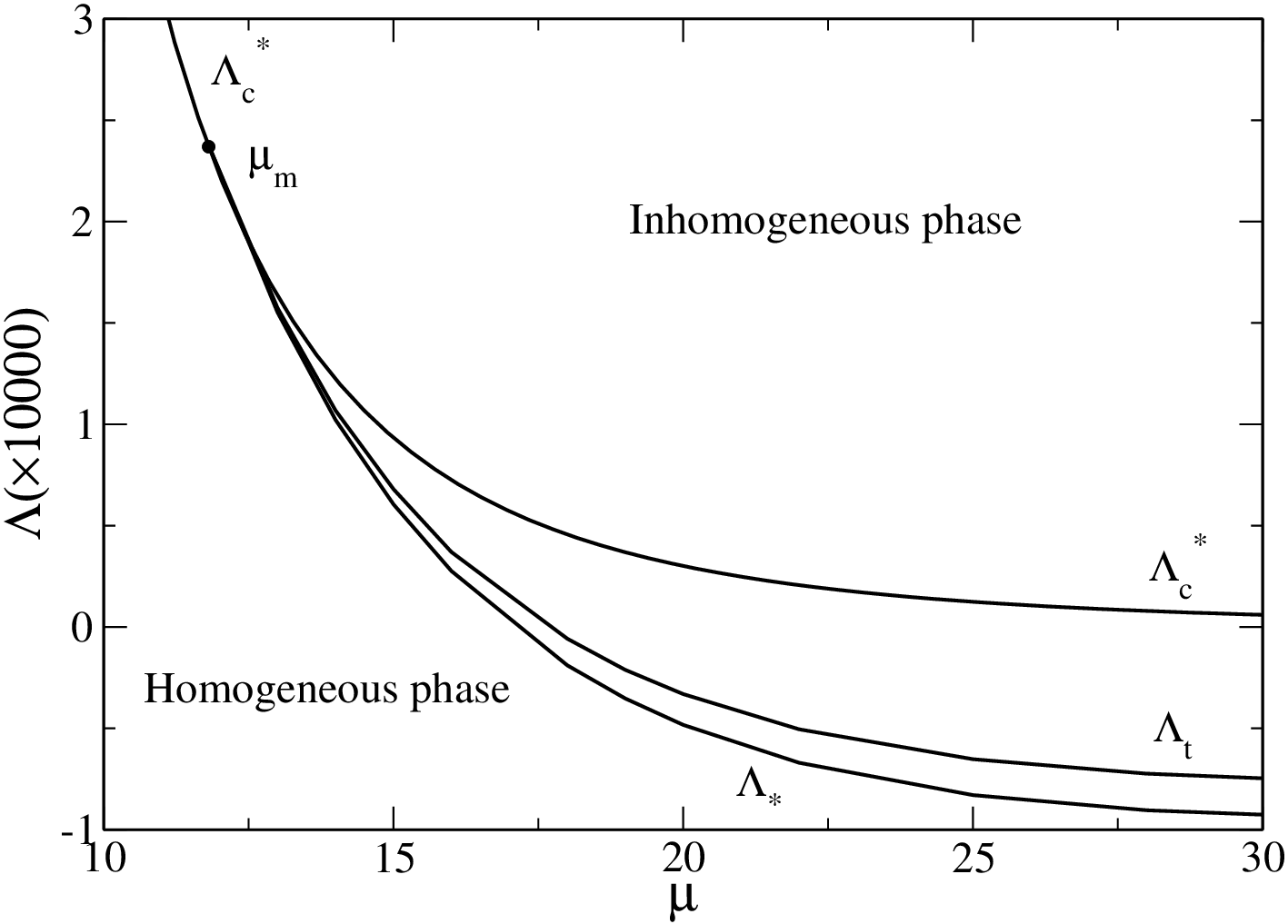}
\caption{Microcanonical phase diagram in $d=1$ exhibiting  a tricritical point at $\mu_m\simeq 11.8$ and $\Lambda\simeq 2.37\, 10^{-4}$. We have represented $\Lambda_c$, $\Lambda_{*}$ and $\Lambda_t$ as a function of $\mu$. The region between $\Lambda_{*}$ and $\Lambda_c^*$ contains stable and metastable states. We again emphasize the small range of energies where this first order phase transition takes place. }
\label{ene-crit-micro}
\end{center}
\end{figure}

\begin{figure}
\begin{center}
\includegraphics[clip,scale=0.3]{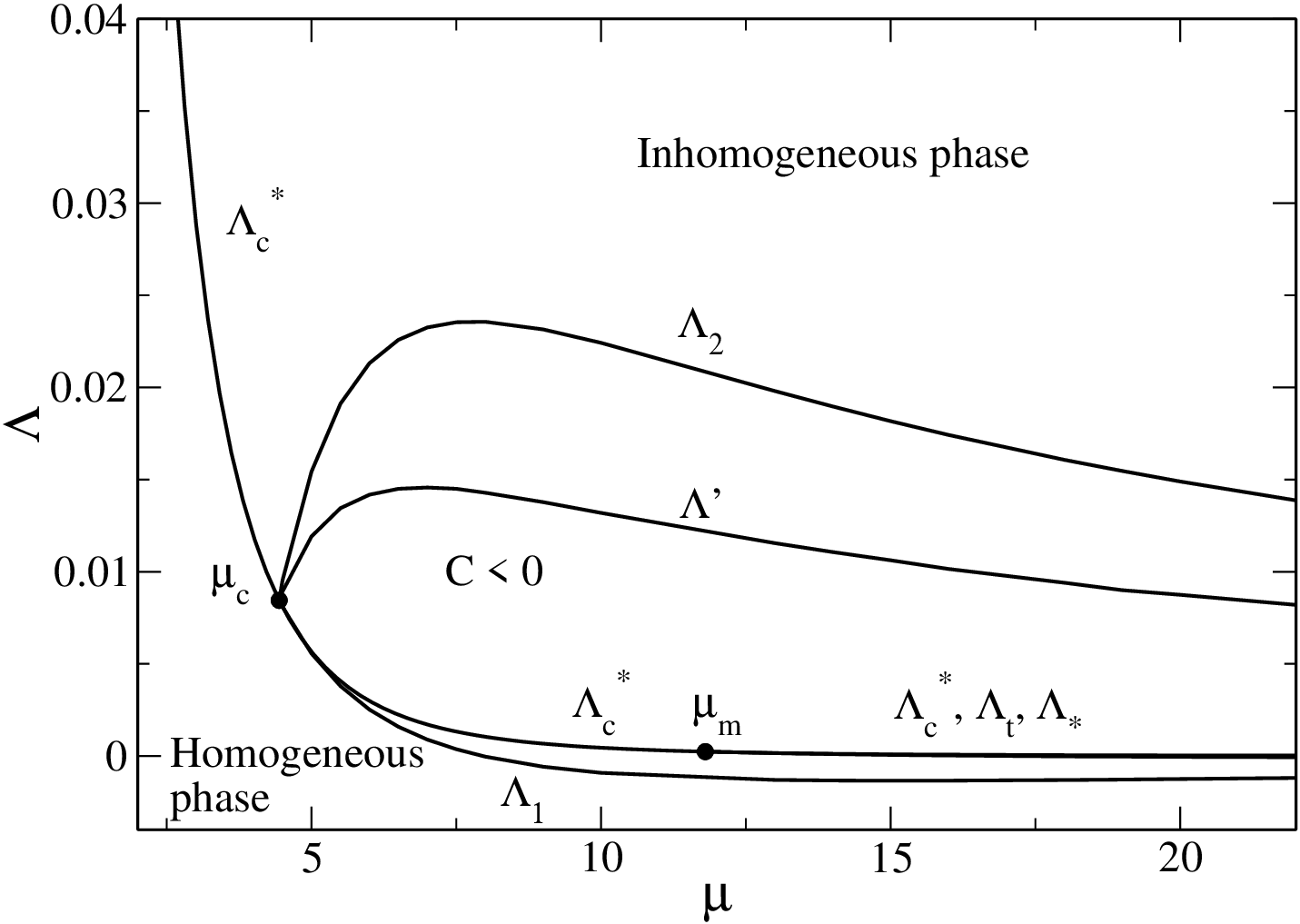}
\caption{Microcanonical phase diagram in $d=1$. We have represented $\Lambda_{c}^{*}$, $\Lambda'$, $\Lambda_1$ and $\Lambda_2$ as a function of $\mu$. These energies coincide for $\mu_c=\sqrt{2}\pi\simeq 4.44$ and $\Lambda\simeq 0.0084$. These energies delimitate respectively the region of  negative specific heats and the region of strict ensembles inequivalence (see main text): the energies in these regions cannot be reached in the canonical ensemble.}
\label{ene-crit}
\end{center}
\end{figure}

Let us now consider $\mu>\mu_c$ (see Fig. \ref{fig5}). The first
branch $n=1$ of inhomogeneous states exists only for
$\eta>\eta_*(\mu)$. The caloric curve displays a canonical first order
phase transition at $\eta_t(\mu)$ marked by the discontinuity of the
energy $E$ (see Fig. \ref{canonic-mod2}). The temperature of
transition $\eta_t(\mu)$ can be obtained by plotting the free energy
of the two phases as a function of the temperature and determining at
which temperature they become equal (see
Fig. \ref{free1-mu10-mod2}). Equivalently, it can be obtained by
performing a horizontal Maxwell construction \cite{ijmpb}. The homogeneous phase is
fully stable for $\eta<\eta_t$, metastable for $\eta_t<\eta<\eta_c^*$
and unstable for $\eta>\eta_c^*$. The right branch of the
inhomogeneous phase is fully stable for $\eta>\eta_t$ and metastable
for $\eta_*<\eta<\eta_t$. The left branch is unstable. Note that this
branch has negative specific heats which is not permitted in the
canonical ensemble. Secondary branches appear for smaller values of
the temperature but they have smaller values of free energy $J$ and
they are unstable. We also note that the branch of inhomogeneous
states is degenerate since the curves $\alpha<\alpha_c$ and
$\alpha>\alpha_c$ coincide. In conclusion: (i) for $\eta<\eta_*$,
there is only one stable state (homogeneous); (ii) for
$\eta_*<\eta<\eta_c^*$, there are three stable states (one homogeneous
and two inhomogeneous) and two unstable states (inhomogeneous); (iii)
for $\eta>\eta_c^*$, there are two stable states (inhomogeneous) and
one unstable state (homogeneous). The pairs of inhomogeneous states
have the same free energy. Therefore, the central density $\alpha$
plays the role of an order parameter (see Fig.
\ref{eta1-mu10-mod2}).

The canonical phase diagram is represented in Fig. \ref{tricrit} where
we have plotted $\eta_c^*$, $\eta_{*}$ and $\eta_t$ as a function of
$\mu$. The three temperatures coincide at the
tricritical point $\mu=\mu_c$. At that point, the
phase transition goes from second order ($\mu<\mu_c$) to first order
($\mu>\mu_c$).

The strict caloric curve (see Figs. \ref{fig4} and  \ref{canonic-mod2}), corresponding to the fully
stable states, is denoted (S). The physical caloric curve should take
into account the metastable states (M) because they are
long-lived. The states (U) are unstable. We see that there exists a
fully stable equilibrium state for any temperature and any screening
length. This is consistent with the usual Newtonian model in $d=1$
\cite{kl,sc}. This is also consistent with the results of chemotaxis
since it has been established rigorously that there is no blow up in
$d=1$
\cite{nagai}.

Let us finally describe the microcanonical ensemble (MCE). The control
parameter is the energy $\Lambda$ and the stable states are maxima of
entropy $S$ at fixed mass $M$ and energy $E$. The homogeneous phase
exists for any $\Lambda<\Lambda_{max}=d/(2\mu^2)$. It is stable for
$\Lambda<\Lambda_c^*$ and unstable for $\Lambda>\Lambda_c^*$ (see
Sec. \ref{sec_stab}).  Comparing Figs. \ref{micro-mod2} and
\ref{fig-mu15}, we see that there exists a microcanonical tricritical
point at $\mu_m\simeq 11.8$ and $\Lambda\simeq 2.37\, 10^{-4}$
(corresponding to $\eta\simeq 149.1096$). For $\mu<\mu_m$ the phase
transition is second order and for $\mu>\mu_m$ the phase transition is
first order.

Let us first consider $\mu<\mu_m$ (see Figs. \ref{fig4} and \ref{micro-mod2}). 
The first branch $n=1$ of inhomogeneous states exists for
$\Lambda_c^*<\Lambda<\Lambda_{max}^{(1)}=1/(2\mu\tanh(\mu))$ (see
Appendix \ref{sec_me}). It
has a higher entropy $S$ than the homogeneous phase and it is
stable. Secondary branches appear for smaller values of the energy but
they have smaller values of entropy and are unstable. The
microcanonical caloric curve displays a second order phase transition
marked by the discontinuity $\frac{\partial\beta}{\partial E}$ at
$E=E_c^*$. For $\mu<\mu_c$, the specific heat is always positive. In
that case, the microcanonical and canonical ensembles are
equivalent. For $\mu>\mu_c$, a region of negative specific heats
appears. This leads to a convex dip in the entropic curve $S(E)$ (see
Fig. \ref{entropy1-mu10-mod2}). In that case, the microcanonical and
canonical ensembles are inequivalent: the states with negative
specific heats are stable in MCE while they are unstable in CE
(compare Figs. \ref{canonic-mod2} and
\ref{micro-mod2}). Therefore, these energies cannot be achieved in a
canonical description.

Let us now consider $\mu>\mu_m$ (see Fig.  \ref{fig-mu15}). The first
branch $n=1$ of inhomogeneous states exists only for
$\Lambda>\Lambda_*(\mu)$. The caloric curve displays a microcanonical
first order phase transition at $\Lambda_t(\mu)$ marked by the
discontinuity of the temperature $T$ and the existence of metastable
states. The energy of transition $\Lambda_t(\mu)$ can be obtained by
plotting the entropy of the two phases as a function of the energy and
determining at which energy they become equal. Equivalently, it can be
obtained by performing a vertical Maxwell construction \cite{ijmpb}.
The discussion is similar to that given in the canonical ensemble except that
the axis are reversed. In conclusion: (i) for $\Lambda<\Lambda_*$,
there is only one stable state (homogeneous); (ii) for
$\Lambda_*<\Lambda<\Lambda_c^*$, there are three stable states (one homogeneous
and two inhomogeneous) and two unstable states (inhomogeneous); (iii)
for $\Lambda>\Lambda_c^*$, there are two stable states (inhomogeneous) and
one unstable state (homogeneous). The pairs of inhomogeneous states
have the same entropy. Therefore, the central density $\alpha$
plays the role of an order parameter (see Fig. \ref{ene-alfa-mu15}).

The microcanonical phase diagram is represented in  Figs. \ref{ene-crit-micro}  and \ref{ene-crit} where
we have plotted $\Lambda_c^*$, $\Lambda_{*}$ and $\Lambda_t$ as a function of
$\mu$. The three energies coincide at the microcanonical
tricritical point $\mu=\mu_m$. At that point, the
phase transition goes from second order ($\mu<\mu_m$) to first order
($\mu>\mu_m$). We have also represented the region of negative specific heats which appears at the canonical tricritical point  $\mu=\mu_c$. For $\mu_c<\mu<\mu_m$, it is delimited by $\Lambda_c^*$ and $\Lambda'$ and for $\mu>\mu_m$, it is delimited by $\Lambda_*$ and $\Lambda'$. This region of negative specific heats also defines  the physical region of ensembles inequivalence, i.e. the states that are stable in MCE but unstable in CE (metastable states are considered here as stable states).  Finally, we have represented the strict region of ensembles inequivalence, i.e. the states that are stable in MCE but unstable or metastable in CE. It is delimited by $\Lambda_1$ and $\Lambda_2$ and, of course, contains the negative specific heats region.

The strict caloric curve (see Figs.  \ref{fig4}, \ref{micro-mod2} and  \ref{fig-mu15}), corresponding to the fully stable states, is denoted (S). The states (U) are unstable. The states (M) are metastable but they are long-lived. We see that there exists a fully stable  equilibrium state for any accessible energy and any screening length. This is consistent with the usual Newtonian model in $d=1$ \cite{kl,sc}.

In conclusion, for $\mu<\mu_c$, the system displays canonical and microcanonical second order phase transitions. For $\mu_c<\mu<\mu_m$ (canonical tricritical point), the system displays canonical first order phase transitions and microcanonical second order phase transitions. For $\mu>\mu_m$ (microcanonical tricritical point), the system displays canonical and microcanonical first order phase transitions. Note that the canonical and microcanonical tricritcal points do not coincide as also observed in other models \cite{bmr,prefermi,tbdr}.

\subsubsection{The dimensions $d=2$ and $d=3$}

\begin{figure}[!ht]
\begin{center}
\includegraphics[clip,scale=0.3]{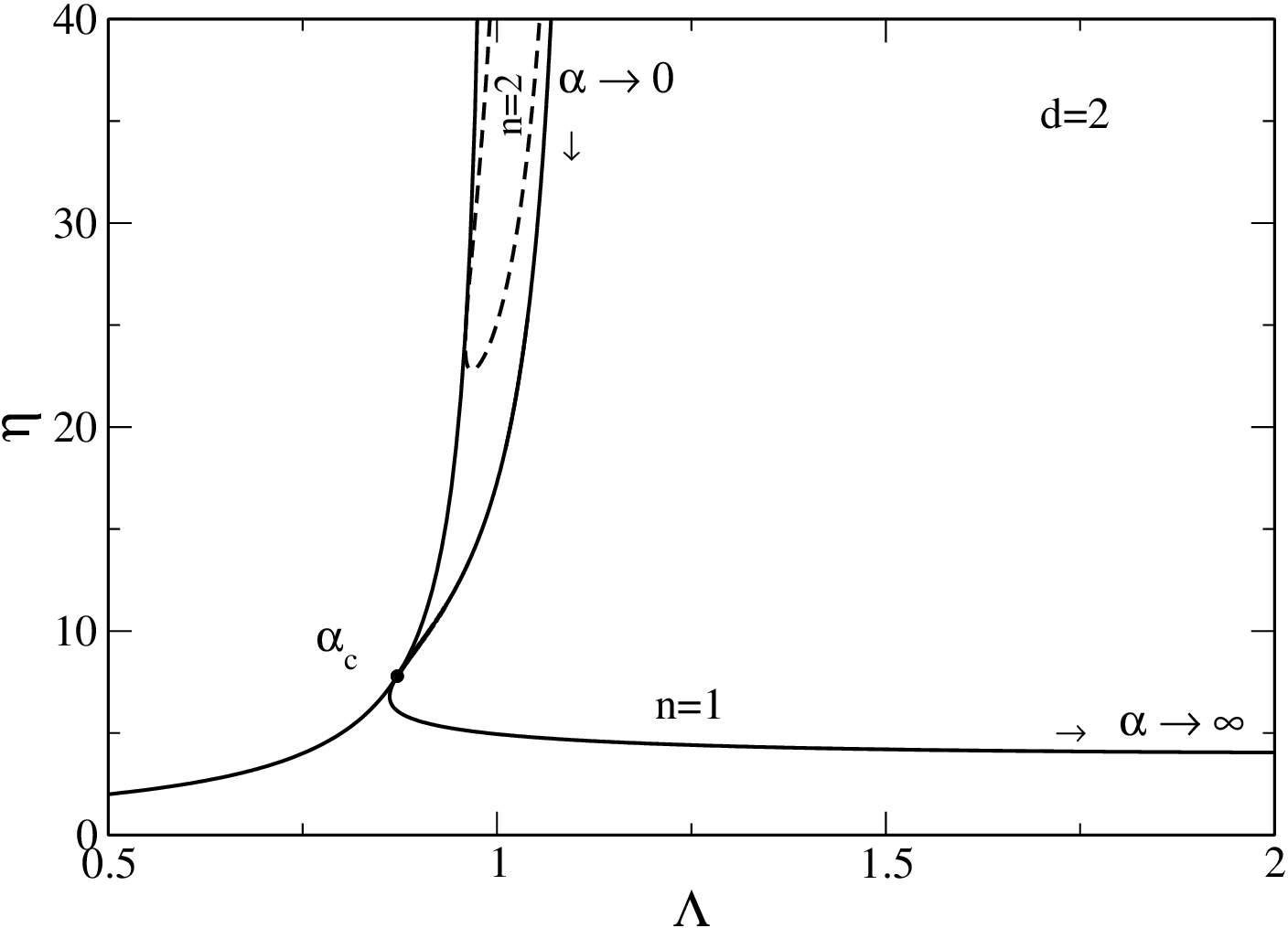}
\caption{Caloric curve in $d=2$ for $\mu=1$.}
\label{fig6}
\end{center}
\end{figure}

\begin{figure}[!ht]
\begin{center}
\includegraphics[clip,scale=0.3]{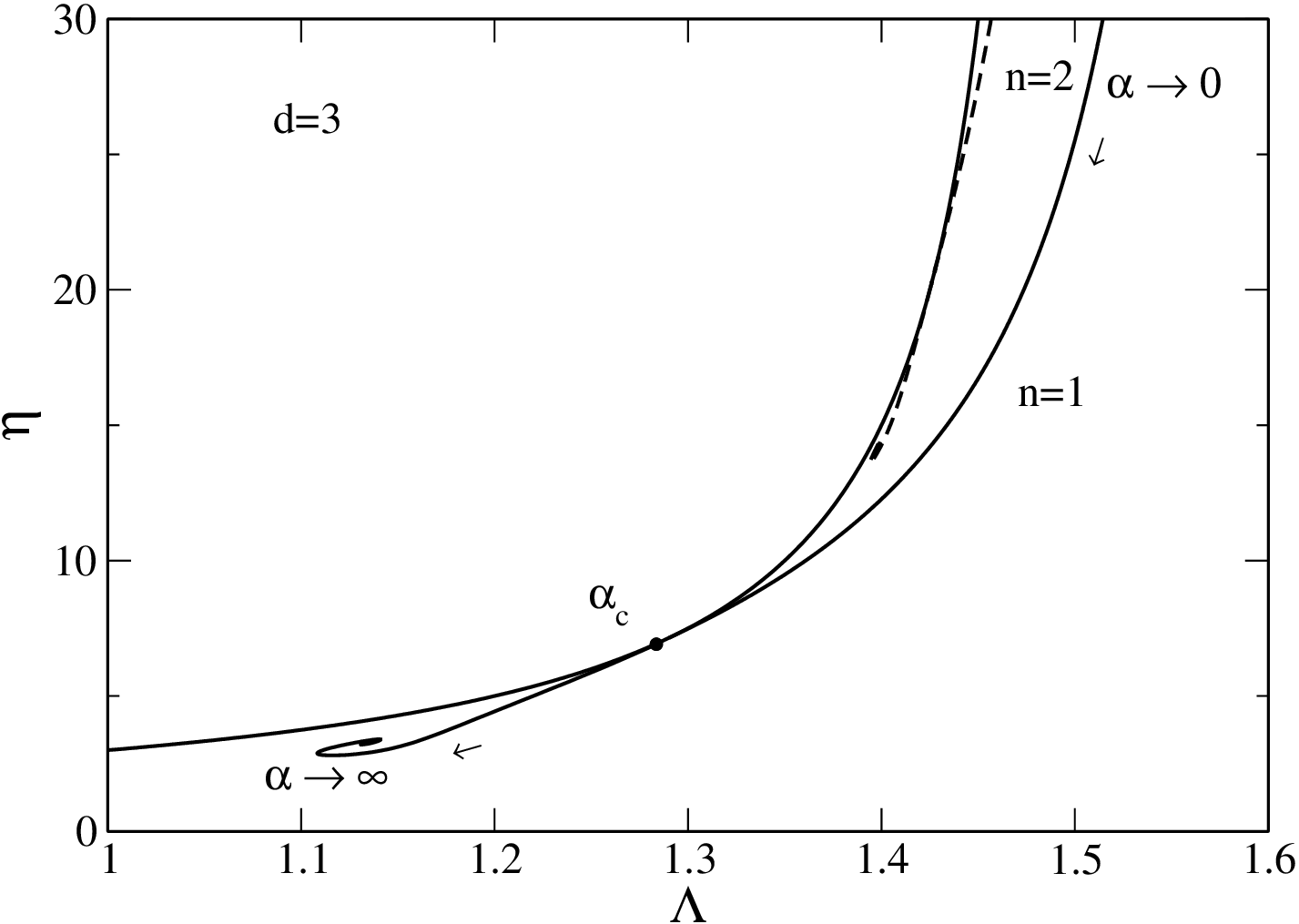}
\caption{Caloric curve in $d=3$ for $\mu=1$.}
\label{fig8}
\end{center}
\end{figure}

In Figs. \ref{fig6} and \ref{fig8}, we plot the series of equilibria
in $d=2$ and $d=3$. We have considered different values of $\mu$ but
only the case $\mu=1$ is shown. We have observed that the shape of the
diagrams does not significantly depend on the value of the screening
parameter $\mu$. Therefore, the description of these diagrams is
similar to the one given in Secs. \ref{sec_modif2} and
\ref{sec_modif3} for the modified Newtonian model.

\section{Stability of the homogeneous phase}
\label{sec_stab}

In this section, we study the stability of the homogeneous phase in the case where the potential satisfies the modified Poisson equation (\ref{mn1}) or the screened Poisson equation (\ref{ay1}). We first consider the spectral stability of the homogeneous phase with respect to the Smoluchowski equation or, equivalently, with respect to the Keller-Segel model. This will allow us to determine the growth rate (unstable case) or the damping rate (stable case) of the perturbation.  Then, we investigate the dynamical and thermodynamical stability of a larger class of systems by determining whether the homogeneous phase is a maximum of entropy at fixed mass and energy in MCE or a minimum of free energy at fixed mass in CE.

\subsection{Spectral stability}
\label{sec_s}

We consider the mean field Smoluchowski equation
\begin{eqnarray}
\frac{\partial \rho}{\partial t}=\nabla \cdot \left\lbrack\frac{1}{\xi} \left (\frac{k_B T}{m}\nabla \rho + \rho \nabla \Phi
\right )\right\rbrack,
\label{s1}
\end{eqnarray}
coupled to the modified Poisson equation (\ref{mn1}) or to the screened Poisson equation (\ref{ay1}). The boundary conditions are given by  Eq. (\ref{ksbcb}). Up to a change of notation, these equations also describe the Keller-Segel model (\ref{ks1})-(\ref{ks5}) or  (\ref{ks1})-(\ref{ks6}). In the modified Newtonian model, the homogeneous steady state  satisfies
\begin{eqnarray}
\rho=\overline{\rho}, \qquad \Phi=0.
\label{s6}
\end{eqnarray}
In the screened Newtonian model, it satisfies
\begin{eqnarray}
-k_{0}^{2}\Phi =S_d G \rho,
\label{s7}
\end{eqnarray}
where $\rho$ and $\Phi$ are uniform. In both models, the linearized equations can be written
\begin{eqnarray}
\label{s7b}
\xi\frac{\partial \delta\rho}{\partial t}=\frac{k_B T}{m}\Delta \delta\rho+ \rho \Delta\delta\Phi,
\end{eqnarray}
\begin{eqnarray}
\Delta \delta\Phi-k_{0}^{2}\delta\Phi =S_d G \delta\rho,
\label{s8}
\end{eqnarray}
where $k_0=0$ in the modified Newtonian model and  $k_0\neq 0$ in the screened Newtonian model.

In an infinite domain, the spectral stability of the homogeneous solutions of the mean field Smoluchowski equation (and its generalizations) coupled to Eqs. (\ref{mn1}) and (\ref{ay1}) has been studied by Chavanis \& Sire \cite{jeansbio} who stressed the analogy with the Jeans instability in astrophysics \cite{jeans}. Here, we describe how the results are modified in a bounded domain with the boundary conditions (\ref{ksbcb}). This problem was considered by Keller \& Segel \cite{ks} in $d=2$. We shall perform the stability analysis in $d$ dimensions. Let us call $\psi_k({\bf r})$ the eigenfunctions of the Laplacian and $-k^2$ the corresponding eigenvalues. They are solution of
\begin{eqnarray}
\Delta \psi_k=-k^{2}\psi_k,
\label{s9}
\end{eqnarray}
with
\begin{eqnarray}
\nabla\psi_k\cdot {\bf n}=0,
\label{s10}
\end{eqnarray}
on the boundary. It is easy to check
that the eigenvalues are necessarily negative (hence the notation $-k^2$). Indeed, multiplying Eq. (\ref{s9})
by $\psi_k$, integrating on the whole domain, and using an integration
by parts, we get $\int(\nabla\psi_k)^2\, d{\bf r}=k^2\int \psi_k^2\, d{\bf
r}$, which proves the result. In a bounded domain, their values are
``quantized'' (see below). The lowest non-zero value of $k$ will play a
particular role as it determines the critical temperature below which the homogeneous phase becomes unstable. The expression of the eigenfunctions and eigenvalues depends on the domain shape and on
the dimension of space. In the following, we shall work in a spherical
box in $d=1,2,3$ dimensions.

$\bullet$ In $d=1$, we have
\begin{eqnarray}
\psi_n=\cos(k_n x),
\label{s11}
\end{eqnarray}
with
\begin{eqnarray}
k_n=n\frac{\pi}{R},
\label{s12}
\end{eqnarray}
where $n$ is an integer. The smallest non zero eigenvalue is $k_1={\pi}/R$.

$\bullet$ In $d=2$, we have
\begin{eqnarray}
\psi_{ni}=J_{n}(k_{ni} r)\cos(n\theta),
\label{s13}
\end{eqnarray}
with
\begin{eqnarray}
k_{ni}=\frac{\gamma_{ni}}{R},
\label{s14}
\end{eqnarray}
where $n$ is an integer and $\gamma_{ni}$ is the $i$-th zero of $J'_n(x)$.
The smallest non zero eigenvalue is $k_{01}={\gamma_{01}}/R$ where $\gamma_{01}=j_{11}=3.83171...$ is the first zero of $J'_0(x)=-J_{1}(x)$. The axisymmetric mode ($n=0$) is
\begin{eqnarray}
\psi_{0i}=J_{0}(k_{0i} r).
\label{s15}
\end{eqnarray}

$\bullet$ In $d=3$, we have
\begin{eqnarray}
\psi_{lmi}=\frac{1}{\sqrt{r}} J_{l+\frac{1}{2}}(k_{li}r)Y_{lm}(\theta,\phi),
\label{s16}
\end{eqnarray}
with
\begin{eqnarray}
k_{li}=\frac{\gamma_{li}}{R},
\label{s17}
\end{eqnarray}
where $l,m$ are integers with $l\ge |m|$ and $\gamma_{li}$ is the $i$-th zero of
\begin{eqnarray}
\frac{x J'_{l+1/2}(x)}{J_{l+1/2}(x)}-\frac{1}{2}=0.
\end{eqnarray}
The smallest non zero eigenvalue is $k_{01}={\gamma_{01}}/{R}$ where $\gamma_{01}=x_1=4.49341...$ is the first root of $\tan(x)=x$.
The spherically symmetric  mode ($l,m=0$) is
\begin{eqnarray}
\psi_{00i}=\frac{\sin(k_{0i}r)}{r}.
\label{s18}
\end{eqnarray}

The solutions of the linearized equations (\ref{s7b}) and (\ref{s8}) can be expanded on the eigenmodes, writing
\begin{eqnarray}
\delta\rho({\bf r},t)=\sum_{k}A_{k}e^{\sigma_k t/\xi}\psi_k({\bf r}),
\label{s19}
\end{eqnarray}
\begin{eqnarray}
\delta\Phi({\bf r},t)=\sum_{k}B_{k}e^{\sigma_k t/\xi}\psi_k({\bf r}),
\label{s20}
\end{eqnarray}
where the sum runs on the (quantized) eigenvalues.    Substituting Eqs. (\ref{s19}) and (\ref{s20}) in Eqs. (\ref{s7b}) and (\ref{s8}), we obtain the algebraic equations
\begin{eqnarray}
\left (\sigma_k+\frac{k_B T}{m}k^2\right )A_k+\rho k^2 B_k=0,
\label{s21}
\end{eqnarray}
\begin{eqnarray}
S_d G A_k+(k^2+k_0^2) B_k=0.
\label{s22}
\end{eqnarray}
There will be non-trivial solutions only if the determinant of this system of equations is zero. This yields the dispersion relation
\begin{eqnarray}
\sigma_k=\left (\frac{S_d G\rho}{k^2+k_0^2}-\frac{k_B T}{m}\right )k^2,
\label{s23}
\end{eqnarray}
relating $\sigma_k$ to the wavenumber $k$. The amplitudes $A_k$ and $B_k$ are determined by the initial condition. We see that $\sigma_k$ is real so that the perturbation either grows or
decays exponentially rapidly. The homogeneous phase will be spectrally stable if $\sigma_k<0$ for all $k$ and it will be spectrally unstable if there exists one or several modes for which $\sigma_k>0$. We note that the dispersion relation (\ref{s23}) is the same as in an
infinite domain \cite{jeansbio}. However, in a finite domain, the allowed wavenumbers $k$ are quantized while they are continuous in an infinite domain.

\begin{figure}[!ht]
\begin{center}
\includegraphics[clip,scale=0.3]{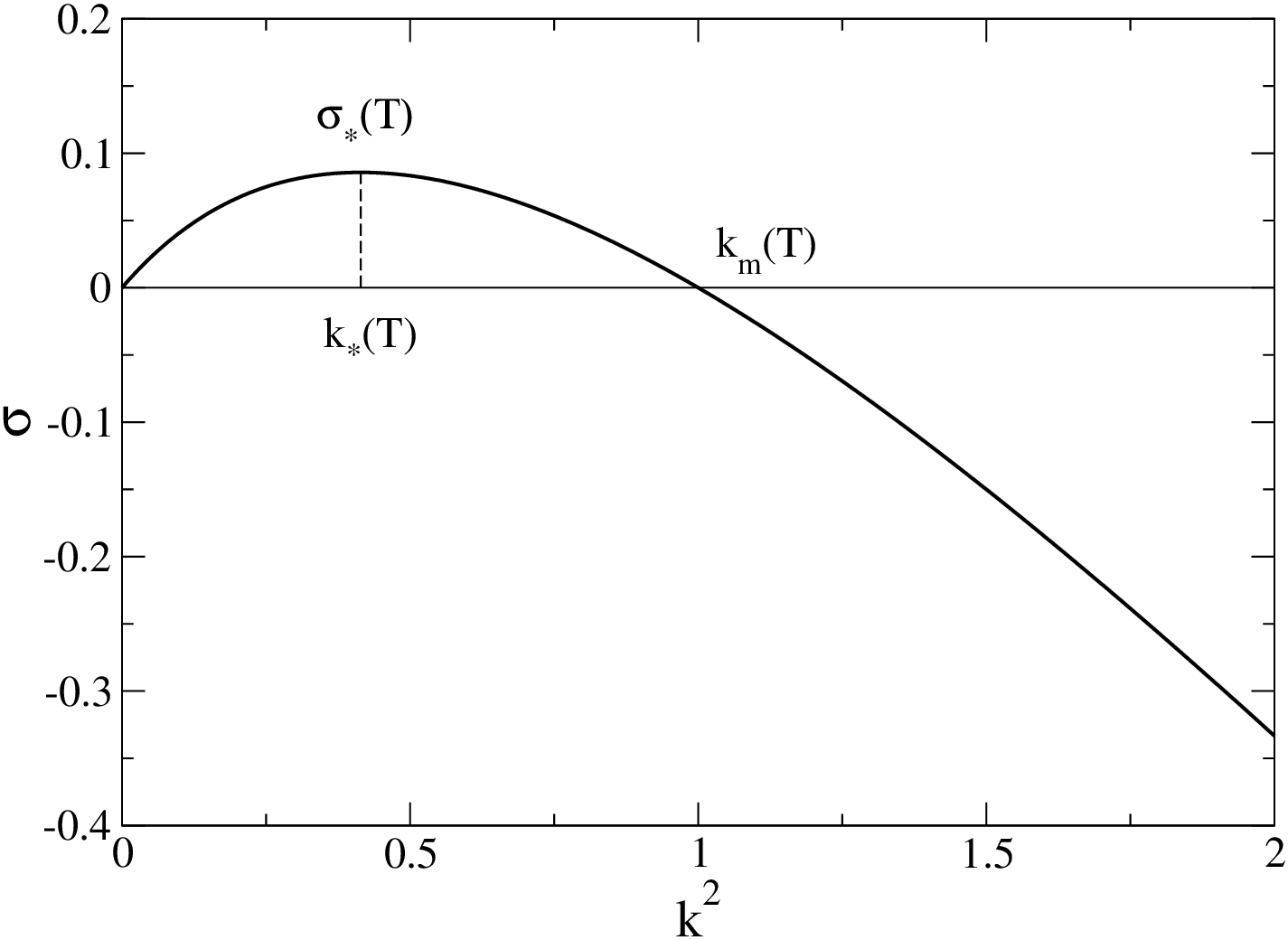}
\caption{Growth ($\sigma>0$) or decay ($\sigma<0$) rate as a function of the wavenumber $k$. The system is unstable for $k<k_m(T)$ and the maximum growth rate is reached for $k=k_{*}(T)$. The parameters have been scaled such that $k_0=1$, $T_c=1$, $T=1/2$. }
\label{sigmak}
\end{center}
\end{figure}

\begin{figure}[!ht]
\begin{center}
\includegraphics[clip,scale=0.3]{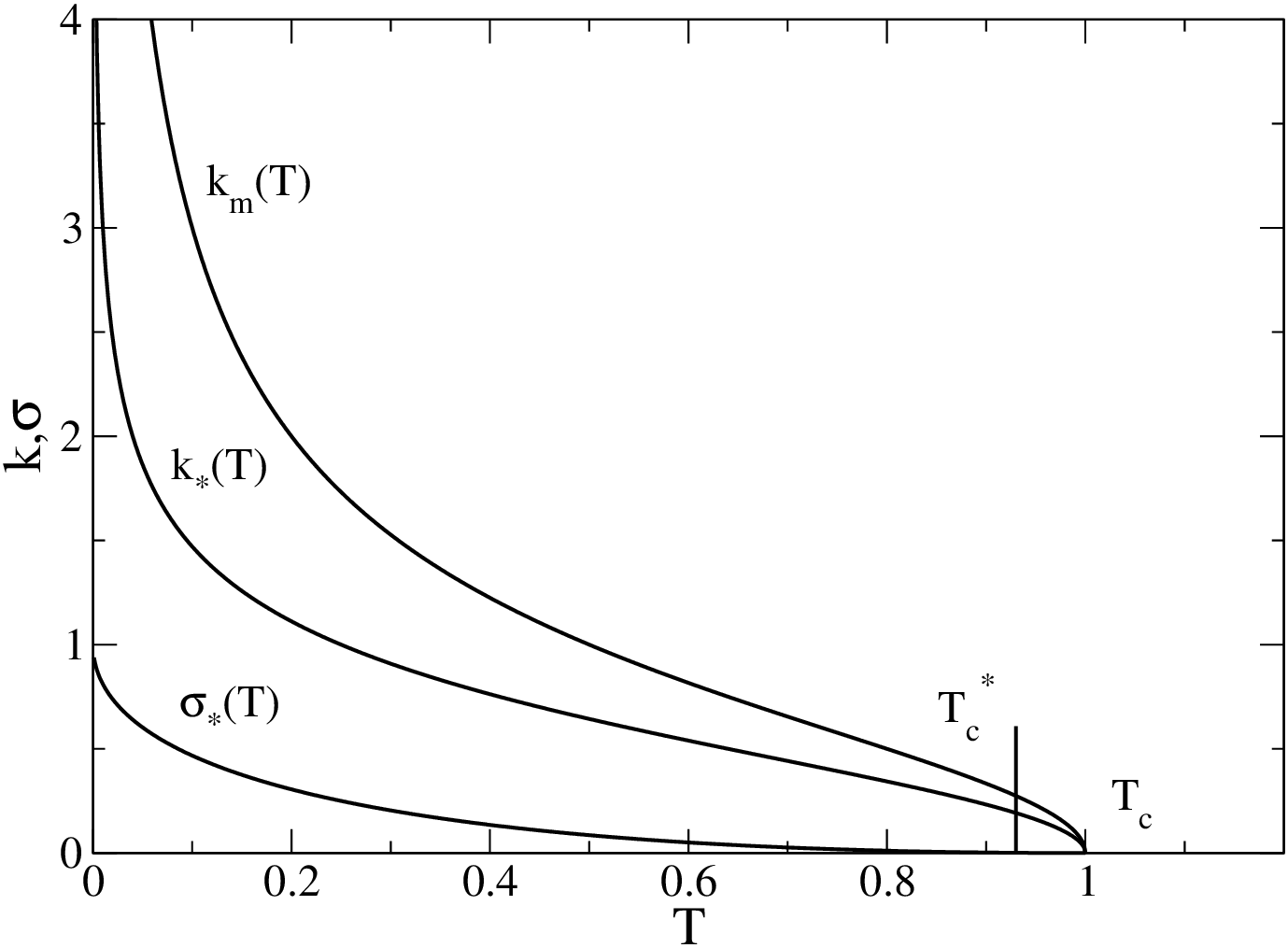}
\caption{Evolution of $k_m$, $k_*$ and $\sigma_*$ as a function of the temperature. The parameters have been scaled such that $k_0=1$ and $T_c=1$. }
\label{kT}
\end{center}
\end{figure}

According to Eq. (\ref{s23}), the system will be unstable if there exists $k\neq 0$ such that
\begin{eqnarray}
\frac{S_d G\rho}{k^2+k_0^2}>\frac{k_B T}{m}.
\label{s24}
\end{eqnarray}
Therefore, a necessary condition of instability is that
\begin{eqnarray}
\frac{k_B T}{m}<\frac{S_d G\rho}{k_f^2+k_0^2}\equiv \frac{k_B T_c^*}{m},
\label{s25}
\end{eqnarray}
where $k_f$ is the smallest non-zero wavenumber. For $T>T_c^*$, the homogeneous distribution is stable for perturbations with arbitrary wavenumbers. For $T<T_c^*$,  the homogeneous distribution is unstable for perturbations with wavenumbers
\begin{eqnarray}
k^2<\frac{S_d G m \rho}{k_B T}-k_{0}^{2}\equiv k_m^2.
\label{s26}
\end{eqnarray}
For $k_0=0$, the critical temperature is
\begin{eqnarray}
\frac{k_B T_c^*}{m}=\frac{S_d G\rho}{k_f^2},
\label{s27}
\end{eqnarray}
and we recover the Jeans criterion
\begin{eqnarray}
k^2<\frac{S_d G m \rho}{k_B T}\equiv k_J^2.
\label{s28}
\end{eqnarray}
In the general case, the instability criterion can be written
\begin{eqnarray}
k^2<k_{J}^2-k_{0}^{2}\equiv k_m^2.
\label{s29}
\end{eqnarray}
We see that, for the screened Newtonian potential ($k_0\neq 0$), the instability occurs for larger wavelengths as compared to the Newtonian model ($k_0=0$). Let us introduce the notation
\begin{eqnarray}
k_B T_c=\frac{S_d G m\rho}{k_0^2},
\label{s30}
\end{eqnarray}
which corresponds to the critical temperature in an infinite domain ($k_f=0$). Since the dispersion relation (\ref{s23}) does not explicitly depend on $k_f$, it is convenient to introduce the notation (\ref{s30}). We have
\begin{eqnarray}
T_c^*=\frac{T_c}{1+(k_f/k_0)^2}.
\label{s31}
\end{eqnarray}
When $T<T_c^*$,  the system is unstable for the modes such that
\begin{eqnarray}
k<k_0\left (\frac{T_c}{T}-1\right )^{1/2}\equiv k_m(T).
\label{s32}
\end{eqnarray}
The growth rate can be written
\begin{eqnarray}
\sigma_k=\frac{k_B T}{m}\frac{k^2(k_m(T)^2-k^2)}{k_0^2+k^2},
\label{s33}
\end{eqnarray}
where
\begin{eqnarray}
k_m(T)=k_0\left (\frac{T_c}{T}-1\right )^{1/2}.
\label{s33b}
\end{eqnarray}
It achieves its maximum value for $k=k_*(T)$ where
\begin{eqnarray}
k_*(T)=k_0\left \lbrack \left (\frac{T_c}{T}\right )^{1/2}-1\right\rbrack^{1/2}.
\label{s34}
\end{eqnarray}
The corresponding value of the growth rate is
\begin{eqnarray}
\sigma_*(T)=\frac{k_B T_c}{m} k_0^2 \left \lbrack 1-\left (\frac{T}{T_c}\right )^{1/2}\right\rbrack^{2}.
\label{s35}
\end{eqnarray}
The number of clusters that is expected to form in the linear regime is $N(T)=R/(2\pi/k_*(T))$. For a fixed value of $k_0$, this number increases as the temperature decreases. The behaviour of the different quantities defined above is represented in Figs. \ref{sigmak} and \ref{kT}.

\begin{figure}[!ht]
\begin{center}
\includegraphics[clip,scale=0.3]{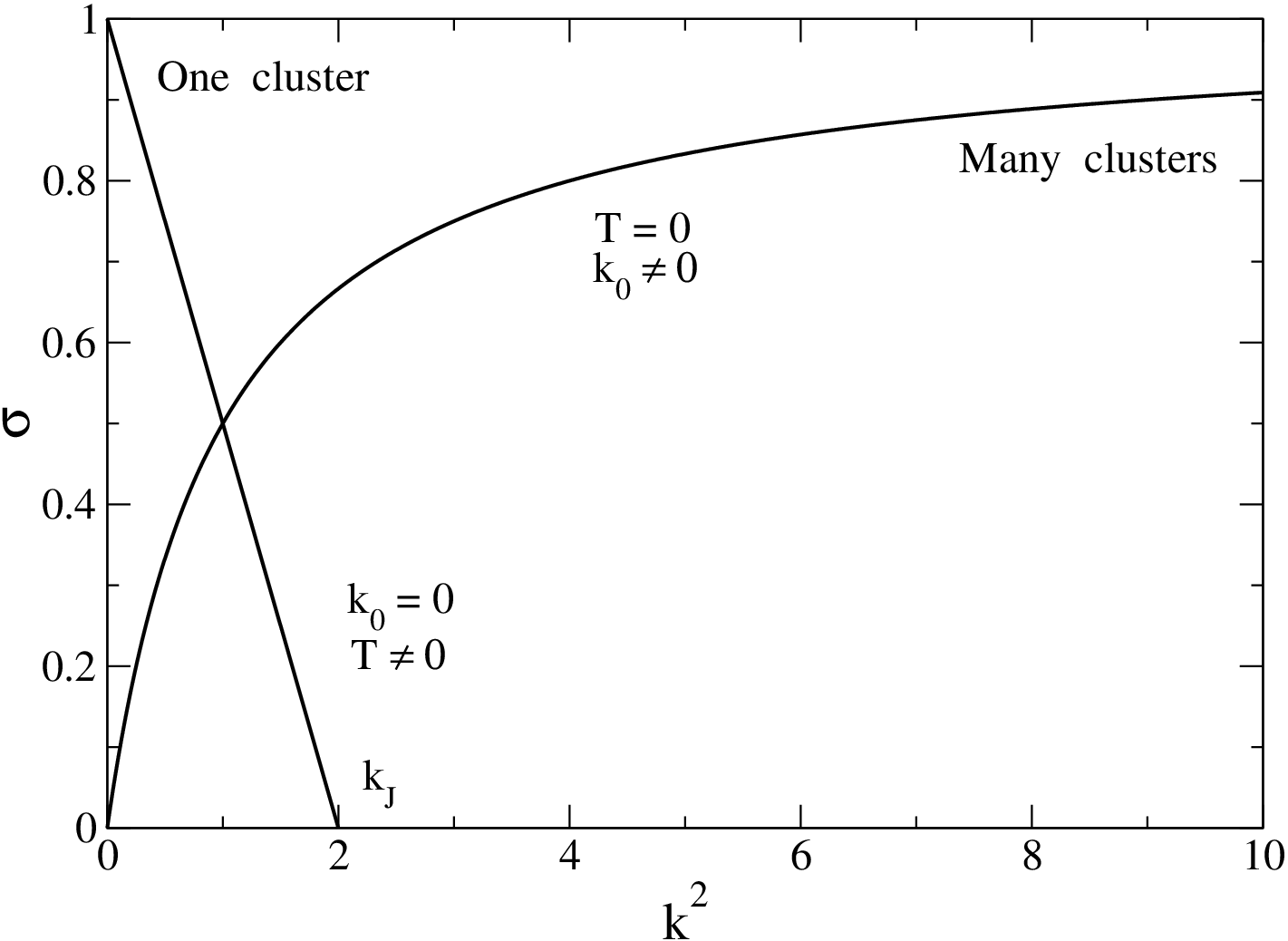}
\caption{Growth ($\sigma>0$) rate as a function of the wavenumber $k$ in two limits: (i) the Newtonian limit $k_0=0$ (and $T\neq 0$) for which the maximum growth rate corresponds to $k_*=k_f<<1$ (large scales), and (ii) the cold limit $T=0$ (and $k_0\neq 0$) for which the maximum growth rate corresponds to $k_*\rightarrow +\infty$ (small scales).}
\label{sigmaklimite}
\end{center}
\end{figure}

Let us consider some particular cases:

$\bullet$ For $T=T_c$, we have $k_m=0$, $k_*=0$, $\sigma_*=0$ and
\begin{eqnarray}
\sigma_k=-\frac{k_B T_c}{m}\frac{k^4}{k_0^2+k^2}<0.
\label{s36}
\end{eqnarray}
Therefore, the system is stable. More generally, for $T\ge T_c$, the system is stable. For $T\rightarrow +\infty$, we have $\sigma_k=-\frac{k_B T}{m}k^2$.

$\bullet$ For $T=0$, we have $k_m\rightarrow +\infty$, $k_*\rightarrow +\infty$, $\sigma_*\rightarrow k_B T_c k_0^2/m$ and
\begin{eqnarray}
\sigma_k=\frac{k_B T_c}{m}\frac{k_0^2k^2}{k_0^2+k^2}.
\label{s37}
\end{eqnarray}
The growth rate is maximum for $k_*\rightarrow +\infty$, i.e. for very small wavelengths $\lambda_*\rightarrow 0$. In that case, we expect a very large number of clusters in the linear regime.

$\bullet$ For $k_0=0$ (modified Newtonian model), we have
\begin{eqnarray}
\sigma_k=S_d G\rho-\frac{k_B T}{m}k^2.
\label{s38}
\end{eqnarray}
The system is unstable for $T<T_c^*$ where the critical temperature is given by Eq. (\ref{s27}). Furthermore,  the unstable wavenumbers correspond to $k<k_J$ where the Jeans wavenumber is given by Eq. (\ref{s28}). The growth rate is maximum for $k_*=k_f$ i.e. for the maximum wavelength $\lambda_f=2\pi/k_f$.  In that case, we have only one cluster. The corresponding value of the growth rate is $\sigma_*=S_d G\rho-k_B Tk_f^2/m$.

The two limit cases discussed above are illustrated in Fig. \ref{sigmaklimite}.

\subsection{Thermodynamical stability}
\label{sec_d}

We now analyze the thermodynamical stability of the homogeneous phase by using variational principles. Basically, we have to solve the maximization problem  (\ref{mce1}) in MCE and the minimization problem  (\ref{ce1}) in CE. However, for spatially homogeneous systems, it is shown in Appendix \ref{sec_simpler} that they are both {\it equivalent}  to the minimization problem  (\ref{ce6}).  Therefore, the system is stable iff the second order variations of free energy (\ref{ce3b}) are positive definite for any perturbations $\delta\rho$ that conserve mass, i.e. $\int\delta\rho\, d{\bf r}=0$. We are led therefore to considering the eigenvalue problem
\begin{eqnarray}
\delta\Phi_{\lambda}+\frac{k_B T}{\rho m}\delta\rho_\lambda=\lambda\delta\rho_\lambda,
\label{d1}
\end{eqnarray}
\begin{eqnarray}
\Delta \delta\Phi_\lambda-k_{0}^{2}\delta\Phi_\lambda =S_d G \delta\rho_\lambda.
\label{d2}
\end{eqnarray}
If all the eigenvalues $\lambda$ are positive, then the system is stable since $\delta^2 F=\frac{1}{2}\sum_\lambda \lambda a_\lambda^2>0$ where the perturbation has been decomposed under the form $\delta\rho=\sum_\lambda a_\lambda\delta\rho_\lambda$. If at least one eigenvalue is negative, the system is unstable since $\delta^2 F=\frac{1}{2}\lambda\int (\delta\rho_\lambda)^2\, d{\bf r}<0$ for that perturbation. It is easy to see that the eigenfunctions are
\begin{eqnarray}
\delta\rho({\bf r})=A_{k}\psi_k({\bf r}),\quad \delta\Phi({\bf r})=-\frac{S_d GA_k}{k^2+k_0^2}\psi_k({\bf r}),
\label{d3}
\end{eqnarray}
and that the corresponding eigenvalues are
\begin{eqnarray}
\lambda(k)=-\frac{S_d G}{k^2+k_0^2}+\frac{k_B T}{\rho m},
\label{d4}
\end{eqnarray}
for all quantized $k$ (see Sec. \ref{sec_s}). We note that $\int \psi_k\, d{\bf r}=-\frac{1}{k^2}\int \Delta\psi_k\, d{\bf r}=-\frac{1}{k^2}\oint \nabla\psi_k\, \cdot d{\bf S}=0$, so that $\int\delta\rho\, d{\bf r}=0$ as required. Regrouping all these results, we conclude that the system is stable iff
\begin{eqnarray}
\frac{S_d G}{k^2+k_0^2}-\frac{k_B T}{\rho m}<0,
\label{d5}
\end{eqnarray}
for all (quantized) $k$. This returns the stability condition obtained in Sec. \ref{sec_s}. Therefore, the system is stable iff $T>T_c^*$. If $T<T_c^*$, the homogeneous phase is an unstable saddle point of free energy at fixed mass.  This method proves the thermodynamical stability of the homogeneous phase in the canonical and microcanonical ensembles. This implies the stability with respect to the mean field Kramers equation (\ref{b2}), with respect to the Smoluchowski equation (\ref{b6}), with respect to the Keller-Segel model (\ref{ks1}) and with respect to the kinetic  equation (\ref{h1}).

We can now determine the values of the normalized inverse temperature $\eta_c^*$ and normalized energy $\Lambda_c^*$ above which the homogeneous phase becomes unstable. Using Eqs. (\ref{mn9}) and (\ref{s25}), we get
\begin{eqnarray}
\eta_c^*=\frac{1}{d}(k_f^2+k_0^2)R^2.
\label{d6}
\end{eqnarray}
We obtain
\begin{eqnarray}
\eta_c^*=\pi^2+\mu^2=9.8696044+\mu^2\qquad (d=1),
\label{d7}
\end{eqnarray}
\begin{eqnarray}
\eta_c^*=\frac{1}{2}(j_{11}^2+\mu^2)=7.3410008+\frac{\mu^2}{2}\qquad (d=2),
\label{d8}
\end{eqnarray}
\begin{eqnarray}
\eta_c^*=\frac{1}{3}(x_{1}^2+\mu^2)=6.7302445+\frac{\mu^2}{3}\qquad (d=3).
\label{d9}
\end{eqnarray}
The corresponding critical energy is given by Eq. (\ref{mn20}) for the modified Newtonian model and by (\ref{ay24}) for the screened Newtonian model.

\section{Conclusion}

In this paper, we have completed the description of phase transitions
in self-gravitating systems and bacterial populations. We have
introduced generalized models in which the ordinary Poisson
equation is modified so as to allow for the existence of a spatially
homogeneous phase. This avoids the Jeans swindle and leads to a great
variety of microcanonical and canonical phase transitions between
homogeneous and inhomogeneous states. These generalized models can
have application in chemotaxis where the degradation of the chemical
leads to a shielding of the interaction and in cosmology where the
expansion of the universe creates a sort of ``neutralizing
background''. In this paper, we have only considered equilibrium
states. In future works, we shall study the dynamics of some simple
models for which the present study can be a useful guide.

Our study also allows to explore the link between cosmology where one
studies the evolution of the universe as a whole \cite{peebles} and
stellar dynamics where one studies the structure of individual galaxies
\cite{bt}. The description of phase transitions in these two
disciplines is usually very different \cite{pad}. However, our study
allows to make some basic connections.  In cosmology, one usually
starts from an infinite homogeneous distribution and study the
appearance of clusters representing galaxies. Our thermodynamical
approach shows that, indeed, the homogeneous phase is unstable for
sufficiently low temperatures and energies and leads to clusters. The
formation of these clusters can be studied by making a linear
stability analysis of the Vlasov or Euler equations. Then, in the
nonlinear regime, the system is expected to achieve a statistical
equilibrium state due to violent relaxation or collisional relaxation
(finite $N$ effects) \footnote{This is, in fact, just a
quasistationary state that forms on a timescale that is short with
respect to the Hubble time so that the expansion of the universe can
be neglected or treated adiabatically. Indeed, if we allow for the
time variation of the scale factor $a(t)$, it is simple to see that
there is no statistical equilibrium state in a strict sense.}. This
corresponds to the inhomogeneous phase.  In $d=1$, there exists an
equilibrium state for any value of energy and temperature. For low
energies and temperatures, it is spatially inhomogeneous. In the core
of the cluster, the density is so high that we can disregard the
effect of the neutralizing background. In that case, the statistical
equilibrium state (representing a ``galaxy'') is described by the Camm
solution like in 1D stellar dynamics.  In $d=3$, there is no
inhomogeneous equilibrium state and, for sufficiently small energies
and temperatures, the system undergoes a gravothermal catastrophe or
an isothermal collapse. In $d=2$, the situation is intermediate. There
exists an equilibrium state in the microcanonical ensemble for all
energies while in the canonical ensemble no equilibrium state exists
at low temperatures. Similar behaviours occur in chemotaxis and will
be investigated in future papers. Note that for self-gravitating
systems, the proper statistical ensemble is the microcanonical
ensemble while in chemotaxis (or for the academic model of
self-gravitating Brownian particles) the proper statistical ensemble
is the canonical one. It is therefore interesting to study these two
systems in parallel to describe the analogies and differences between
statistical ensembles.

\appendix

\section{Equivalent but simpler optimization  problems}
\label{sec_simpler}

In this Appendix, following the approach of Padmanabhan \cite{paddy} and Chavanis \cite{aaiso}, we shall reduce the optimization problems (\ref{mce1}) and (\ref{ce1}) to simpler forms. In particular, we shall show that these optimization problems for $f({\bf r},{\bf v})$ are equivalent to optimization problems for $\rho({\bf r})$.

\subsection{Microcanonical ensemble}

To solve the maximization problem (\ref{mce1}) we can proceed in two steps. We first maximize the entropy at fixed energy, mass {\it and}  density profile $\rho({\bf r})$. Since the specification of $\rho({\bf r})$ determines the mass and the potential energy, this is equivalent to maximizing the entropy at fixed kinetic energy and density profile. Writing
\begin{eqnarray}
\label{mce8b}
\delta S-\beta\delta\left (\int f\frac{v^2}{2}\, d{\bf r}d{\bf v}\right )-\int\lambda({\bf r})\delta\left (\int f\, d{\bf v}\right )\, d{\bf r}=0,\nonumber\\
\end{eqnarray}
this leads to the Maxwellian distribution function
\begin{eqnarray}
\label{mce9}
f({\bf r},{\bf v})=\left (\frac{m}{2\pi k_B T}\right )^{d/2} \, \rho({\bf r})\, e^{-\frac{mv^2}{2k_B T}},
\end{eqnarray}
which is the global entropy maximum with the previous constraints since $\delta^2 S=-\int\frac{(\delta f)^2}{2fm}\, d{\bf r}d{\bf v}<0$ (the constraints are linear in $f$ so that their second variations vanish).
We can now express the mass, the entropy and the energy in terms of $\rho({\bf r})$ and $T$. Up to unimportant constants, we obtain
\begin{eqnarray}
\label{mce10}
S=\frac{d}{2}Nk_B \ln T-k_B\int \frac{\rho}{m}\ln\frac{\rho}{m}\, d{\bf r},
\end{eqnarray}
\begin{eqnarray}
\label{mce11}
E=\frac{d}{2}Nk_B T+\frac{1}{2}\int \rho({\bf r},t) u({\bf r},{\bf r}')\rho({\bf r}',t)\, d{\bf r}d{\bf r}'\nonumber\\
+\int \rho V\, d{\bf r}.\qquad\qquad
\end{eqnarray}
We now have to solve the maximization problem
\begin{eqnarray}
\label{mce13}
\max_\rho\left\lbrace S\lbrack \rho\rbrack\, |\, E\lbrack \rho\rbrack=E, \, M\lbrack \rho\rbrack=M\right\rbrace.
\end{eqnarray}
Finally, the solution of (\ref{mce1}) is given by the distribution function (\ref{mce9}) with the density profile that is solution of (\ref{mce13}). Let us compute the variations of entropy and energy up to second order. We have
\begin{eqnarray}
\Delta S=\frac{d}{2}Nk_B \frac{\delta T}{T}-k_B\int \left (\ln\frac{\rho}{m}+1\right )\frac{\delta\rho}{m}\, d{\bf r}\nonumber\\
-\frac{d}{4}Nk_B\left (\frac{\delta T}{T}\right )^2-k_B \int \frac{(\delta\rho)^2}{2\rho m}\, d{\bf r},
\end{eqnarray}
\begin{eqnarray}
\Delta E=\frac{d}{2}Nk_B \delta T+\int \Phi\delta\rho \, d{\bf r}+\frac{1}{2}\int\delta\rho\delta\Phi\, d{\bf r}.
\end{eqnarray}
Using the conservation of energy $\Delta E=0$ to eliminate $\delta T$, we obtain
\begin{eqnarray}
\Delta S=-\frac{1}{T}\int \Phi\delta\rho \, d{\bf r}-k_B\int \left (\ln\frac{\rho}{m}+1\right )\frac{\delta\rho}{m}\, d{\bf r}\nonumber\\
-\frac{1}{2T}\int\delta\rho\delta\Phi\, d{\bf r}-\frac{1}{dNk_B T^2}\left (\int \Phi\delta\rho\, d{\bf r}\right )^2\nonumber\\
-k_B \int \frac{(\delta\rho)^2}{2\rho m}\, d{\bf r}.
\label{DS}
\end{eqnarray}
Let us  consider the first order variations. At first order, we have
\begin{eqnarray}
\label{mce16}
\delta S=-\frac{1}{T}\int \Phi\delta\rho \, d{\bf r}-k_B\int \left (\ln\frac{\rho}{m}+1\right )\frac{\delta\rho}{m}\, d{\bf r}.
\end{eqnarray}
The conservation of mass can be  taken into account by introducing a Lagrange multiplier. Writing the variational principle as
\begin{eqnarray}
\label{mce17}
\delta S-\alpha\delta M=0,
\end{eqnarray}
we obtain the mean field Boltzmann distribution
\begin{eqnarray}
\label{mce18bw}
\rho=A'e^{-\frac{m\Phi}{k_B T}},
\end{eqnarray}
where $\Phi({\bf r})$ is given by Eq. (\ref{mce7}). Combining Eq. (\ref{mce18bw}) with Eq. (\ref{mce9}), we recover the mean field  Maxwell-Boltzmann distribution (\ref{mce6}). However, the present approach allows us to simplify the condition of thermodynamical stability. Indeed, the system is stable in the microcanonical ensemble  iff the second order variations of entropy (\ref{DS}) are negative definite
\begin{eqnarray}
\label{mce18b}
-k_B \int \frac{(\delta\rho)^2}{2\rho m}\, d{\bf r}-\frac{1}{2T}\int\delta\rho\delta\Phi\, d{\bf r}\nonumber\\
-\frac{1}{dNk_B T^2}\left (\int \Phi\delta\rho\, d{\bf r}\right )^2\le 0,
\end{eqnarray}
for any perturbation $\delta\rho$ that conserves mass at first order, i.e. $\int\delta\rho\, d{\bf r}=0$ (the conservation of energy has automatically been taken into account in the previous derivation). This stability criterion is equivalent to the stability criterion (\ref{mce8}) but it is simpler because it is expressed in terms of the density instead of the distribution function.

\subsection{Canonical ensemble}

To solve the maximization problem (\ref{ce1}) we can proceed in two steps. We first minimize the free energy at fixed mass {\it and} density profile $\rho({\bf r})$. This is equivalent to minimizing the free energy  at fixed density profile. Writing
\begin{eqnarray}
\label{mce8c}
\delta F+T\int\lambda({\bf r})\delta\left (\int f\, d{\bf v}\right )\, d{\bf r}=0,
\end{eqnarray}
this leads to the Maxwellian distribution function
\begin{eqnarray}
\label{ce4}
f({\bf r},{\bf v})=\left (\frac{m}{2\pi k_B T}\right )^{d/2} \, \rho({\bf r})\, e^{-\frac{mv^2}{2k_B T}},
\end{eqnarray}
which is the  global minimum of free energy with the previous constraint since $\delta^2 F=T\int\frac{(\delta f)^2}{2fm}\, d{\bf r}d{\bf v}>0$ (the constraints are linear in $f$ so that their second variations vanish). We can now express the free energy in terms of $\rho({\bf r})$. Up to unimportant constants, we get
\begin{eqnarray}
\label{ce5}
F=\frac{1}{2}\int \rho({\bf r},t) u({\bf r},{\bf r}')\rho({\bf r}',t)\, d{\bf r}d{\bf r}'\nonumber\\
+\int \rho V\, d{\bf r}+k_B T\int \frac{\rho}{m}\ln\frac{\rho}{m}\, d{\bf r}.
\end{eqnarray}
We now have to solve the minimization problem
\begin{eqnarray}
\label{ce6b}
\min_\rho\left\lbrace F\lbrack \rho\rbrack\, |\,  M\lbrack \rho\rbrack=M\right\rbrace.
\end{eqnarray}
Finally, the solution of (\ref{ce1}) is given by the distribution function (\ref{ce4}) with the density profile that is solution of (\ref{ce6b}). The first variations
\begin{eqnarray}
\delta F+\alpha T\delta M=0,
\end{eqnarray}
lead to the mean field Boltzmann distribution
\begin{eqnarray}
\label{mce18bza}
\rho=A'e^{-\frac{m\Phi}{k_B T}},
\end{eqnarray}
where $\Phi({\bf r})$ is given by Eq. (\ref{mce7}). Combining Eq. (\ref{mce18bza}) with Eq. (\ref{ce4}), we recover the mean field  Maxwell-Boltzmann distribution (\ref{mce6}). However, the present approach allows us to simplify the condition of thermodynamical stability. Indeed, the system is stable in the canonical ensemble  iff the second order variations of free energy (\ref{ce5}) are positive definite
\begin{eqnarray}
\label{mce18c}
\frac{1}{2}\int\delta\rho\delta\Phi\, d{\bf r}+\frac{k_B T}{m} \int \frac{(\delta\rho)^2}{2\rho}\, d{\bf r}\ge 0,
\end{eqnarray}
for any perturbation $\delta\rho$ that conserves mass at first order, i.e. $\int\delta\rho\, d{\bf r}=0$. This stability criterion is equivalent to the stability criterion (\ref{ce3bq}) but it is simpler because it is expressed in terms of the density instead of the distribution function.

{\it Remark:} From the stability criteria (\ref{mce18b}) and (\ref{mce18c}), we clearly see that canonical stability implies microcanonical stability (but not the converse). Indeed, since the last term in Eq. (\ref{mce18b}) is negative, it is clear that if inequality (\ref{mce18c}) is satisfied then inequality (\ref{mce18b}) is automatically satisfied. In general, this is not reciprocal and we may have ensembles inequivalence. However, if we consider a spatially homogeneous system for which $\Phi$ is uniform, the last term in Eq. (\ref{mce18b}) vanishes (since the mass is conserved) and the stability criteria (\ref{mce18b}) and (\ref{mce18c}) coincide. Therefore, for spatially homogeneous systems, we have ensembles equivalence.

\section{Explicit expressions of the potential}
\label{sec_exp}

In this Appendix, we limit ourselves to the case $d=1$ although the results can be easily generalized to any dimension.  Using standard methods, we can obtain the Green function associated with the screened Poisson equation (\ref{ay1}) in a box with Neumann boundary conditions. Then, we find that the potential is explicitly given by
\begin{eqnarray}
\label{exp1}
\Phi(x)=\int_{-R}^{R}\rho(x')u(x,x')\, dx'
\end{eqnarray}
with
\begin{eqnarray}
\label{exp2}
&&u(x,x')=-\frac{G}{k_0}\frac{1}{\sinh(2k_0R)}\nonumber\\
&\times& \left (\cosh\left\lbrack k_0 (2R-|x-x'|)\right\rbrack+\cosh\left\lbrack k_0(x+x')\right\rbrack\right ).\nonumber\\
\end{eqnarray}
In an infinite domain ($R\rightarrow +\infty$), we obtain
\begin{eqnarray}
\label{exp3}
u(|x-x'|)=-\frac{G}{k_0}e^{-k_0|x-x'|}.
\end{eqnarray}
Similarly, for the modified Newtonian model (\ref{mn1}) with Neumann boundary conditions, the potential is explicitly given by
\begin{eqnarray}
\label{exp4}
\Phi(x)=G\int_{-R}^{R}(\rho-\overline{\rho})(x')|x-x'|\, dx',
\end{eqnarray}
and this expression remains valid in an infinite domain ($R\rightarrow +\infty$).

\section{External potential for the modified Newtonian model}
\label{sec_ext}

For the modified Newtonian model (\ref{mn1}), the potential can be written
\begin{eqnarray}
\label{ext1}
\Phi({\bf r})=\int (\rho-\overline{\rho})({\bf r}') u({\bf r},{\bf r}')\, d{\bf r}',
\end{eqnarray}
where $u({\bf r},{\bf r}')$ is the Green function of the Poisson equation with Neumann boundary conditions.
Comparing Eq. (\ref{ext1}) with Eq. (\ref{mce7}), we find that the external potential is
\begin{eqnarray}
\label{ext2}
V({\bf r})=-\overline{\rho}\int u({\bf r},{\bf r}')\, d{\bf r}'.
\end{eqnarray}
Using Eq. (\ref{ext2}), the potential energy (\ref{h5}) can be written
\begin{eqnarray}
\label{ext3}
W=\frac{1}{2}\int \rho\Phi\, d{\bf r}-\frac{1}{2}\overline{\rho}\int \rho({\bf r})u({\bf r},{\bf r}')\, d{\bf r}d{\bf r}'.
\end{eqnarray}
Interchanging the dummy variables ${\bf r}$ and ${\bf r}'$ and using the symmetry  $u({\bf r}',{\bf r})=u({\bf r},{\bf r}')$, we get
\begin{eqnarray}
\label{ext4}
W=\frac{1}{2}\int \rho\Phi\, d{\bf r}-\frac{1}{2}\overline{\rho}\int \rho({\bf r}')u({\bf r},{\bf r}')\, d{\bf r}d{\bf r}'.
\end{eqnarray}
Finally, using Eq. (\ref{mce7}), we obtain
\begin{eqnarray}
\label{ext5}
W=\frac{1}{2}\int (\rho-\overline{\rho})\Phi\, d{\bf r}+\frac{1}{2}\overline{\rho}\int V({\bf r})\, d{\bf r}.
\end{eqnarray}
Therefore, the potential energy is given by Eq. (\ref{mn11}) up to an unimportant additive constant
$\frac{1}{2}\overline{\rho}\int V({\bf r})\, d{\bf r}=-\frac{1}{2}\overline{\rho}^2\int u({\bf r},{\bf r}')d{\bf r}d{\bf r}'$.

In $d=1$, according to Eq. (\ref{exp4}), the potential of interaction is $u=G|x-x'|$. Therefore, the external potential is explicitly given by
\begin{eqnarray}
\label{ext6}
V(x)=-\overline{\rho}G(x^2+R^2).
\end{eqnarray}
The additive constant in the energy is
\begin{eqnarray}
\label{ext7}
\frac{1}{2}\overline{\rho}\int_{-R}^{R} V(x)\, dx=-\frac{4}{3}G\overline{\rho}^2 R^3.
\end{eqnarray}

\section{The minimum energy}
\label{sec_me}

Let us consider the modified Newtonian model (\ref{mn1}) in $d=1$. At $T=0$, the density profile is a Dirac peak $\rho=M\delta(x)$ and the energy (\ref{mn14}) is
\begin{eqnarray}
\label{me1}
E=-\frac{1}{4G}\int_{-R}^R \left (\frac{d\Phi}{dx}\right )^2\, dx.
\end{eqnarray}
For a symmetric density profile, the modified Poisson equation can be integrated into
\begin{eqnarray}
\label{me2}
\Phi'(x)=2G\int_0^x \rho(x')\, dx'-2G\overline{\rho}x,
\end{eqnarray}
which is the appropriate Gauss theorem. If all the mass is concentrated at $x=0$, we obtain
\begin{eqnarray}
\label{me3}
\Phi'(x)=GM\left ({\rm sign}(x)-\frac{x}{R}\right ).
\end{eqnarray}
Substituting this expression in Eq. (\ref{me1}), we obtain $E=-GM^2R/6$. The total normalized energy is therefore
\begin{eqnarray}
\label{me4}
\Lambda_{max}^{(1)}=\frac{1}{6}.
\end{eqnarray}
This corresponds to the minimum energy of the branch $n=1$.

Let us consider the screened Newtonian model (\ref{ay1}) in $d=1$. At $T=0$, the density profile is a Dirac peak $\rho=M\delta(x)$ and the energy  (\ref{ay11}) is $E=\frac{1}{2}M\Phi_0$. According to Eq. (\ref{exp2}), the potential created in $x$  by a mass $M$ located at $x'=0$ is
\begin{eqnarray}
\label{me5}
\Phi(x)=-\frac{GM}{k_0}\frac{\cosh\left\lbrack k_0 (R-|x|)\right\rbrack}{\sinh (k_0 R)},
\end{eqnarray}
where we have used elementary trigonometric identities to simplify the expression. This leads to $\Phi_0=-GM/(k_0\tanh(k_0 R))$ and $E=-GM^2/(2k_0\tanh(k_0 R))$. The total normalized energy is therefore
\begin{eqnarray}
\label{me6}
\Lambda_{max}^{(1)}=\frac{1}{2\mu\tanh(\mu)}.
\end{eqnarray}
This corresponds to the minimum energy of the branch $n=1$.

\section{Approximate expressions of the density profile}
\label{sec_app}

In $d=1$, the screened Emden equation (\ref{ay4}) can be written
\begin{eqnarray}
\label{app1}
\frac{d^2\psi}{d\xi^2}=e^{-\psi}-\lambda+\kappa^2\psi=-\frac{dV}{d\psi},
\end{eqnarray}
with
\begin{eqnarray}
\label{app2}
V(\psi)=e^{-\psi}+\lambda\psi-\frac{1}{2}\kappa^2\psi^2.
\end{eqnarray}
This is similar to the equation of motion of a particle of unit mass in a potential $V(\psi)$ where $\psi$ plays the role of position and $\xi$ the role of time. Using the boundary condition $\psi=\psi'=0$ at $\xi=0$, we find that the first integral (pseudo energy) is
\begin{eqnarray}
\label{app3}
{\cal E}\equiv \frac{1}{2}\left (\frac{d\psi}{d\xi}\right )^2+V(\psi)=1.
\end{eqnarray}
This first order differential equation can be easily integrated until $\xi=\alpha$, which formally solves the problem.

Let us consider the limit $\rho_0\rightarrow +\infty$ corresponding to $\alpha\rightarrow +\infty$. In the inner region, the term $e^{-\psi}$ dominates and Eq. (\ref{app1}) reduces to the ordinary Emden equation whose solution is the Camm profile \cite{camm,sc}:
\begin{eqnarray}
\label{app4}
e^{-\psi}=\frac{1}{\cosh^2(\xi/\sqrt{2})}.
\end{eqnarray}
In the outer region, the term  $e^{-\psi}$ can be neglected and Eqs. (\ref{app1}) and (\ref{app3})
reduce to
\begin{eqnarray}
\label{app5}
\frac{d^2\psi}{d\xi^2}=-\lambda+\kappa^2\psi,
\end{eqnarray}
\begin{eqnarray}
\label{app6}
\frac{1}{2}\left (\frac{d\psi}{d\xi}\right )^2+\lambda\psi-\frac{1}{2}\kappa^2\psi^2=1.
\end{eqnarray}
The boundary condition at the wall is $\psi'(\alpha)=0$. Substituting this result in  Eq. (\ref{app6}), we get $\lambda\psi(\alpha)-\frac{1}{2}\kappa^2\psi(\alpha)^2=1$. The physical solution of this equation is $\psi(\alpha)=(\lambda-\sqrt{\lambda^2-2\kappa^2})/\kappa^2$. Solving Eq. (\ref{app5}) with these boundary conditions, we find that
\begin{eqnarray}
\label{app7}
\psi(\xi)=\frac{\lambda}{\kappa^2}-\frac{1}{\kappa^2}\sqrt{\lambda^2-2\kappa^2}\cosh\left\lbrack\kappa (\xi-\alpha)\right\rbrack.
\end{eqnarray}
The matching of the outer solution with the inner solution implies that $\psi_{outer}(0)=0$. Using $\kappa\alpha=\mu$, we obtain
\begin{eqnarray}
\label{app8}
\lambda\sim \frac{\sqrt{2}\mu}{\tanh(\mu)}\frac{1}{\alpha},\qquad (\alpha\rightarrow +\infty).
\end{eqnarray}
Finally, substituting the inner profile (\ref{app4}) in Eq. (\ref{ay10}), we obtain at leading order
\begin{eqnarray}
\label{app9}
\eta\sim \sqrt{2}\alpha, \qquad (\alpha\rightarrow +\infty).
\end{eqnarray}
For $\alpha\rightarrow +\infty$, the density profile tends to a Dirac peak $\rho=M\delta(x)$. The potential energy reduces to $W=\frac{1}{2}M\Phi_0$. Using Eqs. (\ref{ay16}), (\ref{app8}) and $\kappa=\mu/\alpha$, we recover Eq. (\ref{me6}).

The modified Emden equation (\ref{mn4}) can be studied similarly. In fact, most of the preceding results remain valid by taking $\kappa=0$. The potential is $V(\psi)=e^{-\psi}+\lambda\psi$. It has a minimum at $\psi_0=-\ln\lambda$ so that the solution $\psi(\xi)$ of the Emden equation (with energy ${\cal E}=1$)  oscillates around this value. Integrating Eq. (\ref{app3}), the density profiles of the solutions of the branch $n=1$ are given by
\begin{eqnarray}
\label{app10}
\xi=\int_0^\psi \frac{dx}{\sqrt{2(1-e^{-x}-\lambda x)}},
\end{eqnarray}
with $\xi\le\alpha$. The half-period of the oscillations of the function $\psi(\xi)$ is
\begin{eqnarray}
\label{app11}
\frac{L}{2}=\int_0^{\psi(\alpha)} \frac{dx}{\sqrt{2(1-e^{-x}-\lambda x)}},
\end{eqnarray}
where $\psi(\alpha)$ is solution of $e^{-\psi(\alpha)}+\lambda\psi(\alpha)=1$ obtained from Eq. (\ref{app3}) with $\psi'(\alpha)=0$. Let us now consider the limit $\rho_0\rightarrow +\infty$. The inner solution is given by the Camm profile (\ref{app4}) and the outer solution is
\begin{eqnarray}
\label{app12}
\psi(\xi)=\frac{1}{\lambda}-\frac{1}{2}\lambda (\xi-\alpha)^2,
\end{eqnarray}
which is consistent with Eq. (\ref{app7}) when $\kappa\rightarrow 0$. The matching condition $\psi_{outer}(0)=0$ then yields
\begin{eqnarray}
\label{app13}
\lambda\sim \frac{\sqrt{2}}{\alpha},\qquad (\alpha\rightarrow +\infty).
\end{eqnarray}
Using Eq. (\ref{mn10}), we obtain at leading order
\begin{eqnarray}
\label{app14}
\eta\sim \sqrt{2}\alpha, \qquad (\alpha\rightarrow +\infty).
\end{eqnarray}

\section{The bifurcation point}
\label{sec_bif}

In this Appendix, we shall determine the point at which  the spatially homogeneous branch bifurcates to the spatially inhomogeneous branch and show that it coincides with the point at which the spatially homogeneous branch becomes unstable (see Sec. \ref{sec_stab}). For a more detailed theory of bifurcations, we refer to the paper of Schaff \cite{schaaf}.

For the modified Newtonian model, the differential equation determining the field $\Phi({\bf r})$ at statistical equilibrium can be written
\begin{eqnarray}
\label{bif1}
\Delta\Phi=S_d G\left (A e^{-\beta m\Phi}-\overline{\rho}\right ).
\end{eqnarray}
The homogeneous solution corresponds to  $\rho=\overline{\rho}$, $\Phi=0$ and $A={\rho}$. Considering a small perturbation $\Phi=0+\phi({\bf r})$ with $\phi\ll 1$ around the homogeneous solution and linearizing the differential equation (\ref{bif1}), we obtain
\begin{eqnarray}
\label{bif2}
\Delta\phi+S_d G\beta m\rho\phi=0,
\end{eqnarray}
with the boundary conditions $\nabla\phi\cdot {\bf n}=0$ on the boundary. The boundary conditions determine the allowable wavenumbers $k^2\equiv S_d G\beta m\rho$. They take discrete values $k=k_n$ (see Sec. \ref{sec_stab}) which in turn determine discrete values of the temperature $T_n$. The first point of bifurcation corresponds to the smallest wavenumber $k_f$. This is associated with the critical temperature (\ref{s27}) at which the homogeneous branch becomes unstable. Other branches of bifurcations appear at smaller temperatures. They correspond to successive quantized values $k_n$ of the wavenumber.

For the screened Newtonian model, the differential equation determining the field $\Phi({\bf r})$ at statistical equilibrium can be written
\begin{eqnarray}
\label{bif3}
\Delta\Phi-k_0^2\Phi=S_d G A e^{-\beta m\Phi}.
\end{eqnarray}
The homogeneous solution corresponds to  $\rho={\rm const.}$, $\Phi={\rm const.}$ with $-k_0^2\Phi=S_d G\rho$. Considering a small perturbation $\Phi={\rm const.}+\phi({\bf r})$  with $\phi\ll 1$  around the homogeneous solution and linearizing the differential equation (\ref{bif3}), we obtain
\begin{eqnarray}
\label{bif4}
\Delta\phi+(S_d G\beta m\rho-k_0^2)\phi=0,
\end{eqnarray}
with the boundary conditions $\nabla\phi\cdot {\bf n}=0$ on the boundary. The boundary conditions determine the allowable wavenumbers $k^2\equiv S_d G\beta m\rho-k_0^2$. The first point of bifurcation corresponds to the smallest wavenumber $k_f$ (see Sec. \ref{sec_stab}). This is associated with the critical temperature (\ref{s25}) at which the homogeneous branch becomes unstable. Other branches of bifurcations appear at smaller temperatures. They correspond to the successive quantized values $k_n$ of the wavenumber.

\end{document}